# Perfect Secrecy
# under Deep Random assumption

Thibault de Valroger [(*)]

## Abstract

We present a new idea to design information theoretically secure key exchange protocol, based on a new form of randomness, called « Deep Randomness », generated in such a way that probability distribution of the output signal is made unknowledgeable for an observer. By limiting, thanks to Deep Randomness, the capacity of the opponent observer to perform Bayesian inference over public information to estimate private information, we can design protocols, beyond Shannon limit, enabling two legitimate partners, sharing originally no common private information, to exchange secret information with accuracy as close as desired from perfection, and knowledge as close as desired from zero by any unlimited passive opponent. We introduce the Deep Random assumption, based on Prior Probability theory developed by Jaynes. Deep Random assumption gives a rigorous rule to perform Bayesian inference under uncertainty situation. We also present a workable example of protocol with a mathematical proof for information theoretical security under Deep Random assumption. At last we present and develop a method that we present to generate Deep Randomness from classical computing resources.

Modern cryptography mostly relies on mathematical problems commonly trusted as very difficult to solve, such as large integer factorization or discrete logarithm, belonging to complexity theory. No certainty exist on the actual difficulty of those problems, not even the truth of the famous P≠NP conjecture, and, in addition, most of them are not resistant to quantum computing, which should make them useless in the next decades. Some other methods, rather based on information theory, have been developed since early 90's. Those methods relies on hypothesis about the opponent (such as « memory bounded » adversary [6]) or about the communication channel (such as « independent noisy channels » [5]) ; unfortunately, even if they have been proven Information Theoretically secure under given hypothesis, none of those hypothesis are easy to ensure in practice. At last, some other methods based on physical theories like quantum indetermination [3] or chaos generation have been described and experimented, but they are complex to implement, and, again, relies on solid but not proven and still partly understood theories.

(*) See contact and information about the author at last page

Considering this theoretically unsatisfying situation, we propose to explore a new path, where Information Theoretical security against unlimited opponent can be reached, without relying on any assumption about the opponent, that is supposed to have unlimited calculation and storage power, nor about the communication channel, that is supposed to be perfectly public, accessible and equivalent for any playing party (legitimate partners and opponents). The considered opponent is passive, which means that it does not interfere actively in the communication by suppressing, adding or modifying information exchanged between the legitimate partners; it just has a full access to it. Active opponent could be considered by adding authentication schemes, but this is not the core objective of the present work.

## I. Introduction

### Definition of "Information Theoretically Secure" Security Property

The most common Security Property to study Information Theoretically secure protocols is Perfect Secrecy, introduced by Claude Shannon [1]. This Security Property is defined as the condition $P(M|E) = P(M)$, where $M$ designates the private information or "Message" in the protocol, although $E$ designates the public information or "Encrypted information" in the protocol. As a security property, Perfect Secrecy has two drawbacks: (i) it is absolute and thus doesn't give rule to measure the secrecy as soon as $H(M) > H(M|E)$ (where $H(\cdot)$ designates Shannon's entropy), (ii) it does not explicitly differentiate the point of view of the opponent and the point of view of the legitimate receiver, and thus gives no rule to compare them in terms of capacity to access the private information as soon as the above inequality is strict.

In a very general view, in order to define the security property that we aim at satisfying, we consider here communication protocols that can be modelized as proposed by Csiszàr and Körner [15]. The legitimate partners are $A$ and $B$, and $\xi$ is the passive opponent. The purpose of the protocol is for $A$ and $B$ to reach common knowledge, with least possible errors of an output digital information, and to ensure that the passive opponent can learn the lesser possible information about that output digital information. By convention, the reference output digital information is the one reached by $A$ called the Secret Common Output information and denoted $X$. For that reason $B$ will be called the legitimate receiver, even if the protocol is not designed as a one way transmission. $X$ is a discrete random variable taking values over a finite alphabet $\mathcal{X}$. $A$ and $B$ communicate over a Discrete Memoryless Channel (DMC) called main channel and $B$ generates its best version, denoted $Y$, of the Secret Common Output information. $Y$ is a discrete random variable taking values over a finite alphabet $\mathcal{Y}$ (in most cases $\mathcal{Y} = \mathcal{X}$). The opponent $\xi$ receives exact or altered copy (depending on the security model) of all information exchanged on the main channel over another DMC called side channel or wire-tap channel, and then generates its best version of the Secret Common Output information denoted $Z$. $Z$ is a discrete random variable taking values over a finite alphabet $\mathcal{Z}$ (in most cases $\mathcal{Z} = \mathcal{X}$). The alphabets $\mathcal{X}, \mathcal{Y}, \mathcal{Z}$ take values in a sample space $F$. $\xi$'s objective is to obtain the best possible knowledge of $X$. The protocol is a predefined and public communication routine involving public parameters $U$; it is then denoted $\mathcal{P}(U)$. Each stakeholder in the protocol $(A,B,\xi)$ is an entity capable to Generate random bit strings, Publish bit strings on its DMC ($\xi$ is however supposed passive), Read bit strings published on its DMC, Store bit strings, Make calculation on bit strings.

Then, we define that the protocol is "Information Theoretically Secure" if, $\forall \epsilon > 0$, there exists a value of the parameter $U$ such that:

$$H(X|Z) - H(X|Y) \geq (1-\epsilon)H(X) \qquad (SP)$$

It is easy to see that this Security Property tends to Perfect Secrecy when $\epsilon \to 0$. Indeed, for the above inequality to hold when $\epsilon = 0$, it is necessary that (i) $H(X|Y) = 0$ which means that the protocol is errorless, and (ii) that $H(X|Z) = H(X)$ which means that $P(X|Z) = P(X)$. Csiszàr and Körner have shown that the secrecy capacity $C_s$ of the protocol is lower-bounded by:

$$C_s \geq \max_{P_X}\big(H(X|Z) - H(X|Y)\big)$$

which explains the rationale for the choice of $(SP)$ above to characterize the security property. The literature also often reference this property under the name 'Unconditional Security', even if that naming may appear a bit improper due to the fact that some constraint $\mathcal{C}$ is always applied on the stakeholders or on the communication channels, in order to overcome Shannon's impossibility theorem (see next section).

The constraint $\mathcal{C}$ applied to the side channel compared to the main channel may vary with the model. For instance, in *Partially Independent Channels* model introduced by Maurer in [5], the constraint is that the opponent receives the public information over a noisy side BSC (Binary Symmetric Channel) whose noise is at least partially statistically independent of the noise applying on the main BSC. It is then shown that a protocol can be designed to reach $(SP)$ under that constraint, even if the opponent has unlimited calculation and storage power. Another example is BB84, the first Quantum Key Distribution protocol proposed in [3], in which the constraint is that for each digit transmitted thanks to the polarization of a photon, the opponent has to choose a filter among 2 possible ones to measure the polarization wearing the digital information, and if the opponent makes the wrong choice, it obtains a random result rather than the actual result. This constraint legitimately comes from two properties of quantum systems: (i) the Heisenberg uncertainty principle and (ii) the no-cloning theorem. It is then shown as well that BB84 can reach $(SP)$ under that constraint, even if the opponent has unlimited calculation and storage power.

In this article, the constraint of the Security Model is called "Deep Random Assumption". It is detailed, together with the Security Model, in the next sections. It relies on the fact that the probability distribution used by each legitimate partner to generate private information is made 'unknown' for any other party (including the other partner and of course the opponent). We rigorously explain what we mean by 'unknown' probability distribution in the section presenting the Deep Random Assumption hereafter. We present in Section III a protocol that satisfies $(SC)$ under Deep Random Assumption, even if the opponent has unlimited calculation and storage power. Section III also presents the comparison and the similarities between our protocol and both BB84 and Maurer's protocol.

**Shannon and the need for Bayesian inference**

Shannon, in [1], established his famous impossibility result. Shannon defines perfect secrecy of a secrecy system, as its ability to equal the probability of the clear message $P(M)$ and the conditional probability $P(M|E)$ of the clear message knowing the encrypted message. In the case where the encryption system is using a shared secret key $K$ with a public transformation procedure to transform the clear message into the encrypted message, Shannon establishes that perfect secrecy can only be obtained if $H(K) \geq H(M)$.

It is a common belief in the cryptologic community that, in cases where the legitimate partners initially share no secret information (which we can write $H(K) = 0$), the result of Shannon thus means that it is impossible for them to exchange a perfectly secret (or almost perfectly secret) bit of information. The support for that belief is that, in the absence of key entropy, the conditional expectation $E[M|E]$, that is the best possible estimation of $M$ knowing $E$, is completely and equally known by all the parties (legitimate receiver and opponent), as :

$$E[M|E] = \sum_m m \frac{P(E|m)P(m)}{\sum_{m'} P(E|m')P(m')}$$

and thus, that the legitimate receiver cannot gain any advantage over the opponent when he tries to estimate the secret clear message from the public encrypted message. This reasoning however supposes that all the parties have a full knowledge of the distribution $P(M)$, enabling them to perform the above Bayesian inference to estimate $M$ from $E$.

Shannon himself warned the reader of [1] to that regard, but considered that this assumption is fairly reasonable (let's remember that calculaters were almost not yet existing when he wrote his article):

*« There are a number of difficult epistemological questions connected with the theory of secrecy, or in fact with any theory which involves questions of probability (particularly a priori probabilities, Bayes' theorem, etc.) when applied to a physical situation. Treated abstractly, probability theory can be put on a rigorous logical basis with the modern measure theory approach. As applied to a physical situation, however, especially when "subjective" probabilities and unrepeatable experiments are concerned, there are many questions of logical validity. For example, in the approach to secrecy made here, **a priori probabilities of various keys and messages are assumed known by the enemy cryptographer.** »*

The model of security that we develop in this article, by enabling the legitimate partners to use a specific form of randomness where the a priori probabilities of the messages cannot be efficiently known by the opponent, puts this opponent in a situation where the above reasoning based on Bayesian inference no longer stands.

It is remarkable that in his considerations about Perfect Secrecy, and in his proof of the impossibility result, Shannon only considers errorless protocols, meaning protocols such that $H(E|M,K) = H(M|E,K) = 0$. As soon as the protocol is not errorless, and assuming that the protocol has no prior key arrangement ($H(K) = 0$), then the impossibility result is intuitively interpreted as the fact that the legitimate partner cannot get a strictly "lesser" error rate than the opponent; "lesser" meaning as measured in the sample space that is assumed to wear a distance metric. We give in [13] a relation between secrecy measurement based on a distance in the sample space and secrecy measurement based on conditional entropy, as proposed by Csiszàr and Körner, when the sample space is a subset of the fairly general Euclidean space $\mathbb{R}^n$. For non-trivial message distribution, the function $S(E)$ that sets the most probable message given the public information is determined as the unique solution to

$$\text{Inf}_S \|S(E) - M\|$$

That unique solution is of course $E[M|E]$, the conditional expectation of the secret information (generically called by Shannon "message") knowing the public information (generically called by Shannon "cryptogram"). The determination of $E[M|E]$ requires knowledge of the distribution $P(M)$. As soon as the opponent doesn't know the distribution $P(M)$, he cannot determine the unique optimal solution $S(E)$. This does not mean of course that the error rate of the legitimate receiver will become automatically lesser than the error rate of the opponent, but it explains why the proof of Shannon cannot any longer be trivially applied in such cases.

**Prior probabilities theory**

Before presenting the Deep Random assumption, it is needed to introduce Prior probability theory.

The art of prior probabilities consists in assigning probabilities to a random event in a context of partial or complete uncertainty regarding the probability distribution governing that random event. The first author who has rigorously considered this question is Laplace [10], proposing the famous *principle of insufficient reason* by which, if an observer does not know the prior probability of occurrence of 2 events, he should consider them as equally likely. In other words, if a random variable $X$ can take several values $v_1, ..., v_n$, and if no information regarding the prior probabilities $P(X = v_i)$ is available for the observer, he should assign them $P(X = v_i) = 1/n$ in any attempt to produce inference from an experiment of $X$.

Several authors observed that this principle can lead to conclusion contradicting the common sense in certain cases where some knowledge is available to the observer but not reflected in the assignment principle.

If we denote $\mathfrak{I}_<$ the set of all prior information available to observer regarding the probability distribution of a certain random variable ('prior' meaning before having observed any experiment of that variable), and $\mathfrak{I}_>$ any public information available regarding an experiment of $X$, it is then possible to define the set of possible distributions that are compatible with the information $\mathfrak{I} \triangleq \mathfrak{I}_< \cup \mathfrak{I}_>$ regarding an experiment of $X$; we denote this set of possible distributions as:

$$D_\mathfrak{I}$$

The goal of Prior probability theory is to provide tools enabling to make rigorous inference reasoning in a context of partial knowledge of probability distributions. A key idea for that purpose is to consider groups of transformation, applicable to the sample space of a random variable $X$, that do not change the global perception of the observer. In other words, for any transformation $\tau$ of such group, the observer has no information enabling him to privilege $\varphi_\mathfrak{I}(v) \triangleq P(X = v|\mathfrak{I})$ rather than $\varphi_\mathfrak{I} \circ \tau(v) = P(X = \tau(v)|\mathfrak{I})$ as the actual conditional distribution. This idea has been developed by Jaynes [7], in order to avoid the incoherence brought in some cases by the imprudent application of Laplace principle.

We will consider only finite groups of transformation, because one manipulates only discrete and bounded objects in digital communications. We define the acceptable groups $G$ as the ones fulfilling the 2 conditions below:

$(C1)$ Stability - For any distribution $\varphi_\mathfrak{I} \in D_\mathfrak{I}$, and for any transformation $\tau \in G$, then $\varphi_\mathfrak{I} \circ \tau \in D_\mathfrak{I}$

$(C2)$ Convexity - Any distribution that is invariant by action of $G$ does belong to $D_\mathfrak{I}$

It can be noted that the set of distributions that are invariant by action of $G$ is exactly:

$$R_\mathfrak{I}(G) \triangleq \left\{ \frac{1}{|G|} \sum_{\tau \in G} \varphi_\mathfrak{I} \circ \tau \,|\, \forall \varphi_\mathfrak{I} \in D_\mathfrak{I} \right\}$$

The condition $(C2)$ is justified by the fact that in the absence of information enabling the observer to privilege $\varphi_\mathfrak{I}$ from $\varphi_\mathfrak{I} \circ \tau$, he should choose equivalently one or the other distribution, but then of course the average distribution $\frac{1}{|G|}\sum_{\tau \in G}\varphi_{\mathfrak{I}} \circ \tau$ should still belong to the set $D_{\mathfrak{I}}$ of possible distributions knowing $\mathfrak{I}$.

The set of acceptable groups as defined above is denoted:

$$\Gamma_{\mathfrak{I}}$$

Let's consider some examples.

**Example 1:** we consider a 6-sides dice. We are informed that the distribution governing the probability to draw a given side is altered, but we have no information of what that distribution actually is, and we have no available information regarding an experiment. We want nevertheless to assign an a priori probability distribution for the draw of dice. In this very simple case, it seems quite reasonable to assign an a priori probability of $1/6$ to each side. A more rigorous argument to justify this decision, based on the above, is the following: let's consider $G$ the finite group of transformation that let the dice unchanged, this group $G$ is well known, it is generated by the 3 axis $90°$ rotations, and has 24 elements. It is clear here that $G \in \Gamma_{\mathfrak{I}}$. It is also clear that, by considering a given distribution $(p_1, \dots, p_6)$, the a priori information available to the observer gives him no ground to privilege $(p_1, \dots, p_6)$ rather than $(p_{g(1)}, \dots, p_{g(6)})$ for any $g \in G$, and therefore the distribution should be of the form:

$$\left\{ \left( \frac{1}{|G|}\sum_{g \in G} p_{g(i)} \right)_{i \in \{1, \dots, 6\}} \right\}_{\{(p_1, \dots, p_6)\}}$$

It is easy to calculate that whatever is $(p_1, \dots, p_6)$,

$$\frac{1}{|G|}\sum_{g \in G} p_{g(i)} = \frac{1}{6}\sum_{i=1}^{6} p_i = \frac{1}{6}$$

and therefore in this trivial example, the Laplace principle applies nicely.

**Example 2**: let's now suppose that we have the result $J$ of a draw of the dice. Then the symmetry disappears and the opponent may want to assign, as an extreme example, a probability

$$p_j = \Gamma_J(j) \triangleq \begin{cases} 1 \ if \ j = J \\ 0 \ otherwise \end{cases}$$

which does not follow Laplace principle although one could argue that the knowledge of one single draw is not incompatible with the distribution $\{p_j = 1/6\}$. We however clearly don't want to exclude that extreme choice from the set of theoretically valid assignment made by the observer. An applicable group is in this case the sub-group $H$ of $G$ that let the side $J$ invariant. It is easy to see that it is composed with the 4 rotations $(0°, 90°, 180°, 270°)$ whose axis is determined by the centers of the sides $J$ and $7 - J$.

$$\left\{ \left( \frac{1}{|H|} \sum_{h \in H} p_{h(i)} \right)_{i \in \{1,\ldots,6\}} \right\}_{\{(p_1,\ldots,p_6)\}}$$

$$= \left\{ p_J = u; p_{7-J} = v; p_s = w \; \forall s \in \{1,\ldots,6\} \backslash \{J, 7-J\} \; \middle| \; \begin{array}{l} u \geq 0 \\ v \geq 0 \\ w \geq 0 \\ u + v + 4w = 1 \end{array} \right\}$$

One could argue that in a given probabilistic situation, there may exist several groups of transformation in $\Gamma_\mathfrak{I}$, and in that case, the choice of a given such group may appear arbitrary to assign the probability distribution. Although that oddness is not a problem for our purpose, we can solve it when 2 reasonable conditions are fulfilled: (i) we only consider finite sample space (note that objects in digital communication theory are bounded and discrete), and (ii) we assume that $D_\mathfrak{I}$ is convex (which can be ensured by design of the Deep Random source). Under those 2 conditions, all groups of transformation applying on the sample space are sub-groups of the finite (large) group of permutations $\mathfrak{S}$ applying on all the possible states, and if 2 groups of transformations $G$ and $G'$ applying on the sample space are in $\Gamma_\mathfrak{I}$, it is easy to see that $G \vee G'$ is still a finite sub-group of $\mathfrak{S}$, it contains $G$ and $G'$ and it is still in $\Gamma_\mathfrak{I}$. Consequently, for any observer having the same knowledge $\mathfrak{I}$ of the distribution, there exists a unique maximal sub-group $T_\mathfrak{I} \in \Gamma_\mathfrak{I}$, and this one should be ideally applied to obtain a maximal restriction of the set of distribution (because it is easy to check that if $G' \subset G$, then $R_\mathfrak{I}(G) \subset R_\mathfrak{I}(G')$). We can remark also that, if the uniform distribution is in $D_\mathfrak{I}$, then $\bigcap_{G \in \Gamma_\mathfrak{I}} R_\mathfrak{I}(G) \neq \emptyset$ because $R_\mathfrak{I}(G)$ then always contains the uniform distribution.

However, it is not necessary for our purpose to be able to determine such unique maximal sub-group $T_\mathfrak{I}$ if it exists.

Another objection could be that the definition of $D_\mathfrak{I}$ may appear too 'black or white' (a distribution belongs or does not belong to $D_\mathfrak{I}$), although real situations may not be so contrasted. This objection is not applicable to our purpose because in the case of Deep Random Generation, that we will introduce in section IV, we can ensure by design that distributions do belong to a specific set $D_\mathfrak{I}$. But otherwise, generally speaking, it is somehow preferable to draft the condition $(C1)$ as:

$(C1)$ For any distribution $\Phi$, and for any transformation $\tau$ in such group, the observer has no information enabling him to privilege $\Phi$ rather than $\Phi \circ \tau$ as the actual distribution.

**Deep Random assumption**

We can now introduce and rigorously define the Deep Random assumption. The Deep Random assumption imposes a rule enabling to make rigorous inference reasoning about an observer $\xi$ in a context where that observer $\xi$ only has partial knowledge of a probability distribution.

Let $X$ be a private random variable with values in a metric set $F$. Keeping the notations of the previous section, we still denote $\mathfrak{I}_<$ the set of all prior information available to observer regarding the probability distribution of $X$ ('prior' meaning before having observed any experiment of that variable), and $\mathfrak{I}_>$ any public information available regarding an experiment of $X$. Typically, $X$ represents a secret information that $\xi$ has to determine with the highest possible accuracy, and $\mathfrak{I} \triangleq \mathfrak{I}_< \cup \mathfrak{I}_>$ represents the public information available to $\xi$ (like for instance the information about the design of the Deep Random source, plus the information exchanged during a communication protocol as introduced in the following sections).

For any group $G$ of transformations applying on the sample space $F$, we denote by $\Omega_{\mathfrak{I}}(G)$ the set of all possible conditional expectations when the distribution of $X$ courses $R_{\mathfrak{I}}(G)$. In other words:

$$\Omega_{\mathfrak{I}}(G) \triangleq \{Z(\mathfrak{I}) \triangleq E[X|\mathfrak{I}] | \forall \varphi_{\mathfrak{I}} \in R_{\mathfrak{I}}(G)\}$$

Or also:

$$\Omega_{\mathfrak{I}}(G) = \left\{ Z(\mathfrak{I}) = \int_F v \varphi_{\mathfrak{I}}(v) dv \, | \forall \varphi_{\mathfrak{I}} \in R_{\mathfrak{I}}(G) \right\}$$

The **Deep Random assumption** prescribes that, if $G \in \Gamma_{\mathfrak{I}}$, the strategy $Z_\xi$ of the opponent observer $\xi$, in order to estimate $X$ from the public information $\mathfrak{I}$, should be chosen by the opponent observer $\xi$ within the restricted set of strategies:

$$\boldsymbol{Z_\xi \in \Omega_{\mathfrak{I}}(G)} \qquad\qquad (\boldsymbol{A})$$

The Deep Random assumption can thus be seen as a way to restrict the possibilities of $\xi$ to choose his strategy in order estimate the private information $X$ from his knowledge of the public information $\mathfrak{I}$. It is a fully reasonable assumption as exposed in the previous section presenting Prior probability theory, because the assigned prior distribution should remain stable by action of a transformation that let the distribution uncertainty unchanged.

$(A)$ suggests of course that $Z_\xi$ should eventually be picked in $\bigcap_{G \in \Gamma_{\mathfrak{I}}} \Omega_{\mathfrak{I}}(G)$ (that equals to $\Omega_{\mathfrak{I}}(T_{\mathfrak{I}})$ when $T_{\mathfrak{I}}$ exists), but it is enough for our purpose to find at least one group of transformation with which one can apply efficiently the Deep Random assumption to the a protocol in order to measure an advantage distilled by the legitimate partners compared to the opponent.

**The security model**

In our security model, we consider secrecy protocols being specific case of the Csiszàr and Körner model, where all the information exchanged by the legitimate partners $A$ and $B$ over the main channel are fully available to the passive opponent $\xi$. $\xi$ has unlimited computing and storage power. $A$ and $B$ share initially no private information.

Beyond being capable to Generate random bit strings, Publish bit strings on the main channel, Read published bit strings from the main channel, Store bit strings, and Make calculation on bit strings, the legitimate partners $A$ and $B$ are also capable to generate bit strings with Deep Randomness, by using their Deep Random Generator (DRG). A DRG is a random generator that produces an output with a probability distribution that is made unknown, under the meaning given in previous sections, to an external observer. We present in Section IV how a DRG can be implemented with classical computing resources.

$\xi$ is capable to Read all published bit strings from the main channel, Store bit strings, and Make calculation on bit strings, with unlimited computing and storage power. But when $\xi$ desires to infer a private information generated by $A$'s DRG (or by $B$'s DRG) from public information, he can only do it in respect of the Deep Random assumption $(A)$ presented in previous section. This assumption creates a 'virtual' side channel for the opponent conditioning the optimal information $Z$ he can obtain to estimate the Secret Common Output information $X$.

This assumption is fully reasonable, as established in the former sections, under the condition that the DRG of $A$ and the DRG of $B$ can actually produce distributions that are truly undistinguishable and unpredictable among a set $R_{\Im}(G)$.

The protocols that we consider obey the Kerckhoffs's principle by the fact that their specifications are entirely public.

**Information Theoretically Secure Protocols (ITSP)**

The main purpose of this work is to introduce how to design an « Information Theoretically Secure Protocol » (ITSP) under Deep Random assumption, and to introduce how to generate Deep Random from classical computing resources.

We consider protocols as per the Security Model presented in previous section,

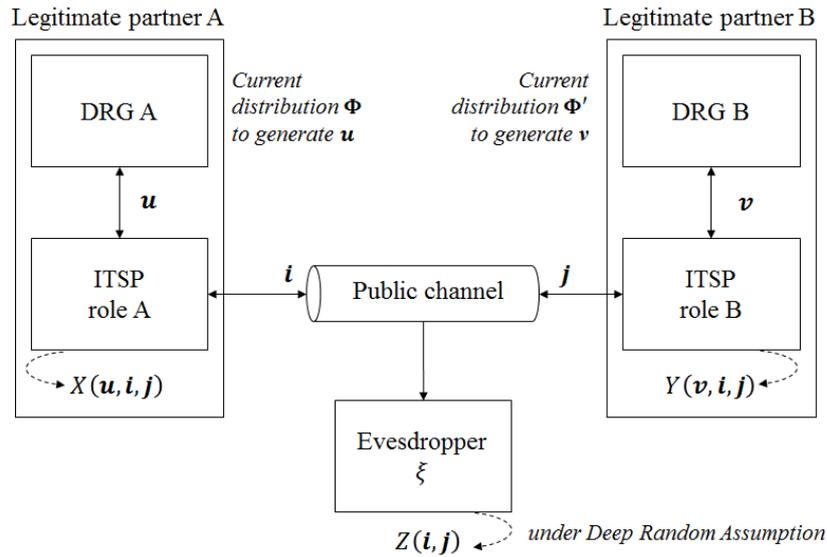

in which $A$ (resp. $B$) requests its Deep Random Generator (DRG) to obtain an experiment $u$ (resp. $v$) of a private random variable with hidden probability distribution $\Phi$ (resp. $\Phi'$); $A$ (resp. $B$) publishes the set of information $i$ (resp. $j$) on the public channel along the protocol. $A$ calculates the Secret Common Output information $X(u, i, j)$, with value in metric space $F$. $B$ calculates its estimation $Y(v, i, j)$ of the Secret Common Output information, also with value in $F$. In this model, $\Im_<$ is the public information available regarding the DRG of $A$ or $B$ (that are supposed to have the same design), and $\Im_> = \{i, j\}$ is the set of information published by the partners along the execution of the protocol.

The DRG of $A$ is run by $A$ completely privately and independently of $B$, and reversely the DRG of $B$ is run by $B$ completely privately and independently of $A$. The 2 DRG are thus not in any way secretly correlated, as one of the assumptions of the security model is that $A$ and $B$ share initially no private information.

The eavesdropping opponent $\xi$ who has a full access to the public information, calculates its own estimation $Z(\Im)$ (that we will also shortly denote $Z(i, j)$). $Z$ is called 'strategy' of the opponent.

As introduced in the 'Deep Random assumption' section, $\mathfrak{I}_<$ designates the public information available about a DRG. We assume here that $\mathfrak{I}_<$ is the same for both DRG of $A$ and $B$, meaning that they have the same design.

From the Deep Random assumption, for any group $G$ in $\Gamma_\mathfrak{I}$, the set of optimal strategies for the opponent can be restricted to:

$$\Omega_\mathfrak{I}(G, \mathcal{P}) = \{Z(\mathfrak{I}) = E[X|\mathfrak{I}] | \forall (\varphi_\mathfrak{I}(u), \varphi'_\mathfrak{I}(v)) \in R_\mathfrak{I}(G) \times R_\mathfrak{I}(G)\}$$

Then, the protocol $\mathcal{P}$ is a ITSP if it verifies the following property $(P)$:

$$\exists G \in \Gamma_\mathfrak{I}, \alpha > 0 | \forall Z \in \Omega_\mathfrak{I}(G, \mathcal{P}) : H(X|Z) - H(X|Y) \geq \alpha \qquad (P)$$

It has been shown ([4], [5]) that when a protocol satisfies $H(X|Z) - H(X|Y) \geq \alpha$, it can be complemented with Reconciliation and Privacy Amplification techniques to satisfy the targeted security property $(SP)$ introduced in the first section with $\epsilon$ as small as desired.

We will also impose a second condition for the definition of an ITSP $\mathcal{P}$. The second condition ensures that one can implement a DRG suitable for the protocol $\mathcal{P}$ thanks to a recursive and continuous generation algorithm that emulates locally the protocol. This approach is presented in section IV and introduced below. The second condition is the following:

$$\exists \alpha > 0 \text{ such that for any strategy } Z(\mathfrak{I}) \in \{E[X|\mathfrak{I}] | \forall \psi(u) \in D_{\mathfrak{I}_<}\}, \text{ there exists an actual distribution}$$
$$\Psi(u) \in D_{\mathfrak{I}_<} \text{ of the variables } X \text{ and } Y \text{ that verifies } H(X|Z) - H(X|Y) \geq \alpha$$

$$(P')$$

The condition $(P')$ means that for any optimal strategy $Z(\mathfrak{I})$ for the distribution $\psi$, there exists a new distribution $\Psi(u) \in D_{\mathfrak{I}_<}$, such that the condition $H(X|Z) - H(X|Y) \geq \alpha$ is satisfied (we take $P_u = P_v = \Psi$ to calculate $H(X|Y)$ because symmetry in the roles $A$ and $B$ can be assumed when their DRG have same design). $\Psi$ does not depend on $\mathfrak{I}_>$ because the generation of a distribution by a DRG takes place before any instantiation of the protocol. The interest of that condition is that it enables to build a Deep Random Generator as follows:

**Deep Random Generation**

A DRG is designed in association with a given ITSP $\mathcal{P}$. The DRG executes continuously the following recursive procedure: at each step $m + 1$, the generator emulates the ITSP internally and picks (through classical randomness) a new couple of identical distributions $\Phi_{m+1}(u) \in D_{\mathfrak{I}_<}, \Phi_{m+1}'(v) = \Phi_{m+1}(v)$ that defeats the optimal strategy (thus belonging to $\{E[X|\mathfrak{I}] | \forall \psi(u) \in D_{\mathfrak{I}_<}\}$) for the past distributions for $t \leq m$. This is always possible for an ITSP as given by condition $(P')$. The source of secret entropy is the current values of the inifinite counters (several can run in parallel) of the continuous recursive process, together with the classical random that is used at each step to pick a defeating distribution.

A detailed presentation of a DRG using a recursive method as introduced above can be found in Section IV.

An example of ITSP is given in Section III, as well as a comparison between that example and well known Bennett and Brassard Quantum Key Agreement protocol BB84.

In Section II, we present the formalism of Bernoulli random variables that we use to design our protocol; notations are defined and first considerations are discussed; Section II can be skipped for a conceptual, non-technical, reading of this article.

In Section III, we present our protocol, an example of ITSP, and we present the main theorem that establishes Information Theoretical Security of the protocol ; a high level argumentation of the proof is given in Section III, while a detailed and fully rigorous proof can be found in Annex 1, with a set of preparatory technical results;

In Section IV, we present some methods using classical computing resources to design a Deep Random Generator;

In Section V, we draw conclusions about the present work and open on further questions to progress towards applications to modern cryptography.

In Annex 1, we give the mathematical proofs that the protocol introduced in Section III verifies properties ($\boldsymbol{P}$) and ($\boldsymbol{P'}$).

In Annex 2, we prove technical results that are needed to configure a Deep Random Generator

## II.    Notations and first considerations

The protocol that we will present in Section III uses Bernoulli random variables, which present the advantage to enable Reconciliation thanks to the law of large numbers.

Let's define some notations. Considering $x = (x_1, \dots, x_n)$ and $y = (y_1, \dots, y_n)$ some parameter vectors in $[0,1]^n$ and $i = (i_1, \dots, i_n)$ and $j = (j_1, \dots, j_n)$ some experiment vectors in $\{0,1\}^n$, $l, r \in \mathbb{N}_n^*$ two integers, and $\theta \in [0,1]$, we define:

$x \cdot y$ (resp. $i \cdot j$) the scalar product of $x$ and $y$ (resp. $i$ and $j$)

$$|x| \triangleq \sum_{s=1}^n x_s \; ; \; |i| \triangleq \sum_{s=1}^n i_s$$

$$i \subset j \Longleftrightarrow \{s | i_s = 1\} \subset \{s | j_s = 1\}$$

$$\bar{i} \triangleq (u_1, \dots, u_n) \mid u_s = 1 \Longleftrightarrow i_s = 0$$

$$i \backslash j \triangleq (u_1, \dots, u_n) \mid u_s = 1 \Longleftrightarrow (i_s = 1) \,\&\, (j_s = 0)$$

$$i \cap j \triangleq (u_1, \dots, u_n) \mid u_s = 1 \Longleftrightarrow (i_s = 1) \,\&\, (j_s = 1)$$

$$i \cup j \triangleq (u_1, \dots, u_n) \mid u_s = 1 \Longleftrightarrow (i_s = 1) \; or \; (j_s = 1)$$

$$\chi_i(x) \triangleq \prod_{s=1}^n \big( i_s x_s + (1 - i_s)(1 - x_s) \big)$$

that represents probability of obtaining $i$ in a draw of a Bernoulli random variable with parameter vector $x$.

$$\Pi_i(x) \triangleq \prod_{s=1}^{n} (i_s x_s + (1 - i_s))$$

$$\psi_{i,r}(x) \triangleq \sum_{j | j.i = r} \chi_j(x) \; ; \; \psi_l(x) = \psi_{(1,\dots,1),l}(x) \triangleq \sum_{j | |j| = l} \chi_j(x)$$

$$\beta_{l,r}(\theta) \triangleq \binom{r}{l} \theta^l (1 - \theta)^{r-l}$$

that represents the Bernoulli coefficient of parameter $\theta$.

We will also manipulate permutation operators over vectors. For $\sigma \in \mathfrak{S}_n$, we write $supp(\sigma) = \ker(\sigma - id_{\mathfrak{S}_n}) = \{i, \sigma(i) \neq i\}$ and $|\sigma| = card(supp(\sigma))$. The permutation of a vector is the following linear application:

$$\forall \sigma \in \mathfrak{S}_n, \sigma(x) \triangleq (x_{\sigma(1)}, \dots, x_{\sigma(n)}) \text{ where } \mathfrak{S}_n \text{ represents the symmetric group}$$

The calculations about Bernoulli random variable take place in the vector space of multinomials with $n$ variables and local degree 1, noted $\mathcal{M}_1(n)$, whose $\{\Pi_i(x)\}_{i \in \{0,1\}^n}$ and $\{\chi_i(x)\}_{i \in \{0,1\}^n}$ are basis. It is easy to justify that $\{\chi_i(x)\}_{i \in \{0,1\}^n}$ is a basis by noting the inversion formula:

$$\sum_{i \in \{0,1\}^n} \chi_i(x) = 1 \Longrightarrow \Pi_j(x) = \sum_{i \supset j} \chi_i(x)$$

To ease the manipulation of upper bounds, we will use the notation $\lhd (M)$ for any quantity with value in $\mathbb{R}$ and whose absolute value is bounded by $M$.

If $A$ and $B$ are two partners, owning private parameter vectors $x/k$ and $y/k$ respectively (where $k$ is a real number parameter $> 1$), and generating vectors $i$ and $j$ respectively through Bernoulli random variables from $x/k$ and $y/k$ respectively, we consider the respective random variables $U_A$ and $U_B$ such that :

$$U_A = \frac{x.j}{n} \; ; \; U_B = \frac{i.y}{n}$$

As shown in Proposition 4 (i):

$$E[(U_B - U_A)^2 | x, y] \leq \frac{2}{nk}$$

As soon as the opponent knows the probability distributions used respectively by $A$ and $B$ to independently generate respectively $x$ and $y$, he can generate a strategy $V_\xi^*(i, j)$ from the public information $(i, j)$, by using Bayesian inference. That strategy this time verifies:

$$E\left[\left(V_\xi^*(i, j) - U_A\right)^2\right] \leq E[(U_B - U_A)^2]$$

For a proof, see Proposition 6 in Annex 1.

$V_\xi^*(i, j)$ is of course the conditional expectation $E[U_A | i, j]$.

If $\Phi$ is the probability distribution of $x$, the optimal strategy $V_\xi^*(i, j)$ is expressed by:

$$V_\xi^*(i,j) = E[U_A|i,j] = \frac{\int_{x\in[0,1]^n} \frac{x.j}{nk} \chi_i\left(\frac{x}{k}\right) \Phi(x)dx}{\int_{x\in[0,1]^n} \chi_i\left(\frac{x}{k}\right) \Phi(x)dx}$$

that clearly depends on the knowledge of $\Phi$.

**DRG's Independence Phenomenon**

However it is possible to find a strategy as efficient as $V_\xi^*$ without having knowledge of $\Phi$, when $\Phi$ or $\Phi'$ is symmetric ($\Phi$ is symmetric meaning that $\Phi \circ \sigma = \Phi \ \forall \sigma \in \mathfrak{S}_n$). We call that situation Independence Phenomenon because no information is given to favor $\Phi$ from $\Phi \circ \sigma$ for any permutation $\sigma \in \mathfrak{S}_n$. In that case, the strategy:

$$V_{\xi,1}(i,j) \triangleq \frac{k|i||j|}{n^2}$$

gives the following inequality (see Proposition 5 in Annex 1), comparable to the performance of $V_\xi^*$:

$$E\left[\left(V_{\xi,1} - U_A\right)^2\right] \leq \frac{2}{nk}$$

That efficient strategy takes advantage of the symmetry resulting from the independence between the choices of the legitimate partners for their respective distributions. We will see in section III that the Independence Phenomenon is an important problem to overcome in the proposed ITSP.

## III. Presentation of the protocol and proof of perfect secrecy

In this Section we present an example of ITSP. Besides being hidden to the opponent, we will see that the probability distribution used by each legitimate partner also need to have specific properties in order to prevent the opponent to evaluate $V_A$ with the same accuracy than the legitimate partners by using the Independence Phenomenon introduced in Section II.

When Bernoulli random variables are considered through quadratic evaluation, the probability distribution $\Phi$ of a legitimate partner can be considered through its quadratic matrix:

$$M_\Phi(u,v) = \int_{[0,1]^n} x_u x_v \Phi(x) dx$$

$M_\Phi$ is a symmetric matrix with elements in $[0,1]$.

$n$ is assumed even. The notation $S_n$ represents the set of all subsets $I$ of $\{1, \ldots, n\}$ containing $n/2$ elements. We remind that for $I \in S_n$, the notation $\bar{I}$ designates $\{1, \ldots, n\}\backslash I$ thus also in $S_n$. Let's first introduce the canonical mid-segment:

$$I_0 \triangleq \left\{1, \ldots, \frac{n}{2}\right\}$$

and the canonical mid-segment permutation $\sigma_0$ that sends $I_0$ in $\bar{I_0}$ and vice versa:

$$\sigma_0 \triangleq \circ_{r \in I_0}\, \tau\left(r, r + \frac{n}{2}\right)$$

With this, we now introduce the « tidied form » of a probability distribution $\Phi$. For any $\Phi$, there exist a (not necessarily unique) pair of permutations $(\sigma_\Phi{}^-, \sigma_\Phi{}^+)$ in $\mathfrak{S}_n$ such that:

$$c_-(\Phi) \triangleq \sum_{u,v \in I_0 \times \bar{I_0}} M_{\Phi \circ \sigma_\Phi{}^-}(u,v) = \min_{\sigma \in \mathfrak{S}_n}\left(\frac{4}{n^2} \sum_{u,v \in I_0 \times \bar{I_0}} M_{\Phi \circ \sigma}(u,v)\right)$$

$$c_+(\Phi) \triangleq \sum_{u,v \in I_0 \times \bar{I_0}} M_{\Phi \circ \sigma_\Phi{}^+}(u,v) = \max_{\sigma \in \mathfrak{S}_n}\left(\frac{4}{n^2} \sum_{u,v \in I_0 \times \bar{I_0}} M_{\Phi \circ \sigma}(u,v)\right)$$

We remark that composing $\Phi$ by $\sigma_0$ does not change a pair $(\sigma_\Phi{}^-, \sigma_\Phi{}^+)$ due to the symmetry of $M_\Phi$. Actually, we will call $\sigma_\Phi$ a tidying permutation of $\Phi$ as being either (with 50% chance) a $\sigma_\Phi{}^-$ or a $\sigma_\Phi{}^+$. In other words, if we designate by $\mathfrak{S}_I$ the sub-group $\mathfrak{S}_I \triangleq \{\sigma \in \mathfrak{S}_n | \forall u \in I, \sigma(u) \in I\}$, then we mean that picking a tidying permutation of $\Phi$ is equivalent to choose a permutation uniformly within $\{\sigma_\Phi{}^+ \circ \sigma, \sigma_\Phi{}^+ \circ \sigma \circ \sigma_0, \sigma_\Phi{}^- \circ \mu, \sigma_\Phi{}^- \circ \mu \circ \sigma_0 | \forall \sigma, \mu \in \mathfrak{S}_{I_0}\}$ (obviously this set does not depend on which possible pair $(\sigma_\Phi{}^-, \sigma_\Phi{}^+)$ is chosen).

We then call a tidied form of $\Phi$, denoted by $\Phi \circ \sigma_\Phi$, a distribution obtained by composing $\Phi$ with any possible tidying permutation picked randomly. In other words, a tidied form of a distribution is a distribution of the form:

$$\bar{\Sigma}(\Phi) \triangleq \frac{1}{4}\left(\delta_{\sigma_\Phi{}^+}(\sigma_\Phi) + \delta_{\sigma_\Phi{}^-}(\sigma_\Phi)\right)\left(\delta_{\sigma_0}(\rho) + \delta_{\mathrm{Id}_{\mathfrak{S}_n}}(\rho)\right)\frac{1}{|\mathfrak{S}_{I_0}|}\sum_{\sigma \in \mathfrak{S}_{I_0}} \Phi \circ \sigma_\Phi \circ \sigma^{-1} \circ \rho^{-1}$$

Then we define the following probability distribution set:

$$\zeta(\alpha) \triangleq \left\{ \Phi \,\middle|\, |c_+(\Phi) - c_-(\Phi)| \geq \sqrt{\alpha} \right\}$$

It can be seen with a bit of attention that $|c_+(\Phi) - c_-(\Phi)| > 0$ if and only if $\Phi$ is different from its symmetric projection

$$\bar{\Phi} \triangleq \frac{1}{n!} \sum_{\sigma \in \mathfrak{S}_n} \Phi \circ \sigma$$

The parameter $\alpha$ is a measure of a distance between $\Phi$ and $\bar{\Phi}$.

Tidied form are interesting because 2 distributions in $\zeta(\alpha)$ that are put in their tidied form $\Phi \circ \sigma_\Phi$ and $\Phi' \circ \sigma_{\Phi'}$ then avoid the DRG's Independence Phenomenon introduced in Section II. Indeed, comparing inequality obtained at Proposition 4 (ii) with result of Lemma 1, both shown in Annex 1, gives:

$$E\left[(V_{\xi,1} - V_A)^2\right]_{\Phi \circ \sigma_\Phi, \Phi' \circ \sigma_{\Phi'}} \geq \frac{C(\alpha)}{k^2} \gg_{n \gg k} E[(V_B - V_A)^2]_{\Phi \circ \sigma_\Phi, \Phi' \circ \sigma_{\Phi'}} = O\left(\frac{1}{nk}\right)$$

The property of a distribution defined as '*belonging to $\zeta(\alpha)$*' will represent the public prior information $\mathfrak{I}_<$ (as defined in Section I) for distributions chosen in the protocol presented below.

Now we can explicit our example of ITSP protocol. It will be demonstrated in Theorem 1 and Theorem 2 that the presented ITSP fulfills the 2 conditions $(P)$ and $(P')$ defining an ITSP. The protocol doesn't have purpose to be optimal in terms of implementation. We will discuss quantitative aspects in Section V, and in related works.

Here are the steps of the proposed ITSP:

$A$ and $B$ are the legitimate partners of the Security Model presented in Section I. The steps of the protocol $\mathcal{P}(\alpha, n, k, K, L)$ are the followings:

***Step 1 – Deep Random Generation***: *A and B use their respective DRG to pick independently the respective probability distributions $\Phi$ and $\Phi' \in \zeta(\alpha)$. $\Phi$ (resp. $\Phi'$) is then secret (under Deep Random assumption) for any observer other than A (resp. B) beholding all the published information. A (resp. B) calculates a tidying permutation $\sigma_\Phi$ (resp. $\sigma_{\Phi'}$) of $\Phi$ (resp. $\Phi'$). A draws the parameter vector $x \in \{0,1\}^n$ from $\Phi$. B draws the parameter vector $y \in \{0,1\}^n$ from $\Phi'$.*

***Step 2 – Degradation***: *$k > 1$ is a public Degradation parameter; A generates a Bernoulli experiment vectors $i \in \{0,1\}^n$ from the parameter vector $x/k$. A publishes i. B generates a Bernoulli experiment vectors $j \in \{0,1\}^n$ from the parameter vector $y/k$. B publishes j.*

***Step 3 – Dispersion:*** *A and B also pick a second probability distribution respectively $\Psi$ and $\Psi' \in \zeta(\alpha)$ such that it is also secret (under Deep Random assumption) for any observer other than A (resp. B). $\Psi$ is selected also such that $\int_{|x| \in [k|i| - \sqrt{n}, k|i| + \sqrt{n}]} \Psi(x)dx \geq \frac{1}{2\sqrt{n}}$ in order to ensure that $|i|$ is not an* *unlikely value for ≈ |x/k| (same for Ψ′ by replacing x by y and i by j). Ψ (resp. Ψ′) is used to scramble the publication of the tidying permutation of A (resp. B). A (resp. B) calculates a permutation $\sigma_d[i]$ (resp. $\sigma'_d[j]$) representing the reverse of the most likely tidying permutation on Ψ (resp. Ψ′) to produce i (resp. j). In other words, with i, $\sigma_d[i]$ realizes :*

$$\max_{\sigma \in \mathfrak{S}_n} \int_X P(i|x) \Psi \circ \sigma_\Psi \circ \sigma^{-1}(x) dx$$

*Then A (resp. B) draws a boolean $b \in \{0,1\}$ (resp. b′) and publishes in a random order $(\mu_1, \mu_2) = t^b(\sigma_d[i], \sigma_\Phi)$, (resp. $(\mu'_1, \mu'_2) = t^{b'}(\sigma'_d[j], \sigma_{\Phi'})$) where t represents the transposition of elements in a couple.*

*(we remark that it is possible to publish only $(\mu_1(I_0), \mu_2(I_0))$ and $(\mu'_1(I_0), \mu'_2(I_0))$ rather than $(\mu_1, \mu_2)$ and $(\mu'_1, \mu'_2)$, which leads to quantity of bits of 4n instead of 4n log n).*

**Step 4 – Synchronization:** *A (resp. B) chooses randomly $\sigma_A$ (resp. $\sigma_B$) among $(\mu'_1, \mu'_2)$ (resp. $(\mu_1, \mu_2)$).*

**Step 5 – Advantage Distillation:** *A calculates $V_A = \frac{\sigma_\Phi^{-1}(x).\sigma_A^{-1}(j)}{n}$, B calculates $V_B = \frac{\sigma_B^{-1}(i).\sigma_{\Phi'}^{'-1}(y)}{n}$. $V_A$ and $V_B$ are then transformed respectively by A and B in binary output thanks to the sampling method:*

$$\widetilde{e_A} = \left\lfloor \frac{(V_A - \rho)\sqrt{nk}}{K} \right\rfloor \bmod 2, \quad \widetilde{e_B} = \left\lfloor \frac{(V_B - \rho)\sqrt{nk}}{K} \right\rfloor \bmod 2$$

*(where ρ is a translation parameter randomly picked in $\left[0, \frac{2K}{\sqrt{nk}}\right[$ at each instance of the protocol). The multiplicative factor K is chosen such that:*

$$\frac{1}{\sqrt{nk}} \ll \frac{K}{\sqrt{nk}} \ll \frac{1}{k}$$

*(K is picked according to Proposition 8 (iii)). We introduce the following notations corresponding to the canonical form of the opponent strategy to estimate $V_A$ (obtained in Proposition 9):*

$$V_\xi(i, j, \sigma_{\xi,A}, \sigma_{\xi,B}) \triangleq \frac{2k\left(\left(\sum_{r \in \sigma_{\xi,B}(I_0)} i_r\right)\left(\sum_{r \in \sigma_{\xi,A}(I_0)} j_r\right) + \left(\sum_{r \in \overline{\sigma_{\xi,B}(I_0)}} i_r\right)\left(\sum_{r \in \overline{\sigma_{\xi,A}(I_0)}} j_r\right)\right)}{n^2}$$

$$\widetilde{e_\xi}(\sigma_{\xi,A}, \sigma_{\xi,B}) \triangleq \left\lfloor \frac{(V_\xi(i, j, \sigma_{\xi,A}, \sigma_{\xi,B}) - \rho)\sqrt{nk}}{K} \right\rfloor \bmod 2$$

$$T_m = \left\{ \widetilde{e_\xi}(\mu'_1, \mu_1)_m, \widetilde{e_\xi}(\mu'_1, \mu_2)_m, \widetilde{e_\xi}(\mu'_2, \mu_1)_m, \widetilde{e_\xi}(\mu'_2, \mu_2)_m \right\}$$

*where $\sigma_{\xi,B}$ is the permutation chosen by ξ among $(\mu_1, \mu_2)$ as $\sigma_\Phi$ and $\sigma_{\xi,A}$ is the permutation chosen by ξ among $(\mu'_1, \mu'_2)$ as $\sigma_A$. The partners discard then the instance of the protocol if $|T_m| \neq 2$.*

**Step 6:** *classical Information Reconciliation and Privacy Amplification (IRPA) techniques then lead to get accuracy as close as desired from perfection between estimations of legitimate partners, and knowledge as close as desired from zero by any unlimited opponent, as shown in [4].*

The choice of the parameters $(\alpha, n, k, K)$ will be discussed in proof of main Theorem. They are set to make steps 5 and 6 possible, which is the main achievement of this work.

The protocol can be heuristically analyzed as follows:

Regarding the legitimate partners:

(1)     when $B$ picks $\sigma_B = \sigma_\Phi$ and $A$ picks $\sigma_A = \sigma_{\Phi'}$ (1/4 of cases), the choice of $\sigma_A$ and $\sigma_B$ remain independant from $i, j$, so that $i$ and $j$ remain draws of independent Bernoulli random variables, then allowing to apply Chernoff-style bounds with accuracy $O\left(\frac{1}{\sqrt{nk}}\right)$ ($E\left[\left(V_{A|\sigma_A=\sigma_{\Phi'}} - V_{B|\sigma_B=\sigma_\Phi}\right)^2\right]^{1/2} = O\left(\frac{1}{\sqrt{nk}}\right)$ as shown in Proposition 4 (ii) in Annex 1).

(2)     When $A$ picks $\sigma_A = \sigma'_d[j]$ and/or $B$ picks $\sigma_B = \sigma_d[i]$ (3/4 of cases), the Chernoff bound no longer applies and instead, and $V_B$ or $V_A$ become erratic, which will lead to an error probability of $\approx 1/2$.

On the other hand, the discarding condition $|T_m| = 2$ forced at step 5 imposes that the opponent has no better choice than to choose its bit $\tilde{e}_\xi$ randomly within $\{0,1\}$ with 50% chance. Thus, as soon as we can prove that condition $|T_m| = 2$ and condition (1) above coexist with a fair probability $\napprox 0$, we have shown that an Advantage is created for the partners over the opponent. This fact is proven in Theorem 1. The main argument of the proof is that, in the estimation $V_\xi(i, j, \sigma_{\xi,A}, \sigma_{\xi,B})$ of $V_A$ by opponent, the opponent's choices among $(\sigma_\Phi, \sigma_d[i])$ for $\sigma_{\xi,B}$ and among $(\sigma_{\Phi'}, \sigma'_d[i])$ for $\sigma_{\xi,A}$ are indistinguishable under Deep Random Assumption, and therefore the choice $(\sigma_{\xi,A}, \sigma_{\xi,B}) \neq (\sigma_{\Phi'}, \sigma_\Phi)$, which has a fair occurrence probability (3/4), leads to opponent's erratic behavior creating a fair occurrence probability that condition $|T_m| = 2$ and condition (1) above coexist.

Below we explain the role of each step:

The Degradation transformations $x \mapsto \frac{x}{k}$ and $y \mapsto \frac{y}{k}$ with $k > 1$ at step 2 are necessary so that $\sigma_d[i](I_0) \neq \sigma_\Phi(I_0)$ (resp. $\sigma'_d[i](I_0) \neq \sigma_{\Phi'}(I_0)$). If $k = 1$, then $i = x$ and $j = y$ and this would cause $\sigma_d[i](I_0) \approx \sigma_\Phi(I_0)$ (resp. $\sigma'_d[i](I_0) \approx \sigma_{\Phi'}(I_0)$).

The Deep Random Generation at step 1 prevents the use of Bayesian inference based on the knowledge of the probability distribution. In particular, if $\Phi, \Psi$ (resp. $\Phi', \Psi'$) were not managed by Deep Randomness, $\xi$ would be able discriminate $\sigma_\Phi$ among $(\mu_1, \mu_2)$ (resp. $\sigma_{\Phi'}$ among $(\mu'_1, \mu'_2)$) by Bayesian inference.

Dispersion step 3 mixes $\sigma_\Phi$ within $(\mu_1, \mu_2)$ with another permutation $\sigma_d[i]$ (and $\sigma_{\Phi'}$ within $(\mu'_1, \mu'_2)$ with another permutation $\sigma'_d[j]$). $i$ is entirely determined by $|i|$ and a permutation, which explains the constraint and transformation applied on $\Psi$ in step 3 to make $\sigma_\Phi$ and $\sigma_d[i]$ indisguishable knowing $i$ (same with $\sigma'_d[j]$, $\sigma_{\Phi'}$, and $j$). $\sigma_d[i]$ is (1) indistinguishable from $\sigma_\Phi$ knowing $\mathfrak{I}_> = \{i, j, (\mu_1, \mu_2), (\mu'_1, \mu'_2)\}$, and (2) causes that $\sigma_d[i]^{-1}(i)$ and $\sigma_\Phi^{-1}(i)$ behave very differently due to the fact that $\sigma_d[i]^{-1}(i)$ only depends on $|i|$; indeed we remark that $\forall \mu \in \mathfrak{S}_n \; \sigma_d[\mu(i)]^{-1} \circ \mu = \sigma_d[i]^{-1}$, and therefore, $\sigma_d[i]^{-1}(i)$ is stable by action of $\mathfrak{S}_n$ on $i$. (resp. $\sigma'_d[j]^{-1}(j)$ only depends on $|j|$). That indistinguishability can be expressed under Deep Random assumption by the symmetry $\Sigma_1$ associated to the group $\{\mathrm{Id}, t\}$ ($t$ being the transposition of pairs) applied to $(\mu_1, \mu_2)$ and $(\mu'_1, \mu'_2)$.

Associating $\Sigma_1$ with other symmetries presented in the proof of Theorem 1, we manage to show that the best strategy of the opponent under Deep Random assumption to estimate $V_A$ and then as a consequence $\widetilde{e_A}$, are:

$$V_\xi(i, j, \sigma_{\xi,A}, \sigma_{\xi,B}) \triangleq \frac{2k\left(\left(\sum_{r \in \sigma_{\xi,B}(I_0)} i_r\right)\left(\sum_{r \in \sigma_{\xi,A}(I_0)} j_r\right) + \left(\sum_{r \in \overline{\sigma_{\xi,B}(I_0)}} i_r\right)\left(\sum_{r \in \overline{\sigma_{\xi,A}(I_0)}} j_r\right)\right)}{n^2}$$

$$\widetilde{e_\xi}(\sigma_{\xi,A}, \sigma_{\xi,B}) \triangleq \left\lfloor \frac{(V_\xi(i, j, \sigma_{\xi,A}, \sigma_{\xi,B}) - \rho)\sqrt{nk}}{K} \right\rceil \bmod 2$$

where $\sigma_{\xi,B}$ is the permutation chosen by $\xi$ among $(\mu_1, \mu_2)$ as $\sigma_\Phi$ and $\sigma_{\xi,A}$ is the permutation chosen by $\xi$ among $(\mu'_1, \mu'_2)$ as $\sigma_A$.

The synchronization step 4 is designed to overcome the Independence Phenomenon. It needs that the distributions to have special properties ($\in \zeta(\alpha)$) in order to efficiently play their role. It is efficient in $1/4$ of cases (when $B$ picks $\sigma_B = \sigma_\Phi$ and $A$ picks $\sigma_A = \sigma_{\Phi'}$, which we will call 'favorable cases'). If $\sigma_\Phi$ and $\sigma_{\Phi'}$ were chosen randomly by $A$ and $B$ instead of being the respective tidying permutations of $\Phi$ and $\Phi'$, then $\xi$ could, even without knowing $\Phi$ and $\Phi'$, use $V_{\xi,1}(i, j) \triangleq \frac{k|i||j|}{n^2}$ in order to estimate say $V_A$ with accuracy $O\left(\frac{1}{\sqrt{nk}}\right)$, in which case $\widetilde{e_\xi} = \left\lfloor \frac{(V_{\xi,1} - \rho)\sqrt{nk}}{K} \right\rceil$ would be as close to $\widetilde{e_A}$ as $\widetilde{e_B}$ in favorable cases.

The step 5 is called Advantage Distillation because at this step, thanks to the discarding condition $|T_m| = 2$, we have managed to create, under Deep Random Assumption, a protocol in which the error rate for the legitimate receiver is strictly lower than the error rate for the opponent.

We remark that the condition $|T_m| = 2$ plays also another role which is to protect against the unexpected predictability effects of using an error correcting code at Reconciliation phase. Indeed, unlike QKD protocols (protected by no cloning theorem) or Maurer's satellite protocol presented in [5], in our case the opponent can try the various possible combinations of choice for $\sigma_{\xi,A}$ and $\sigma_{\xi,B}$, and compare them in terms of matching a code word. To illustrate the issue, let's consider the non-optimal 'bit repeating' error correction method described by Maurer in [5]: the codeword $v_A$ chosen by $A$ can only be $(0,0,...,0)_L$ or $(1,1,...,1)_L$ depending on $e_A = 0$ or $e_A = 1$. $B$ could then publicly discards all decoded sequence $v_B$ that is not $(0,0,...,0)_L$ or $(1,1,...,1)_L$ and obviously decodes accordingly $e_B = 0$ if $|v_B| = 0$, and $e_B = 1$ if $|v_B| = L$. In such scenario, if, for one of the instance $m$ within the sequence, we have $\sigma_A = \sigma_{\Phi'}$ and $T_m = (0,0,0,0)$ (or $T_m = (1,1,1,1)$), then it is clear that the code word is $(0,0,...,0)_L$ (or $(1,1,...,1)_L$). This example shows how, even with Deep Randomness indistinguishability applied locally to each instance, the opponent can still take advantage of an imprudent implementation of the error correcting method to break the secrecy. The condition $|T_m| = 2$ also have the advantage to avoid such predictability flaw.

**Theorem 1.**

*Under the Deep Random assumption, $\mathcal{P}(\alpha, n, k, K)$ satisfies condition* $(P)$

The details of the proof are given in Annex 1.

We also have with a very similar proof (also found in Annex 1):

**Theorem 2.**

$\mathcal{P}(\alpha, n, k, K)$ *satisfies condition* $(P')$

**Parallel with BB84**

The protocol $\mathcal{P}$ presents some interesting similarities with the Quantum Key Distribution protocol BB84 presented by Bennet and Brassard in 1984 [3].

Let's first briefly remind the principles of BB84. $A$ is secretly polarizing photons; for each photon, the polarization is chosen randomly by $A$ between $\{0°, 90°\}$ (right polarization) or between $\{45°, 135°\}$ (inclined polarization); the choice between right polarization type or inclined polarization type is also made randomly by $A$. $B$ can measure the polarization by choosing a right filter (that will give the good answer if the polarization is right), or an inclined filter (that will give the good answer if the polarization is inclined). If $B$ chooses the wrong filter, the result of the measurement is random. The exact same situation applies for the opponent $\xi$.

The polarization in BB84 can be compared to the Dispersion step in $\mathcal{P}$ where 2 indistinguishable permutations, associated to a same Bernoulli vector, are published.

The choice of filter by $B$ (or $E$) in BB84, imposed by the Heisenberg uncertainty principle, can be compared to the Synchronization step in $\mathcal{P}$ where $B$ (or $E$) has no other choice, as per Deep Random Assumption, than choosing randomly one of the 2 permutations because they are indistinguishable knowing the Bernoulli vector.

In case of wrong choice of filter, the measurement step in BB84 will result into a random result for the measuring party. Similarly, the choice of the wrong permutation ($\sigma_d$ instead of $\sigma_\Phi$) in the Synchronization step, can be shown (see Annex 1 for the detailed proof of Theorem 1) to result into erratic variable. However, in $\mathcal{P}$, it is necessary that both partners make the good choice to avoid random behavior, although in BB84, it is sufficient that they make the same incline choice, whatever it is.

Then the protocols differ because BB84 can more easily exploit the partial independence created between choices of $B$ and $E$ thanks to the no-cloning theorem. Indeed, $B$ can publicly disclose its filter choice so that $A$ can publicly discard the wrong choices; $\xi$ will not be able to exploit this information because it cannot redo another measurement on a 'copy' of the polarized photon. Contrariwise BB84, $\mathcal{P}$ only manipulates conventional digital information, that can be copied and accessed anytime by the unlimited opponent. Therefore, the discarding condition $|T_m| = 2$ is used to distill an Advantage.

## IV. Discussion about Deep Random generation

In the present Section, we discuss methods to generate Deep Random from a computing source. It may appear as a impossible at first sight to generate Deep Random from a deterministic computable program. In the real world, even a computer may access sources of randomness whose probability distribution is at least partly unknown, but it doesn't mean that we can build from it Deep Random reliable for cryptographic applications.

### Generation process and strategies

We present a theoretically valid method to generate Deep Random from a computing source. That method mainly relies on infinite counters $\in \mathbb{N}^+$ as unpredictable randomness source, associated with a Cantor' style diagonal constructing argument.

If one doesn't know the date of beginning and the speed of an infinite counter, no probability distribution can be even approximated about the value of the counter at a given time, because of the unlimited nature of a counter. If performed in a physical computing source, the actual speed of the counter is impacted by all external tasks of the processor, for which no probability distribution can be estimated; the only thing that an opponent can do is estimate a rough upper bound of that speed, which, as we will see, is far from being enough to breach the Deep Randomness of the proposed method, and thus to break the secrecy of a protocol such as the one presented in Section III.

We have introduced in Section I the principle of continuous and recursive constructing process, designed in association with a given ITSP. At each step $m + 1$, the generator emulates the ITSP internally and picks (through classical randomness) a new couple of distributions $\Phi(u) \in D_{\mathfrak{I}_<}, \Phi'(v) = \Phi(v)$ that defeats the best possible strategy knowing all the past distributions for $t \leq m$, which is always possible for an ITSP as given by condition $(P')$. In this Section, we will present rigorously a method based on that principle applied to the protocol $\mathcal{P}$ presented in Section III.

Let's consider a strategy $\omega$ of the opponent to estimate $V_A$. As explained in the proof of Theorem 2 presented in Annex 1, $\omega$ depends only on $(i, j)$ and not on $(\mu_1, \mu_2, \mu'_1, \mu'_2)$.

When emulating the protocol, we will thus also place ourselves in the situation $S$ where the legitimate partners make both the good choice of permutation

$$S \triangleq [(\sigma_A = \sigma_{\Phi'}) \,\&\, (\sigma_B = \sigma_\Phi)]$$

If the protocol now executes with a same given distribution $\Psi \in D_{\mathfrak{I}_<}$ on both sides of $A$ and $B$, we will denote:

$$
\begin{aligned}
\langle \omega, \Psi \rangle &\triangleq E[(\omega - V_A)^2 | S] \\
&= \sum_{i,j} \int_{x,y} \Bigg( \omega(i,j) \\
&\quad - \frac{\sigma_\Psi^{-1}(x) \cdot \sigma_\Psi^{-1}(j)}{n} \Bigg)^2 \chi_i\left(\frac{x}{k}\right) \chi_j\left(\frac{y}{k}\right) \tilde{\Sigma}(\Psi)(\sigma_\Psi^{-1}(x)) \tilde{\Sigma}(\Psi)(\sigma_\Psi^{-1}(y)) dx dy \\
&= \sum_{i,j} \int_{x,y} \left( \omega(\sigma_\Psi(i), \sigma_\Psi(j)) - \frac{x \cdot j}{n} \right)^2 \chi_i\left(\frac{x}{k}\right) \chi_j\left(\frac{y}{k}\right) \tilde{\Sigma}(\Psi)(x) \tilde{\Sigma}(\Psi)(y) dx dy
\end{aligned}
$$

Let's consider $\omega_m \in \{E[V_A|S, \Im]|\forall \psi(u) \in D_{\Im_<}\}$. By definition of $\{E[V_A|S, \Im]|\forall \psi(u) \in D_{\Im_<}\}$, there exists a distribution $\Phi_m \in D_{\Im_<}$ such that $\omega_m = E[V_A|S, \Im]_{\Phi_m}$. It has been shown in the proof of Theorem 2 that there exists a distribution $\Phi_{m+1} \in D_{\Im_<}$, such that

$$\langle \omega_m, \Phi_{m+1} \rangle \geq \frac{C_1(\alpha)}{k^2}$$

Let's present first a heuristic argument. With an infinite counter privately executed by a DRG, the moments $m$ and $m + 1$ are indistinguishable for the opponent $\xi$. If a best winning strategy at the moment $m$ exists for $\xi$, then by choosing at moment $m + 1$ the probability distribution $\Phi_{m+1}$ verifying the inequality above, the DRG guarantees that no absolute winning strategy exist, because moment of observation cannot be determined by opponent as rather being $m$ or $m + 1$.

This heuristic argument does not explain how to practically build a Deep Random generator with classical computing resources, but it introduces the diagonal constructing process inspired from non-collaborative games' theory. Of course, no Nash' style equilibrium can be found as the DRG is continuously deviating its distribution strategy to avoid forecast from opponent.

Another interesting remark at this stage is that, in the proof of Theorem 2, we highlighted low entropy subsets of probability distributions (typically $\{\Psi \circ \mu | \forall \mu \in \mathfrak{S}_n\}$) in which a distribution can always be found to scramble a given strategy $\omega$. Such subsets have entropy of order $O(n \ln n)$ although the whole set of possible strategies $\{\omega(i, j)\}$ has an entropy of order $O(2^n)$, which is not manageable by classical computing resources.

Let's now explicit a constructing process, inspired from the above heuristic argument.

Let's first consider the two following recursive constructing process:

Process 1 :

$\Phi_0$ is a distribution in $D_{\Im_<} = \bar{\Sigma}(\zeta(\alpha))$. In the following, we will simply write $\zeta(\alpha)$ instead of $\bar{\Sigma}(\zeta(\alpha))$ for more simplicity. At step $m$:

    i)      $\widehat{\omega}_m$ is performing a minimum value of $\min_\omega \langle \omega, \Phi_m \rangle$ which is $E[V_A|S, \Im]_{\Phi_m}$

    ii)     $\Psi_{m+1}$ is chosen in $\zeta(\alpha)$

    iii)    $\sigma_{m+1}$ in chosen such that $\langle \widehat{\omega}_m, \Psi_{m+1} \circ \sigma_{m+1} \rangle \geq \frac{C_1(\alpha)}{k^2}$ which is always possible as per proof of Theorem 2

    iv)    $\Phi_{m+1} = \Psi_{m+1} \circ \sigma_{m+1}$

Process 2:

$\Phi_0$ is a distribution in $\zeta(\alpha)$; at step $m$:

    i)      $\widehat{\omega}_m$ is performing a minimum value of $\min_\omega \langle \omega, \frac{1}{m}\sum_{s=1}^{m} \Phi_s \rangle$

    ii)     $\Psi_{m+1}$ is chosen in $\zeta(\alpha)$

    iii)    $\sigma_{m+1}$ in chosen such that $\langle \widehat{\omega}_m, \Psi_{m+1} \circ \sigma_{m+1} \rangle \geq \frac{C_1(\alpha)}{k^2}$ which is always possible as per proof of Theorem 2

    iv)    $\Phi_{m+1} = \Psi_{m+1} \circ \sigma_{m+1}$

The process 1 involves fast variations, but, if the choice of $\Psi_{m+1}$ and $\sigma_{m+1}$ is deterministic at each step $m$, it can have short period (typically period of 2), which makes it unsecure. The process 2 has no period but its variations are slowing down when $m$ is increasing. We can justify that the process 2 has no period by seeing that, if it would have a period, then $\frac{1}{m}\sum_{s=1}^{m}\Phi_s$ would converge, which would contradict the fact that $\langle \widehat{\omega}_m, \Phi_{m+1} \rangle \geq \frac{c_1(\alpha)}{k^2} \gg \langle \widehat{\omega}_m, \Phi_m \rangle = O\left(\frac{1}{nk}\right)$ as shown in Proposition 6 of Annex 1.

Thus, by combining the two process, we get a sequence with both fast variations and no periodic behavior. A method to combine both process is the following:

The Recursive Generation Process:

$\Phi_0$ is a distribution in $\zeta(\alpha)$ ; at step $m$ :

i)      $\widehat{\omega}_m$ is performing a minimum value of $\min_\omega \langle \omega, \Phi_m \rangle$

ii)      $\widehat{\omega}'_m$ is performing a minimum value of $\min_\omega \langle \omega, \frac{1}{m}\sum_{s=1}^{m}\Phi_s \rangle$

iii)      $\Psi_{m+1}$ and $\Psi'_{m+1}$ are chosen in $\zeta(\alpha)$

iv)      $\sigma_{m+1}$ is chosen such that $\langle \widehat{\omega}_m, \Psi_{m+1} \circ \sigma_{m+1} \rangle \geq \frac{c_1(\alpha)}{k^2}$ which is always possible as per proof of Theorem 2

v)      $\sigma'_{m+1}$ is chosen such that $\langle \widehat{\omega}'_m, \Psi'_{m+1} \circ \sigma'_{m+1} \rangle \geq \frac{c_1(\alpha)}{k^2}$ which is always possible as per proof of Theorem 2

vi)      $\Phi_{m+1} = \alpha_{m+1}\Psi_{m+1} \circ \sigma_{m+1} + (1-\alpha_{m+1})\Psi'_{m+1} \circ \sigma'_{m+1}$ where $\alpha_{m+1} \in [0,1]$ can be picked randomly

Such a process is secure « against the past », even if the choices at each step are deterministic, but it is not secure « against the future » if the choices at each step are deterministic. Being secure « against the future » means that if the opponent runs the recursive process on its own and is « in advance » compared to the legitimate partner, it still cannot obtain knowledge about the current value of $\Phi_m$. So, for the process to remain secure also against the future, it is necessary that the choices at each step involve randomness (but not necessarily Deep Randomness) with maximum possible entropy among $\zeta(\alpha)$.

It is also important that the DRG is capable to pick at each step a new distribution within the widest possible sub-set of $\zeta(\alpha)$, otherwise, any restriction in such possible 'picking set' would add prior information for the opponent about the possible distribution, and would therefore reduce $\mathfrak{I}_< \supseteq \zeta(\alpha)$.

To that regards, it can be noted that, in the above algorithm, the distribution $\Psi_{m+1}$ (resp. $\Psi'_{m+1}$) can actually be chosen tidied in $I_0$ (e.g. $\sigma_{\Psi_{m+1}}(I_0) = \sigma_{\Psi'_{m+1}}(I_0) = I_0$) because anyway it is further recomposed by a permutation $\sigma_{m+1}$ (resp. $\sigma'_{m+1}$). From there, we also remark that one can reduce the entropy of choosing a new distribution tidied in $I_0$ by observing that, for any such distribution $\Phi$ and for any $\sigma, \mu \in S_{I_0} \times S_{\overline{I_0}}$, a partner has no reason to choose $\Phi$ rather than $\Phi \circ \sigma \circ \mu$, and therefore any such distribution can be considered of the form:

$$F_{DRG} \triangleq \left\{ \frac{1}{\left((n/2)!\right)^2} \sum_{\substack{\sigma \in S_{I_0} \\ \mu \in S_{\overline{I_0}}}} \Phi \circ \sigma \circ \mu \mid \Phi \in \zeta(\alpha) \right\}$$

Then, by denoting

$$L_\Phi(r,s) \triangleq \int_{\substack{|x_{|I_0}|=r \\ |x_{|\overline{I_0}}|=s}} \Phi(x)dx$$

it is clear with a bit of attention that:

$$\frac{1}{\left((n/2)!\right)^2} \sum_{\substack{\sigma \in S_{I_0} \\ \mu \in S_{\overline{I_0}}}} \Phi \circ \sigma \circ \mu(x) = \frac{2^n}{\binom{n/2}{|x_{|I_0}|}\binom{n/2}{|x_{|\overline{I_0}}|}} L_\Phi\left(|x_{|I_0}|, |x_{|\overline{I_0}}|\right)$$

and thus $F_{DRG}$ is isomorph to a subset of the 2 dimensional set:

$$\left\{ f : \mathbb{N}_{n/2} \mapsto [0,1] \,\middle|\, \sum_{r,s \in \mathbb{N}_{n/2}} f(r,s) = 1 \right\}$$

The exact subset is conditioned by the constraint $\Phi \in \zeta(\alpha)$. Let's express this exact subset. By Lemma 1, we know that, for $\Phi \in \zeta(\alpha)$

$$\int_{\{0,1\}^n} \left( \frac{|x||y|}{n^2} - \frac{x \cdot y}{n} \right)^2 \Phi(x)\Phi(y)dxdy \geq C(\alpha) + O\left(\frac{1}{n}\right)$$

Now, if $\Phi$ also belongs to $F_{DRG}$ we can directly calculate the left term and obtain:

$$\int_{\{0,1\}^n} \left( \frac{|x||y|}{n^2} - \frac{x \cdot y}{n} \right)^2 \Phi(x)\Phi(y)dxdy = \left( \sum_{r,s \in \mathbb{N}_{n/2}} \frac{(r-s)^2}{n^2} L_\Phi(r,s) \right)^2 + O\left(\frac{1}{n}\right)$$

Finally, we obtain the simpler expression of $F_{DRG}$ that clearly represents $F_{DRG}$ a 2-dimensional set with reduced entropy of $O(n^2)$:

$$F_{DRG} = \left\{ \Phi(x) = \frac{2^n}{\binom{n/2}{|x_{|I_0}|}\binom{n/2}{|x_{|\overline{I_0}}|}} L_\Phi\left(|x_{|I_0}|, |x_{|\overline{I_0}}|\right) \middle| L_\Phi : \mathbb{N}_{n/2} \mapsto [0,1], \sum_{r,s \in \mathbb{N}_{n/2}} \frac{(r-s)^2}{n^2} L_\Phi(r,s) \right.$$

$$\left. \geq \sqrt{C(\alpha)} + O\left(\frac{1}{n}\right) \right\}$$

**Argument about the minimum number of steps to reach the maturity period of the Recursive Generation Process**

We consider a convex and compact subset $\Omega$ of possible strategies for the opponent (for any subset, its closed convex envelop corresponds to that characteristic). We consider in the followings the sequence of probability distributions $\{\Phi_m\}_{m \in \mathbb{N}^*}$ constructed as follows :

$$\Phi_1 = \Phi$$

where $\Phi$ is a sleeked distribution in $\zeta(\alpha)$ as defined in Annex 2 (Proposition 12), called the seed of the sequence. This time $\widehat{\omega}_m$ is performing a minimum value in:

$$\min_{\omega \in \Omega} \frac{1}{m} \sum_{s=1}^{m} \langle \omega, \Phi_s \rangle$$

$\Phi_{m+1}$ is chosen such that:

$$\langle \widehat{\omega}_m, \Phi_{m+1} \rangle \geq \frac{C_3}{n}$$

which is always possible, as stated by Theorem 3 (introduced and proved in Annex 2).

**Diagonal Sequence Property.**

*There exists two constants $C'$ and $C''$ such that, for any $N$ verifying $\frac{N}{\ln(N)} \geq C' \dim(\Omega)$,*

$$\min_{\omega \in \Omega} \frac{1}{N} \sum_{s=1}^{N} \langle \omega, \Phi_s \rangle \geq \frac{C''}{n}$$

∎

More precisely, we chose $\Phi_{m+1}$ as :

$$\Phi_{m+1} = \frac{1}{2}(1 + \Phi \circ \sigma_m) \qquad (IV.1)$$

where $\sigma_m$ is a permutation such that $\langle \widehat{\omega}_m, \Phi \circ \sigma_m \rangle \geq \frac{C_3}{n}$ as allowed by Theorem 3 (i).

Let's set:

$$\widehat{Q}_m(\omega - \widehat{\omega}_m) \triangleq \frac{1}{m} \sum_{s=1}^{m} \langle \omega, \Phi_m \rangle - \frac{1}{m} \sum_{s=1}^{m} \langle \widehat{\omega}_m, \Phi_m \rangle$$

$$\widehat{\Sigma}_m \triangleq \frac{1}{m} \sum_{s=1}^{m} \langle \widehat{\omega}_m, \Phi_s \rangle$$

$$\langle \omega, \Phi_m \rangle = Q_m(\omega - \omega_m) + \theta_m$$

$\widehat{Q}_m$ and $Q_m$ are positive quadratic forms over $\omega \in \Omega$ derived by respective distributions $\frac{1}{m} \sum_{s=1}^{m} \Phi_s(x) \Phi_s(y)$ and $\Phi_m(x) \Phi_m(y)$. $(IV.1)$ ensures that $\widehat{Q}_m$ and $Q_m$ are non-degenerate in $\Omega$. By considering a basis that is simultaneously $\widehat{Q}_m$-orthonormal and $Q_m$-orthogonal, we obtain the 2 following relations (remind that $\widehat{\omega}_m$ is the minimum over the quadratic form $\widehat{Q}_m$):

$$\frac{1}{m} \sum_{s=1}^{m} \langle \omega, \Phi_s \rangle - \frac{1}{m} \sum_{s=1}^{m} \langle \widehat{\omega}_m, \Phi_s \rangle = \|\omega - \widehat{\omega}_m\|_{\widehat{Q}_m}^{\ 2}$$

$$|\langle \omega, \Phi_{m+1} \rangle - \langle \widehat{\omega}_m, \Phi_{m+1} \rangle| \leq 2\sqrt{\dim(\Omega)} \|\omega - \widehat{\omega}_m\|_{\widehat{Q}_m}$$

We have then:

$$\frac{1}{m+1} \sum_{s=1}^{m+1} \langle \omega, \Phi_s \rangle$$

$$= \frac{m}{m+1} \left( \widehat{\Sigma}_m + \left( \frac{1}{m} \sum_{s=1}^{m} \langle \omega, \Phi_s \rangle - \frac{1}{m} \sum_{s=1}^{m} \langle \widehat{\omega}_m, \Phi_s \rangle \right) \right)$$

$$+ \frac{1}{m+1} \left( \langle \omega, \Phi_{m+1} \rangle - \langle \widehat{\omega}_m, \Phi_{m+1} \rangle \right) + \frac{1}{m+1} \langle \widehat{\omega}_m, \Phi_{m+1} \rangle \qquad (IV.2)$$

Applying the 2 above relations to $(IV.2)$, we get:

$$\widehat{\Sigma}_{m+1} \geq \frac{m}{m+1} \widehat{\Sigma}_m + \frac{m}{m+1} \|\omega - \widehat{\omega}_m\|_{\widehat{\Omega}}^2 - \frac{2\sqrt{\dim(\Omega)}}{m+1} \|\omega - \widehat{\omega}_m\|_{\widehat{\Omega}} + \frac{1}{m+1} \frac{C_3}{4n}$$

$$\geq \frac{m}{m+1} \widehat{\Sigma}_m + \frac{1}{m+1} \frac{C_3}{4n} - \frac{\dim(\Omega)}{m(m+1)}$$

where the last inequality is obtained by taking the minimum of $\frac{m}{m+1} X^2 - \frac{2\sqrt{\dim(\Omega)}}{m+1} X$ over $X$.

Let's consider the recurring sequence:

$$(m+1) X_{m+1} = m X_m + \frac{C_3}{4n} - \frac{\dim(\Omega)}{m}$$

$$X_1 = \widehat{\Sigma}_1$$

By recurring argument, $\widehat{\Sigma}_m \geq X_m$ and thus:

$$\widehat{\Sigma}_m \geq \frac{C_3}{4n} - \dim(\Omega) \, O\left( \frac{\ln(m)}{m} \right)$$

which achieves the proof. ∎

If we consider first $\Omega$ as the full set of strategies (e.g. $\left[0, \frac{1}{k}\right]^{2^{2n}}$), the dimension is $2^{2n}$ and thus the constructing process must be iterated exponential times to reach positive lower bound. On the other hand, we make the excessive assumption that the opponent has a full knowledge of the distributions chosen by the DRG at each step $m$ (of the sequences $\{\Phi_s\}_{s \leq m}$) to be able to build optimal $\widehat{\omega}_m$. In practice, if we consider the protocol $\mathcal{P}$ introduced in Section III, we can see that $\xi$ has never more knowledge about a distribution than a draw of 1 degraded vector $i$, which means that we can restrict the set of strategies to a subset $\Omega^*$ with dimension $O(n)$, and thus the recommended number of iterations for a recursive constructing process before being able to trustedly pick a distribution is $O(n \ln(n))$. This result also means that the generated sequence is $O(k)$-undistinguishable under the definition given in Section I, for for $\frac{N}{\ln(N)} \geq C' \dim(\Omega)$.

With the same reasoning than the one of Diagonal Sequence Property, we can obtain a more general result (with notations introduced in Lemma 3):

**Generalized Diagonal Sequence Property.**

*There exists two constants $C'(\alpha)$ and $C''(\alpha)$ such that, for any $N$ verifying $\frac{N}{\ln(N)} \geq C' \dim(\Omega)$, and for any $\{\Phi_1, \dots, \Phi_N\} \in \zeta(\alpha)$ and $\{\Phi'_1, \dots, \Phi'_N\} \in \zeta(\alpha)$, and any permutation $\sigma_s$ synchronizing $\Phi_s$ and $\Phi'_s$ :*

$$\min_{\omega \in \Omega} \frac{1}{N} \sum_{s=1}^{N} \langle \omega, \Phi_s, \Phi'_s \circ \sigma_s \rangle \geq \frac{C''}{n}$$

This result also means that when $\{\Phi_1, \dots, \Phi_N\}$ and $\{\Phi'_1, \dots, \Phi'_N\}$ are generated by DRGs, from the principle of objectivity explained in Section I, an opponent should choose the same strategy at each step $s$, and therefore the above inequality implies that whatever is that strategy, in average we have $\langle \omega, \Phi_s, \Phi'_s \circ \sigma_s \rangle \gg_k E[(V_A - V_B)^2 | S] = O(1/nk)$ ; which gives Advantage Distillation for $\frac{N}{\ln(N)} \geq C' \dim(\Omega)$.

## V.    Conclusion and further directions

We proved that Information Theoretical Security could be reached under Deep Random assumption.

Let's remark that when the alphabets $\mathcal{X}$, $\mathcal{Y}$ and $\mathcal{Z}$ introduced in Section I are all equal to $\{0,1\}$, and if each digit issued by the protocol is managed independently of the others, then the Security Property:

$$R^*(\mathcal{P}, Z) \triangleq H(X|Z) - H(X|Y) \geq (1-\epsilon)H(X) \qquad (SP)$$

introduced in Section I can equivalently be written as: whatever are $\varepsilon, \varepsilon' > 0$, there exists a set of parameters $U$ for the protocol $\mathcal{P}$ such that the two conditions above are verified:

(i)     $2\left(\min\big(P(e_A \neq e_B), 1 - P(e_A \neq e_B)\big)\right) \leq \varepsilon$

(ii)    $2\left(\max\big(P(e_\xi = e_B), 1 - P(e_\xi = e_B)\big) - \frac{1}{2}\right) \leq \varepsilon'$

where the opponent $\xi$ is passive, with unlimited access to the public information and with unlimited power of storage and calculation; $e_A, e_B, e_\xi$ being estimations of an average digit of information exchanged between $A$ and $B$ through $\mathcal{P}$ respectively from $A, B, \xi$.

The obtained result, apparently contradicting Shannon's pessimistic theorem, is possible thanks to the nature of Deep Randomness, that prevents the use of Bayesian inference from public information, because not only draws but also probability distributions themselves are unknown to the beholding opponent $\xi$.

We also proposed a method to generate Deep Randomness (Section IV) from classical calculation resources, thanks to recursive and continuously executed algorithm, based on Cantor' style diagonal constructing method.

In this work, our main objective was to expose this new idea and to prove the existence of a working protocol. But of course, once done, the next question is: if such protocol exists, what is the best one? In cryptology and communication sciences, there exist many criteria of quality for a protocol. We will concentrate here on the most critical one for practical implementation which is the burden of exchanged data needed to obtain an Information Theoretically secure digit of information.

Considering a ITSP $\mathcal{P}$, we define

$$\mathcal{Q}_{X,Y}(\mathcal{P})$$

as the quantity of digit exchanged by $A$ and $B$ on the main channel through the protocol $\mathcal{P}$ so that $A$ and $B$ can collect their respective value of $X$ and $Y$.

The question raised above comes down to determine the maximum:

$$C = \sup_{\mathcal{P}, G \in \Gamma_3} \left( \frac{\inf_{Z \in \Omega_3(G,\mathcal{P})} R^*(\mathcal{P}, Z)}{\mathcal{Q}_{X,Y}(\mathcal{P})} \right)$$

The quantity $C$ is called the Cryptologic Limit. It is then a constant in $[0,1]$. The Theorem 1 can then be formulated simply as:

**Theorem:**

*Under the Deep Random assumption, $C > 0$*

The research of the Cryptologic Limit opens, to my opinion, an interesting field of investigations. Empirical research of specifically designed protocols may give rapidly interesting improvements to the presently proposed example; estimation of performance can be easily done by simulation rather than painful theoretical examination of upper-bounds. But the theoretical full resolution of the cryptologic limit question will need new ideas.

I hope this challenge will create enthusiasm for the largest possible number of curious minds interested in cryptology, information theory and communication sciences.

**Who is the author ?**

I have been an engineer in calculater science for 20 years. My professional activities in private sector are related to IT Security and digital trust, but have no relation with my personal research activity in the domain of cryptology. If you are interested in the topics introduced in this article, please feel free to establish first contact at tdevalroger@gmail.com

## VI. Annex 1: Proofs of Theorem 1 & 2

The Annex contains 4 parts:

- A first part with basic results regarding Bernoulli random variables (Propositions 1 to 7).
- A second part with basic results regarding sampled binary random variables (Propositions 8).
- A third part with preliminary results regarding the approximated behavior of the opponent under Deep Random assumption (Propositions 9 and 10).
- And lastly the proofs of Theorem 1 and 2, corresponding to the properties $(P)$ and $(P')$ to be satisfied by the protocol $\mathcal{P}$.

### Basic results with Bernoulli random variables

**Proposition 1.**

> For all $i \in \{0,1\}^n$ and $l \in \mathbb{N}_n^*$, stands the relation : $\psi_{i,l}\left(\frac{x}{k}\right) = \sum_{r=l}^{|i|} \beta_{l,r}\left(\frac{1}{k}\right)\psi_{i,r}(x)$

■ Let's start by expressing the multinomial $\chi_i\left(\frac{x}{k}\right)$ in the basis $\{\chi_i(x)\}_{i \in \{0,1\}^n}$ :

$$
\begin{aligned}
\chi_j\left(\frac{x}{k}\right) &= \frac{1}{k^{|j|}}\sum_{u \supset j}\frac{(-1)^{|u\setminus j|}}{k^{|u\setminus j|}}\Pi_u(x) \\
&= \frac{1}{k^{|j|}}\sum_{u \supset j}\frac{(-1)^{|u\setminus j|}}{k^{|u\setminus j|}}\sum_{u' \supset u}\chi_{u'}(x) \\
&= (-1)^{|j|}\sum_{u \supset j}\left(\sum_{j \subset v \subset u}\left(-\frac{1}{k}\right)^{|v|}\right)\chi_u(x) \\
&= (-1)^{|j|}\sum_{u \supset j}\left(\sum_{r=|j|}^{|u|}\binom{|u|-|j|}{r-|j|}\left(-\frac{1}{k}\right)^r\right)\chi_u(x) \\
&= \left(\frac{1}{k}\right)^{|j|}\sum_{u \supset j}\left(1-\frac{1}{k}\right)^{|u|-|j|}\chi_u(x)
\end{aligned}
$$

We thus have :

$$
\begin{aligned}
\psi_{i,l}\left(\frac{x}{k}\right) &= \sum_{j|i.j=l}\left(\frac{1}{k}\right)^{|j|}\sum_{s \supset j}\left(1-\frac{1}{k}\right)^{|s|-|j|}\chi_s(x) \\
&= \sum_{s|i.s \geq l}\left(\sum_{\substack{j.i=l \\ j \subset s}}\left(\frac{1}{k}\right)^{|j|}\left(1-\frac{1}{k}\right)^{|s|-|j|}\right)\chi_s(x) \\
&= \sum_{s|i.s \geq l}\left(1-\frac{1}{k}\right)^{|s|}\left(\sum_{\substack{j.i=l \\ j \subset s}}\left(\frac{1}{k-1}\right)^{|j|}\right)\chi_s(x)
\end{aligned}
$$

Then :

$$
\begin{aligned}
\sum_{\substack{j.i=l \\ j \subset s}}\left(\frac{1}{k-1}\right)^{|j|} &= \sum_{\substack{j.i=l \\ j \subset s}}\left(\frac{1}{k-1}\right)^{i.j+|j\cap(s\setminus i)|} = \sum_{j \subset (s\setminus i)}\sum_{\substack{j' \subset (s\cap i) \\ |j'|=l}}\left(\frac{1}{k-1}\right)^{l+|j|} \\
&= \left(\frac{1}{k-1}\right)^l\binom{|s\cap i|}{l}\sum_{j \subset (s\setminus i)}\left(\frac{1}{k}\right)^{|j|} = \left(\frac{1}{k-1}\right)^l\binom{|s\cap i|}{l}\sum_{t=0}^{|s\setminus i|}\left(\frac{1}{k-1}\right)^t\binom{|s\setminus i|}{t}
\end{aligned}
$$

$$= \left(\frac{1}{k-1}\right)^l \binom{|s \cap i|}{l} \left(\frac{k}{k-1}\right)^{|s|-i.s}$$

And finally,

$$\psi_{i,l}\left(\frac{x}{k}\right) = \sum_{s|i.s \geq l} \left(\frac{1}{k-1}\right)^l \binom{|s \cap i|}{l} \left(\frac{k-1}{k}\right)^{i.s} \chi_s(x) = \sum_{r=l}^{|i|} \beta_{l,r}\left(\frac{1}{k}\right) \psi_{i,r}(x) \qquad \blacksquare$$

Proposition 2 gives an Chernoff-style upper bound:

**Proposition 2.**

(i)   *Tail inequality: for all $\Delta$, we have the upper bound :* $\beta_{l,kl+\Delta}\left(\frac{1}{k}\right) \leq e^{-\frac{\Delta^2}{2k^2 l}}$

(ii)  *Central approximation: $\forall\, u, |u| \leq u_m \ll \theta w$, there exists two constants $C_1, C_2$ such that :* $\beta_{w\theta+u,w}(\theta) = N(u, w\sigma^2)\left(1 + \triangleleft\left(C_1 \frac{u_m}{\theta w}\right)\right) e^{\triangleleft\left(C_2 \frac{u_m{}^3}{w^2 \theta^2}\right)}$ *with variance*
$$\sigma^2 \triangleq \theta(1-\theta)$$

■ (i) We can write the quotient:

$$\frac{\beta_{l,kl+\Delta}\left(\frac{1}{k}\right)}{\beta_{l,kl}\left(\frac{1}{k}\right)} = \prod_{s=1}^{\Delta} \frac{(k-1)(kl+s)}{k((k-1)l+s)} = \prod_{s=1}^{\Delta} \frac{1 + \frac{(k-1)s}{k(k-1)l}}{1 + \frac{ks}{k(k-1)l}}$$

$$= exp\left(\sum_{s=1}^{\Delta}\left(ln\left(1 + \frac{(k-1)s}{k(k-1)l}\right) - ln\left(1 + \frac{ks}{k(k-1)l}\right)\right)\right)$$

Considering that $f(x) = \ln(1+x)$ is 1-lipschitz over $\mathbb{R}^+$ and increasing, we then conclude easily :

$$\frac{\beta_{l,kl+\Delta}\left(\frac{1}{k}\right)}{\beta_{l,kl}\left(\frac{1}{k}\right)} \leq exp\left(-\sum_{s=1}^{\Delta} \frac{s}{k(k-1)l}\right) \leq e^{-\frac{\Delta^2}{2k^2 l}}$$

(ii) We can consider $\beta_{w\theta+u,w}(\theta)$ as continuous in $u$ by extending $n!$ to $\Gamma(n)$. By Stirling formula, there exists $C$ such that :

$$\Gamma(x) = \sqrt{2\pi} x^{x+\frac{1}{2}} e^{-x}\left(1 + \frac{C(x)}{x}\right) \text{ with } C(x) \leq C$$

On the other hand, we have, by Taylor's inequality, for any , $|y| \leq y_m < 1$ :

$$\left|-\ln(1-y) - \left(y + \frac{y^2}{2}\right)\right| \leq \frac{y_m{}^3}{3(1-y_m)^3}$$

We then develop $\beta_{w\theta+u,w}(\theta)$ using the two above inequalities, which gives, by noting $= w\theta + u$ :

$$\beta_{w\theta+u,w}(\theta) = \sqrt{\frac{w}{2\pi r(w-r)}} \frac{1 + \frac{C(w)}{w}}{\left(1 + \frac{C(r)}{r}\right)\left(1 + \frac{C(w-r)}{w-r}\right)} exp\left(-\frac{u^2}{2w\theta(1-\theta)}\right.$$

$$\left. + \frac{u^3}{2w^2}\left(\frac{1}{\theta^2} - \frac{1}{(1-\theta)^2}\right) + O\left(\theta w \frac{u_m{}^3}{w^3\theta^3}\right)\right)$$

There exists two constants $C_1, C_2$ such that :

$$\sqrt{\frac{w}{2\pi r(w-r)}} \frac{1 + \frac{C(w)}{w}}{\left(1 + \frac{C(r)}{r}\right)\left(1 + \frac{C(w-r)}{w-r}\right)} = \sqrt{\frac{1}{2\pi w\theta(1-\theta)}}\left(1 + \triangleleft\left(C_1\frac{u_m}{\theta w}\right)\right)$$

$$exp\left(\frac{u^3}{2w^2}\left(\frac{1}{\theta^2} - \frac{1}{(1-\theta)^2}\right) + O\left(\theta w \frac{u_m{}^3}{w^3\theta^3}\right)\right) = e^{\triangleleft\left(C_2\frac{u_m{}^3}{w^2\theta^2}\right)}$$

which ends the proof.

∎

As a consequence of Propositions 1 & 2, we obtain in Proposition 3 below a Chernoff-style bound for the legitimate partners. The classical Chernoff bound results are not efficient at neighborhood of zero (which, in our context, is due to the increasing of the degradation parameter $k$), this is why we need to adapt them.

In the following, $A$ and $B$ are two partners, owning private parameter vectors $x/k$ and $y/k$ respectively (where $k$ is a real number parameter $> 1$), and generating vectors $i$ and $j$ respectively through Bernoulli random variables from $x/k$ and $y/k$ respectively.

**Proposition 3.**

*Let $a$ be a positive integer such that $a < \frac{x.y}{k}$, then we have the following upper bound for the legitimate partner evaluation gap (with notations of section II):*

$$P(|U_A - U_B| \geq 2a|x,y) \leq 2ne^{-\frac{na^2}{2E[U_A]}}$$

■ We will suppose in the following that $x, y \in \{0,1\}^n$. We then have:

$$P(|U_A - E[U_A]| \geq a|x,y) = \sum_{|r-nE[U_A]|\geq na} \sum_{x.j=r}{}_j \chi_j\left(\frac{y}{k}\right)$$

$$= \sum_{|r-nE[U_A]|\geq na} \sum_{s=r}^{n} \beta_{r,s}\left(\frac{1}{k}\right) \psi_{x,s}(y) \text{ from Proposition 1.}$$

and noticing that $\psi_{x,s}(y) = \begin{cases} 1 \ if \ x.y = s \\ 0 \quad else \end{cases}$ :

$$= \sum_{|r-nE[U_A]|\geq na} \beta_{r,x.y}\left(\frac{1}{k}\right) \leq ne^{-\frac{na^2}{2E[U_A]}} \text{ from Proposition 2 (i).}$$

(notice also that $\left(\left(r \leq \frac{x.y}{k} - na\right) or \left(r \geq \frac{x.y}{k} + na\right)\right) \Rightarrow \left((x.y \leq kr - kna) or (x.y \geq kr + kna)\right)$. We conclude by writing that:

$$E\left[\frac{x.j}{n}|x,y\right] = E\left[\frac{i.y}{n}|x,y\right] = \frac{x.y}{nk}$$

$$P(|U_A - U_B| \geq 2a|x,y) \leq P(|U_A - E[U_A]| \geq a|x,y) + P(|U_B - E[U_B]| \geq a|x,y)$$

■

Proposition 4 below, gives an upper-bound of gap between the legitimate partners' respective estimations in the favorable cases where $(\sigma_A = \sigma_{\Phi'})$ & $(\sigma_B = \sigma_\Phi)$.

**Proposition 4.**

(i)    $E\left[\left(\frac{x.j}{n} - \frac{i.y}{n}\right)^2 |x,y\right] \leq \frac{2}{nk}$

(ii)    With the notations of the protocol $\mathcal{P}$, $E[(V_B - V_A)^2|(\sigma_A = \sigma_{\Phi'})$ & $(\sigma_B = \sigma_\Phi)] \leq \frac{2}{nk}$

■ (i) By direct calculation:

$$E\left[\left(\frac{x.j}{n} - \frac{i.y}{n}\right)^2 |x,y\right] = \frac{1}{n^2 k}\sum_{r=1}^{n} x_r y_r \left(x_r + y_r - \frac{2x_r y_r}{k}\right) \leq \frac{2}{nk}$$

Then (ii) comes from (i) by seeing that:

$$E[(V_B - V_A)^2|(\sigma_A = \sigma_{\Phi'})$ & $(\sigma_B = \sigma_\Phi)]$$
$$= \int_{x,y\in[0,1]^n}\sum_{i,j\in\{0,1\}^n} E\left[\left(\frac{x.j}{n} - \frac{i.y}{n}\right)^2 |x,y\right]\Phi \circ \sigma_\Phi(x)\Phi' \circ \sigma_{\Phi'}(y)dxdy$$

■

Proposition 5 below, gives an upper-bound of gap between estimations of 2 stakeholders in case where both partners' distribution is symmetric. $\Phi$ is symmetric means that $\Phi \circ \sigma = \Phi$ for any permutation $\sigma \in \mathfrak{S}_n$. We remind that $V_{\xi,1}(i,j) \triangleq \frac{k|i||j|}{n^2}$.

**Proposition 5.**

$\Phi$ and $\Phi'$ are two distributions. If $\Phi$ or $\Phi'$ is symmetric, then:

$$E\left[\left(V_{\xi,1}(i,j) - \frac{x.j}{n}\right)^2\right] \leq \frac{2}{nk}$$

■ By direct calculation we get:

$$E\left[\left(V_{\xi,1}(i,j) - \frac{x.j}{n}\right)^2\right]$$

$$= \frac{n-1}{n^2}\left(\left(v'\left(\frac{w}{k} - \frac{v}{k^2}\right) + v\left(\frac{w'}{k} - \frac{v'}{k^2}\right)\right) + \frac{(u-v)(u'-v')}{k^2}\right) + \frac{1}{n^2}\left(ww' - \frac{vv'}{k^2}\right) =$$

$$\lhd \left(\frac{2}{nk}\left(1 - \frac{1}{2k}\right)\right)$$

$$\text{with} \begin{cases} u = \int_{[0,1]^n} x_r x_s \Phi(x)dx \; ; \; u' = \int_{[0,1]^n} x_r x_s \Phi'(x)dx \; ; \; \forall r \neq s \\ v = \int_{[0,1]^n} x_r{}^2 \Phi(x)dx \; ; \; v' = \int_{[0,1]^n} x_r{}^2 \Phi'(x)dx \\ w = \int_{[0,1]^n} x_r \Phi(x)dx \; ; \; w' = \int_{[0,1]^n} x_r \Phi'(x)dx \end{cases}$$

■

Proposition 6 below confirms the obvious intuition that, if the opponent knows the distributions, then at least in the favorable cases, his estimation is at least as good as the one of the legitimate receiver:

**Proposition 6.**

*With the notations of Section II, we have :* $\inf_{\omega} E\left[\left(\omega_{i,j} - \frac{x.j}{n}\right)^2\right] \leq E\left[\left(\frac{x.j}{n} - \frac{i.y}{n}\right)^2\right].$

■ we write $E\left[\left(\frac{x.j}{n} - \frac{i.y}{n}\right)^2\right]$ and simply apply Schwarz inequality:

$$E\left[\left(\frac{x.j}{n} - \frac{i.y}{n}\right)^2\right] = \int_{x,y\in[0,1]^n} \sum_{i,j\in\{0,1\}^n} \left(\frac{x.j}{n} - \frac{i.y}{n}\right)^2 \chi_i\left(\frac{x}{k}\right)\chi_j\left(\frac{y}{k}\right)\Phi(x)\Phi'(y)dxdy$$

$$= \int_y \sum_{i,j}\left[\int_x \left(\frac{x.j}{n} - \frac{i.y}{n}\right)^2 \chi_i\left(\frac{x}{k}\right)\Phi(x)dx\right]\chi_j\left(\frac{y}{k}\right)\Phi'(y)dy$$

$$\geq \int_y \sum_{i,j}\left(\int_x \chi_i\left(\frac{x}{k}\right)\Phi(x)dx\right)\left(\frac{\int_x \frac{x.j}{n}\chi_i\left(\frac{x}{k}\right)\Phi(x)dx}{\int_x \chi_i\left(\frac{x}{k}\right)\Phi(x)dx} - \frac{i.y}{n}\right)^2 \chi_j\left(\frac{y}{k}\right)\Phi'(y)dy$$

By remarking that $\frac{\int_x \frac{x.j}{n}\chi_i\left(\frac{x}{k}\right)\Phi(x)dx}{\int_x \chi_i\left(\frac{x}{k}\right)\Phi(x)dx}$ is of the form $\omega(i,j)$ the result follows. The result also applies equivalently to $E\left[\left(V_\xi^t(i,j) - V_A\right)^2\right] \leq E[(V_B - V_A)^2]$ by taking distributions $\Phi \circ \sigma_\Phi$ and $\Phi' \circ \sigma_{\Phi'}$ ■

Lastly, the following is a very simple result for quadratic upper bounds:

**Proposition 7.**

*Let $f$ and $g : I \to \mathbb{R}$ be two square-integrable functions, $\alpha \in ]0,1[$, and $\phi$ be a probability distribution function over $I$, with $\int_I f(x)^2\phi(x)dx \geq A$ and $\int_I g(x)^2\phi(x)dx \leq a \leq A\left(\frac{\alpha}{1-\alpha}\right)^2$, then:*

$$\int_I (\alpha f(x) + (1-\alpha)g(x))^2\phi(x)dx \geq \alpha^2 A\left(1 - \frac{1-\alpha}{\alpha}\sqrt{\frac{a}{A}}\right)^2$$

■ Obvious from Cauchy-Schwarz inequality:

$$\int_I (\alpha f(x) + (1-\alpha)g(x))^2 \phi(x)dx$$

$$\geq \alpha^2 \int_I f(x)^2 \phi(x)dx + (1-\alpha)^2 \int_I g(x)^2 \phi(x)dx - 2\alpha(1-\alpha) \int_I |f(x)||g(x)|\phi(x)dx$$

$$\geq \alpha^2 \int_I f(x)^2 \phi(x)dx + (1-\alpha)^2 \int_I g(x)^2 \phi(x)dx - 2\alpha(1-\alpha) \left(\int_I f(x)^2 \phi(x)dx\right)^{1/2} \left(\int_I g(x)^2 \phi(x)dx\right)^{1/2}$$

■

## Basic results on sampled binary random variables

The following propositions is useful to establish independence of binary random variables when they are generated by sampling from 2 independent random variables with value in an Euclid space, typically $U$ and $V = U - E[U|I]$. We denote $\Gamma$ the sampling function:

$$\Gamma_\theta(x) \triangleq \left\lfloor \frac{x}{\theta} \right\rfloor \bmod 2$$

where $\theta$ is a real number such that $0 < \theta \ll 1$.

### Proposition 8.

Let $U$ and $V$ be 2 independent random variables with values in $[0,1]$. Let also $\rho$ be a random variable independent of $U$ and $V$, and equi-distributed over $[0,2\theta[$. Then:

(i) the variable $e_V = \Gamma_\theta(U + V + \rho) \oplus \Gamma_\theta(U + \rho)$ is independent of the variable $e_U = \Gamma_\theta(U + \rho)$.

(ii) If $|U - V| < \theta$, then $E[\Gamma_\theta(U - \rho) \oplus \Gamma_\theta(V - \rho)|U,V,\theta] = \frac{|U-V|}{\theta}$

(iii) Let $\theta_0 > 0$, $\lambda > 1$ be two real positive numbers, and $\theta$ be a random variable, independent of $U$ and $V$, and distributed according to the distribution $g(\theta)d\theta = \frac{\lambda \theta_0}{(\lambda-1)} \frac{d\theta}{\theta^2}$ over $\theta \in [\theta_0, \lambda\theta_0]$; then

$$E[\Gamma_\theta(U - \rho) \oplus \Gamma_\theta(V - \rho)|U,V] = \frac{1}{2} + \triangleleft \left(\frac{2\lambda\theta_0}{(\lambda-1)|U-V|}\right)$$

(iv) If $E[(U - V)^2] = m^2 > 0$ and $\sup|U - V| = M > m$, then

$$E[\Gamma_\theta(U - \rho) \oplus \Gamma_\theta(V - \rho)] \geq \sup_{c \in [0,m^2/M]} \left(\left(\frac{1}{2} - \frac{2\lambda\theta_0}{(\lambda-1)c}\right) \frac{m^2 - Mc}{M^2 - Mc}\right)$$

■ (i) Let's denote $f(u)$ and $g(v)$ the respective distributions of $U$ and $V$ over $[0,1]$. We also denote the set:

$$\Gamma_\theta{}^+ \triangleq \bigcup_{p \in \mathbb{Z}} [(2p-1)\theta, 2p\theta[$$

$$\Gamma_\theta{}^- \triangleq \bigcup_{p \in \mathbb{Z}} [2p\theta, (2p+1)\theta[$$

In order to establish independence between $e_U$ and $e_V$, we can write $P\big((e_U = 1) \ \& \ (e_V = 1)\big)$ and $P(e_V = 1)$:

$$P\big((e_U = 1) \ \& \ (e_V = 1)\big) = \int_{\rho=0}^{2\theta} \int_{u,v=0}^{1} P\left((u+\rho \in \Gamma_\theta{}^+) \ \& \ (u+v+\rho \in \Gamma_\theta{}^+)\right) f(u)g(v)dudvd\rho$$

$$P(e_V = 1) = \int_{\rho=0}^{2\theta} \int_{u,v=0}^{1} P\left((u+\rho \in \Gamma_\theta{}^+) \ \& \ (u+v+\rho \in \Gamma_\theta{}^+)\right) f(u)g(v)dudvd\rho$$
$$+ \int_{\rho=0}^{2\theta} \int_{u,v=0}^{1} P\big((u+\rho \in \Gamma_\theta{}^-) \ \& \ (u+v+\rho \in \Gamma_\theta{}^-)\big) f(u)g(v)dudvd\rho$$

By doing the change of variable $\rho' = (\rho - \theta) \bmod 2\theta$ on the second term of the right side of the above equality, we get

$$P(e_V = 1) = 2P\big((e_U = 1) \ \& \ (e_V = 1)\big)$$

And of course we have $\frac{1}{2} = P(e_U = 1)$, which means eventually that

$$P\big((e_U = 1) \ \& \ (e_V = 1)\big) = P(e_U = 1)P(e_V = 1)$$

(ii) $\rho$ can be considered as equidistributed over $[\min(U,V), \min(U,V) + \theta[$ because $\Gamma_\theta(U-\rho) \oplus \Gamma_\theta(V-\rho)$ is unchanged by translation of $\theta$. Then the result is obvious by observing that $\rho$ need to be within $[\min(U,V), \max(U,V)[$ so that $\Gamma_\theta(U-\rho) \oplus \Gamma_\theta(V-\rho) = 1$.

(iii) We can write:

$$E[\Gamma_\theta(U-\rho) \oplus \Gamma_\theta(V-\rho)|U,V] = \int_{\theta_0}^{\lambda\theta_0} E[\Gamma_\theta(U-\rho) \oplus \Gamma_\theta(V-\rho)|U,V,\theta]g(\theta)d\theta$$

$$= \int_{\theta_0}^{\lambda\theta_0} \left( \frac{|U-V|}{\theta} - \left\lfloor \frac{|U-V|}{\theta} \right\rfloor \right) g(\theta)d\theta = \frac{1}{|U-V|}\frac{\lambda\theta_0}{(\lambda-1)} \int_{\frac{|U-V|}{\lambda\theta_0}}^{\frac{|U-V|}{\theta_0}} (u - \lfloor u \rfloor)du$$

with the change of variable $u = \frac{|U-V|}{\theta}$. We remark, by considering the surface covered by the function $f(u) \triangleq u - \lfloor u \rfloor$ over $\left[ \frac{|U-V|}{\lambda\theta_0}, \frac{|U-V|}{\theta_0} \right]$, that:

$$\frac{1}{|U-V|}\frac{\lambda\theta_0}{(\lambda-1)} \int_{\frac{|U-V|}{\lambda\theta_0}}^{\frac{|U-V|}{\theta_0}} (u - \lfloor u \rfloor)du = \frac{1}{2} + \triangleleft \left( \frac{2\lambda\theta_0}{(\lambda-1)|U-V|} \right)$$

(iv) Let's remark first that, for a variable $X \in [0, M]$ such that $E[X] = m < M$, we have, for any $c \leq m$ :

$$E[X] = m = \int_0^c xP(X = x) + \int_c^M xP(X = x) \leq cP(X \leq c) + M\big(1 - P(X \leq c)\big)$$

which implies that

$$P(X \geq c) \geq \frac{m - c}{M - c} \qquad\qquad (p8.1)$$

We can then apply $(p8.1)$ to $X = |U - V|$ and write:

$$E[\Gamma_\theta(U - \rho) \oplus \Gamma_\theta(V - \rho)] \geq \left(\frac{1}{2} - \frac{2\lambda\theta_0}{(\lambda - 1)c}\right) P(|U - V| \geq c) \geq \left(\frac{1}{2} - \frac{2\lambda\theta_0}{(\lambda - 1)c}\right) \frac{E[|U - V|] - c}{M - c}$$

We can conclude by seeing that $E[|U - V|] = ME[|U - V|/M] \geq E[(U - V)^2]/M = m^2/M$.

∎

### Considerations on the opponent's strategy

The following propositions justify the form of the canonical strategy.

For $I, i \in \{0,1\}^n$, we denote $\mathfrak{S}_{I,i}$ is the sub-group of $\mathfrak{S}_n$ that let stable $i \cap I$, $i \cap \bar{I}$, $\bar{i} \cap I$ and $\bar{i} \cap \bar{I}$: $\mathfrak{S}_{I,i} \triangleq \{\sigma \in \mathfrak{S}_n | \forall \{u, v\} \in I \times i, \{\sigma(u), \sigma(v)\} \in I \times i\}$. We remind that the notation $\bar{I}$ designates the complement of $I$ in $\{0,1\}^n$, and the notation $|I|$ designates the cardinality of $I$. If $\Phi$ is a hidden distribution generated by $A$'s Deep Random Generator, and if the observer only knows $\mu$, the assumed value of a synchronization permutation $\sigma_\Phi$ of $\Phi$ and $i$ issued from a Bernoulli trial of parameter vector $x$ generated by $\Phi$, all happen for the observer as if $A$ would perform the sequence below:

$$\Phi \circ \sigma_\Phi : x'' \xrightarrow{\sigma \in \mathfrak{S}_{I_0, \mu^{-1}(i)}} x' \xrightarrow{\mu} x$$

At the first step $x''$ is generated by a certain distribution $\Phi \circ \sigma_\Phi$ (synchronized on $I_0$), at second step $x' = \sigma^{-1}(x'')$ corresponds to a mixing simultaneously within $I_0 \cap \mu^{-1}(i)$, $I_0 \cap \overline{\mu^{-1}(i)}$, $\bar{I_0} \cap \mu^{-1}(i)$, and $\bar{I_0} \cap \overline{\mu^{-1}(i)}$ and at third step, $\mu$ produces the final $x$. The resulting distribution to be considered is denoted:

$$\Sigma_i(\Phi, \mu) \triangleq \frac{1}{|\mathfrak{S}_{I_0, \mu^{-1}(i)}|} \sum_{\sigma \in \mathfrak{S}_{I_0, \mu^{-1}(i)}} \Phi \circ \sigma_\Phi \circ \sigma \circ \mu^{-1} = \frac{1}{|\mathfrak{S}_{\mu(I_0), i}|} \sum_{\sigma \in \mathfrak{S}_{\mu(I_0), i}} \Phi \circ \sigma_\Phi \circ \mu^{-1} \circ \sigma$$

In the following, $x|_I$ designates the restriction of the vector $x$ to its components having their index within $I$.

**Proposition 9.**

$\mu$, $\mu'$ being two permutations, $i, j$ being the two observable vectors in $\{0,1\}$, and $\hat{\Sigma}_i(\Phi, \mu)$ (resp. $\hat{\Sigma}_j(\Phi', \mu')$) being a distribution that is symmetric within the segments $\mu(I_0) \cap i$, $\mu(I_0) \cap \bar{\imath}$, $\overline{\mu(I_0)} \cap i$, and $\overline{\mu(I_0)} \cap \bar{\imath}$ (resp. $\mu'(I_0) \cap j$, $\mu'(I_0) \cap \bar{\jmath}$, $\overline{\mu'(I_0)} \cap j$, and $\overline{\mu'(I_0)} \cap \bar{\jmath}$). Then we have the canonical form approximation:

$$E\left[\frac{\mu^{-1}(x) \cdot \mu'^{-1}(y)}{nk} \Big| i, j\right]_{\hat{\Sigma}_i(\Phi, \mu), \hat{\Sigma}_j(\Phi', \mu')}$$
$$= \frac{2k}{n^2}\left((\mu^{-1}(i) \cdot I_0)(\mu'^{-1}(j) \cdot I_0) + (\mu^{-1}(i) \cdot \overline{I_0})(\mu'^{-1}(j) \cdot \overline{I_0})\right) + O\left(\frac{1}{k^2}\right)$$

■ The set of distributions verifying symmetry within the segments $\mu(I_0) \cap i$, $\mu(I_0) \cap \bar{\imath}$, $\overline{\mu(I_0)} \cap i$, and $\overline{\mu(I_0)} \cap \bar{\imath}$ are exactly these of the form:

$$\hat{\Sigma}_i(\Phi, \mu) = \frac{1}{\left|\mathfrak{S}_{I_0, \mu^{-1}(i)}\right|} \sum_{\sigma \in \mathfrak{S}_{I_0, \mu^{-1}(i)}} \Phi \circ \sigma \circ \mu^{-1} = \frac{1}{\left|\mathfrak{S}_{\mu(I_0), i}\right|} \sum_{\sigma \in \mathfrak{S}_{\mu(I_0), i}} \Phi \circ \mu^{-1} \circ \sigma \ , \forall \Phi$$

From Bayes:

$$E\left[\frac{\mu^{-1}(x) \cdot \mu'^{-1}(y)}{nk} \Big| i, j\right]_{\hat{\Sigma}_i(\Phi, \mu), \hat{\Sigma}_j(\Phi', \mu')}$$

$$= \frac{1}{nk} \frac{\sum_{\sigma, \sigma' \in \mathfrak{S}_{I_0, \mu^{-1}(i)} \times \mathfrak{S}_{I_0, \mu'^{-1}(j)}} \int_{x, y \in \{0,1\}^n} x \cdot y \chi_{\mu^{-1}(i)}\left(\frac{x}{k}\right) \chi_{\mu'^{-1}(j)}\left(\frac{y}{k}\right) \Phi \circ \sigma(x) \Phi' \circ \sigma'(y) dx dy}{\sum_{\sigma, \sigma' \in \mathfrak{S}_{I_0, \mu^{-1}(i)} \times \mathfrak{S}_{I_0, \mu'^{-1}(j)}} \int_{x, y \in \{0,1\}^n} \chi_{\mu^{-1}(i)}\left(\frac{x}{k}\right) \chi_{\mu'^{-1}(j)}\left(\frac{y}{k}\right) \Phi \circ \sigma(x) \Phi' \circ \sigma'(y) dx dy}$$

In all the following, in order to simplify notations, we will generically calculate

$$E\left[\frac{x \cdot y}{nk} \Big| i, j\right]_{\hat{\Sigma}_i(\Phi, \mu), \hat{\Sigma}_j(\Phi', \mu')}$$

with assumption that $\mu = \mu' = \mathrm{Id}_{\mathfrak{S}_n}$, and $\hat{\Sigma}_i(\Phi, \mu)$ will be then more shortly denoted $\Sigma_i(\Phi)$. The above equality shows how to transpose the generic calculation to the desired result.

We introduce the following subsets of $\{0,1\}^n$:

$$J_1 = \bar{\imath} \cap \bar{\jmath}, J_2 = i \cap \bar{\jmath}, J_3 = \bar{\imath} \cap j, J_4 = i \cap j$$

$$L_s = J_s \cap I_0, \qquad s \in \{1,2,3,4\}$$

$$L_s = J_{s-4} \cap \overline{I_0}, \qquad s \in \{5,6,7,8\}$$

and the restricted distributions:

$$\Sigma_{i,s}(\Phi)\left(x|_{L_s}\right) \triangleq \left(\int_{x|_{\overline{L_s}}} \chi_{i|_{\overline{L_s}}}\left(\frac{x|_{\overline{L_s}}}{k}\right) \Sigma_i(\Phi)(x) dx|_{\overline{L_s}}\right) \Big/ \left(\int_x \chi_{i|_{\overline{L_s}}}\left(\frac{x|_{\overline{L_s}}}{k}\right) \Sigma_i(\Phi)(x) dx|_{\overline{L_s}}\right)$$

Let $\tau$ be any permutation invariant over $\overline{L_s}$, then it is clear that $\tau \in \mathfrak{S}_{I_0, i}$, and therefore $\Sigma_i(\Phi)(\tau(x)) = \Sigma_i(\Phi)(x)$. Consequently, $\forall s$, $\Sigma_{i,s}(\Phi)\big(x|_{L_s}\big)$ are distributions that are symmetric over respectively $x|_{L_s}$. By splitting the sum of the scalar product $x \cdot y$ as $\sum_{s=1}^{8} \sum_{t \in L_s} x_t y_t$, we obtain:

$$E\left[\frac{x \cdot y}{nk}\Big| i, j\right]_{\Sigma_i(\Phi), \Sigma_j(\Phi')} = \frac{1}{nk} \sum_{s=1}^{8} \frac{\int_{x, y \in \{0,1\}^n} x|_{L_s} \cdot y|_{L_s} \chi_i\left(\frac{x}{k}\right) \chi_j\left(\frac{y}{k}\right) \Sigma_i(\Phi)(x) \Sigma_j(\Phi')(y) dx dy}{\int_{x, y \in \{0,1\}^n} \chi_i\left(\frac{x}{k}\right) \chi_j\left(\frac{y}{k}\right) \Sigma_i(\Phi)(x) \Sigma_j(\Phi')(y) dx dy}$$

$$= \frac{1}{n} \sum_{s=1}^{8} |L_s| \frac{\int_{x|_{L_s}, y|_{L_s} \in \{0,1\}^{|L_s|}} \frac{x|_{L_s} \cdot y|_{L_s}}{k|L_s|} \chi_{i|_{L_s}}\left(\frac{x|_{L_s}}{k}\right) \chi_{j|_{L_s}}\left(\frac{y|_{L_s}}{k}\right) \Sigma_{i,s}(\Phi)(x|_{L_s}) \Sigma_{j,s}(\Phi')(y|_{L_s}) dx|_{L_s} dy|_{L_s}}{\int_{x|_{L_s}, y|_{L_s} \in \{0,1\}^{|L_s|}} \chi_{i|_{L_s}}\left(\frac{x|_{L_s}}{k}\right) \chi_{j|_{L_s}}\left(\frac{y|_{L_s}}{k}\right) \Sigma_{i,s}(\Phi)(x|_{L_s}) \Sigma_{j,s}(\Phi')(y|_{L_s}) dx|_{L_s} dy|_{L_s}}$$

$$(p9.1)$$

We can easily see that $i|_{L_s} \in \{L_s, \emptyset\}$ and $j|_{L_s} \in \{L_s, \emptyset\}$ depending on $s$. By replacing $i|_{L_s}$ (resp. $j|_{L_s}$) by its value in $\{L_s, \emptyset\}$ in each fraction of the sum of the last term of $(p9.1)$, we can apply Proposition 10 to each fraction and we obtain the result ($L_4$ and $L_8$ being the only segments where $i|_{J_s} \neq \emptyset$ and $j|_{J_s} \neq \emptyset$ simultaneously). ∎

**Proposition 10.**

where $\overline{\Phi}$ and $\overline{\Phi}'$ are symmetric distribution over $\{0,1\}^n$, we have:

$$\frac{1}{nk} \frac{\int_{x, y \in \{0,1\}^n} x \cdot y \chi_i\left(\frac{x}{k}\right) \chi_j\left(\frac{y}{k}\right) \overline{\Phi}(x) \overline{\Phi}'(y) dx dy}{\int_{x, y \in \{0,1\}^n} \chi_i\left(\frac{x}{k}\right) \chi_j\left(\frac{y}{k}\right) \overline{\Phi}(x) \overline{\Phi}'(y) dx dy} = \frac{k|i||j|}{n^2} + O\left(\frac{1}{k^2}\right)$$

■ we keep notations of Proposition 9. By considering $\mathfrak{S}_{i,j}$ the subgroup of $\mathfrak{S}_n$ whose permutations let $J_1, J_2, J_3, J_4$ stable, $\overline{\Phi}$ and $\overline{\Phi}'$ are obviously invariant by action of $\mathfrak{S}_{i,j}$, and thus we can write:

$$\frac{1}{nk} \frac{\int_{x, y \in \{0,1\}^n} x \cdot y \chi_i\left(\frac{x}{k}\right) \chi_j\left(\frac{y}{k}\right) \overline{\Phi}(x) \overline{\Phi}'(y) dx dy}{\int_{x, y \in \{0,1\}^n} \chi_i\left(\frac{x}{k}\right) \chi_j\left(\frac{y}{k}\right) \overline{\Phi}(x) \overline{\Phi}'(y) dx dy}$$

$$= \frac{1}{nk} \sum_{s=1}^{4} \sum_{t \in J_s} \left(\frac{\int_{x \in \{0,1\}^n} x_t \chi_i\left(\frac{x}{k}\right) \overline{\Phi}(x) dx}{\int_{x \in \{0,1\}^n} \chi_i\left(\frac{x}{k}\right) \overline{\Phi}(x) dx}\right) \left(\frac{\int_{y \in \{0,1\}^n} y_t \chi_j\left(\frac{y}{k}\right) \overline{\Phi}'(y) dy}{\int_{y \in \{0,1\}^n} \chi_j\left(\frac{y}{k}\right) \overline{\Phi}'(y) dy}\right)$$

$$= \frac{1}{nk} \sum_{s=1}^{4} \frac{1}{|J_s|} \left(\frac{\int_{x \in \{0,1\}^n} x \cdot J_s \chi_i\left(\frac{x}{k}\right) \overline{\Phi}(x) dx}{\int_{x \in \{0,1\}^n} \chi_i\left(\frac{x}{k}\right) \overline{\Phi}(x) dx}\right) \left(\frac{\int_{y \in \{0,1\}^n} y \cdot J_s \chi_j\left(\frac{y}{k}\right) \overline{\Phi}'(y) dy}{\int_{y \in \{0,1\}^n} \chi_j\left(\frac{y}{k}\right) \overline{\Phi}'(y) dy}\right) \quad (p10.2)$$

Let's study the first bracket. For $s \in \{2, 4\}$, due to the fact that $x_t \in \{0,1\}$:

$$\frac{\int_{x \in \{0,1\}^n} x \cdot J_s \chi_i\left(\frac{x}{k}\right) \overline{\Phi}(x) dx}{\int_{x \in \{0,1\}^n} \chi_i\left(\frac{x}{k}\right) \overline{\Phi}(x) dx} = |J_s|$$

For $s \in \{1, 3\}$:

$$\frac{\int_{x\in\{0,1\}^n} x \cdot J_s \chi_i\left(\frac{x}{k}\right)\overline{\Phi}(x)dx}{\int_{x\in\{0,1\}^n}\chi_i\left(\frac{x}{k}\right)\overline{\Phi}(x)dx} = \frac{|J_s|}{|\bar{\iota}|}\frac{\int_{x\in\{0,1\}^n} x \cdot \bar{\iota}\chi_i\left(\frac{x}{k}\right)\overline{\Phi}(x)dx}{\int_{x\in\{0,1\}^n}\chi_i\left(\frac{x}{k}\right)\overline{\Phi}(x)dx}$$

$$= \frac{|J_s|}{n-|i|}\left(\frac{\int_{x\in\{0,1\}^n}|x|\chi_i\left(\frac{x}{k}\right)\overline{\Phi}(x)dx}{\int_{x\in\{0,1\}^n}\chi_i\left(\frac{x}{k}\right)\overline{\Phi}(x)dx} - |i|\right)$$

We denote $\forall u \in \mathbb{N}_n$, $P_u \triangleq \int_{|x|=u}\overline{\Phi}(x)dx$. Then let's study:

$$\frac{\int_{x\in\{0,1\}^n}|x|\chi_i\left(\frac{x}{k}\right)\overline{\Phi}(x)dx}{\int_{x\in\{0,1\}^n}\chi_i\left(\frac{x}{k}\right)\overline{\Phi}(x)dx} = \frac{\int_{x\in\{0,1\}^n}|x|\psi_{|i|}\left(\frac{x}{k}\right)\overline{\Phi}(x)dx}{\int_{x\in\{0,1\}^n}\psi_{|i|}\left(\frac{x}{k}\right)\overline{\Phi}(x)dx}$$

$$= \frac{\int_{x\in\{0,1\}^n}|x|\sum_{\Delta=(1-k)|i|}^{n/2-k|i|}\beta_{|i|,k|i|+\Delta}\left(\frac{1}{k}\right)\overline{\Phi}(x)dx}{\int_{x\in\{0,1\}^n}\sum_{\Delta=(1-k)|i|}^{n/2-k|i|}\beta_{|i|,k|i|+\Delta}\left(\frac{1}{k}\right)\overline{\Phi}(x)dx}$$

$$= \frac{\sum_{\Delta=(1-k)|i|}^{n/2-k|i|}(k|i|+\Delta)\beta_{|i|,k|i|+\Delta}\left(\frac{1}{k}\right)P_{k|i|+\Delta}}{\sum_{\Delta=(1-k)|i|}^{n/2-k|i|}\beta_{|i|,k|i|+\Delta}\left(\frac{1}{k}\right)P_{k|i|+\Delta}}$$

$K$ being a constant that we will calibrate appropriately. We also denote:

$$U_K \triangleq \{r \in \mathbb{Z} | \Delta^2 \le 2Kk^2|i|\}$$

$$\overline{U_K} \triangleq \left\{(1-k)|i|, \dots, n/2 - k|i|\right\} \backslash U_K$$

Thanks to Proposition 1, we can upper-bound the difference:

$$\left|\frac{\int_{x\in\{0,1\}^n}(|x|-k|i|)\chi_i\left(\frac{x}{k}\right)\overline{\Phi}(x)dx}{\int_{x\in\{0,1\}^n}\chi_i\left(\frac{x}{k}\right)\overline{\Phi}(x)dx}\right| \le \frac{\sum_{\Delta=(1-k)|i|}^{n/2-k|i|}|\Delta|\beta_{|i|,k|i|+\Delta}\left(\frac{1}{k}\right)P_{k|i|+\Delta}}{\sum_{\Delta=(1-k)|i|}^{n/2-k|i|}\beta_{|i|,k|i|+\Delta}\left(\frac{1}{k}\right)P_{k|i|+\Delta}}$$

$$\le \frac{\sum_{\Delta\in U_K}|\Delta|\beta_{|i|,k|i|+\Delta}\left(\frac{1}{k}\right)P_{k|i|+\Delta} + \sum_{\Delta\in\overline{U_K}}|\Delta|\beta_{|i|,k|i|+\Delta}\left(\frac{1}{k}\right)P_{k|i|+\Delta}}{\sum_{\Delta\in U_K}\beta_{|i|,k|i|+\Delta}\left(\frac{1}{k}\right)P_{k|i|+\Delta}}$$

$$\le \frac{\sqrt{2Kk^2|i|}\sum_{\Delta\in U_K}\beta_{|i|,k|i|+\Delta}\left(\frac{1}{k}\right)P_{k|i|+\Delta} + n^2 e^{-K}}{\sum_{\Delta\in U_K}\beta_{|i|,k|i|+\Delta}\left(\frac{1}{k}\right)P_{k|i|+\Delta}}$$

$$\le 2\sqrt{2Kk^2|i|} \qquad\qquad\qquad (p10.3)$$

if $K$ is calibrated such that:

$$\frac{n^2}{2}e^{-K} \le \sqrt{2Kk^2|i|}\sum_{\Delta\in U_K}\beta_{|i|,k|i|+\Delta}\left(\frac{1}{k}\right)P_{k|i|+\Delta} \qquad (C)$$

This is always possible as soon as $|i| \ge 1$; and when $\Phi$ (resp. $\Phi'$) is sufficiently regular (typically by being transformed by a permutative sleeking transform introduced in Annex 2), there exist a constant $\gamma$ such that:

$$P_{k|\dot{i}|} \geq \frac{\gamma}{n}$$

Therefore, there exist a constant $C_0 > 0$ such that

$$\sum_{\Delta \in U_K} \beta_{|\dot{i}|, k|\dot{i}| + \Delta} \left(\frac{1}{k}\right) P_{k|\dot{i}| + \Delta} \geq \beta_{|\dot{i}|, k|\dot{i}|} \left(\frac{1}{k}\right) P_{k|\dot{i}|} \geq \frac{C_0}{n\sqrt{n}}$$

and thus $(C)$ implies that $K = O(\ln n)$.

By injecting the upper-bounds obtained at $(p10.3)$ in $(p10.2)$, we finally get the expected result.

■

Proposition 9 means in all cases,

$$V_\xi(i, j, \sigma_{\xi,A}, \sigma_{\xi,B}) = \frac{2k\left(\left(\sum_{r \in \sigma_{\xi,B}(I_0)} i_r\right)\left(\sum_{r \in \sigma_{\xi,A}(I_0)} j_r\right) + \left(\sum_{r \in \overline{\sigma_{\xi,B}(I_0)}} i_r\right)\left(\sum_{r \in \overline{\sigma_{\xi,A}(I_0)}} j_r\right)\right)}{n^2}$$

is the unique canonical approximation strategy for the opponent (where $\sigma_{\xi,B}$ is the permutation chosen by the opponent among $(\mu_1, \mu_2)$ and $\sigma_{\xi,A}$ is a permutation chosen by the opponent among $(\mu'_1, \mu'_2)$).

**Proof of Theorem 1 and 2**

The quadratic matrix $M_\Phi$ introduced in Section I is a symmetric matrix with elements in $[0,1]$. For such matrix, we define the permutation as, $\forall \sigma \in \mathfrak{S}_n$:

$$\sigma(M_\Phi) = M_{\Phi \circ \sigma} = \{M_\Phi(\sigma(u), \sigma(v))\}_{u,v \in \{1,\dots,n\}}$$

and the scalar product, in the vector space of $n \times n$ matrix:

$$M_\Phi \cdot M_{\Phi'} = Tr(M_\Phi M_{\Phi'}) = \sum_{u,v \in \{1,\dots,n\}} M_\Phi(u,v) M_{\Phi'}(u,v)$$

We can now explicit the Separable Criteria lemma. This lemma is a useful tool to build probability distributions that are not sensitive to the Independence Phenomenon introduced in Section I. The notations are the one introduced in Section III.

**Lemma 1. (Separable Criteria lemma)**

*For 2 probability distributions $\Phi$ and $\Phi'$ in $\zeta(\alpha)$, and for any $g \colon \mathbb{R}^2 \to [0, 1/k]$, we then have the inequality:*

$$E\left[\left(g(|\dot{i}|, |j|) - \frac{x \cdot y}{nk}\right)^2\right]_{\widetilde{\Sigma}(\Phi), \widetilde{\Sigma}(\Phi')} \geq \frac{\alpha^2}{4k^2} - \lhd \left(\frac{2}{nk^2}\right) \triangleq \frac{C(\alpha)}{k^2}$$

■ We introduce the quadrant sum, for $\xi, \xi' \in \{+, -\}$:

$$\Delta_{\xi,\xi'}(x,y) \triangleq \frac{1}{4}\Big(\delta_{\sigma_0}(\rho) + \delta_{\mathrm{Id}_{\mathfrak{S}_n}}(\rho)\Big)\Big(\delta_{\sigma_0}(\rho')$$

$$+ \delta_{\mathrm{Id}_{\mathfrak{S}_n}}(\rho')\Big)\frac{1}{|\mathfrak{S}_{I_0}|^2}\sum_{\sigma,\mu\in\mathfrak{S}_{I_0}}\left(\frac{\rho\circ\sigma\circ\sigma_\Phi^{\xi^{-1}}(x)\cdot\rho'\circ\mu\circ\sigma_{\Phi'}^{\xi'^{-1}}(y)}{n}\right)^2$$

$$\Delta_{\xi,\xi'}(\Phi,\Phi') \triangleq \int_{x,y\in\{0,1\}^n}\Delta_{\xi,\xi'}(x,y)\Phi(x)\Phi'(y)dxdy$$

$$= \frac{1}{4n^2}\Big(\delta_{\sigma_0}(\rho) + \delta_{\mathrm{Id}_{\mathfrak{S}_n}}(\rho)\Big)\Big(\delta_{\sigma_0}(\rho') + \delta_{\mathrm{Id}_{\mathfrak{S}_n}}(\rho')\Big)\frac{1}{|\mathfrak{S}_{I_0}|^2}\sum_{\sigma,\mu\in\mathfrak{S}_{I_0}}\rho\circ\sigma\circ\sigma_\Phi^{\xi^{-1}}(M_\Phi)$$

$$\cdot\,\rho'\circ\mu\circ\sigma_{\Phi'}^{\xi'^{-1}}(M_{\Phi'})$$

By using the trivial inequality $a,b\in[0,1]\implies(a-b)^2\geq\frac{1}{4}(a^2-b^2)^2$, we can write:

$$E\left[\left(g(|i|,|j|)-\frac{x\cdot y}{nk}\right)^2\right]_{\tilde{\Sigma}(\Phi),\tilde{\Sigma}(\Phi')}$$

$$= \frac{1}{k^2}\sum_{i,j\in\{0,1\}^n}\int_{x,y\in\{0,1\}^n}\Big(kg(|i|,|j|)-\frac{x\cdot y}{n}\Big)^2 P(i,j|x,y)\tilde{\Sigma}(\Phi)(x)\tilde{\Sigma}(\Phi')(y)dxdy$$

$$\geq \frac{1}{4k^2}\sum_{i,j\in\{0,1\}^n}\int_{x,y\in\{0,1\}^n}\Big(k^2 g(|i|,|j|)^2$$

$$-\Big(\frac{x\cdot y}{n}\Big)^2\Big)^2 P(i,j|x,y)\tilde{\Sigma}(\Phi)(x)\tilde{\Sigma}(\Phi')(y)dxdy$$

By applying Schwarz inequality, and observing that (i) $(i,j|\sigma(x),\mu(y)) = P(\sigma^{-1}(i),\mu^{-1}(j)|x,y)$ $\forall\sigma,\mu\in\mathfrak{S}_n$, (ii) $g(|\sigma(i)|,|\mu(j)|) = g(|i|,|j|)$ $\forall\sigma,\mu\in\mathfrak{S}_n$, and (iii) $\left(\frac{a^2+b^2+c^2+d^2}{4}\right)\geq\left(\frac{a-b-c+d}{4}\right)^2$, we get from the above inequality:

$$E\left[\left(g(|i|,|j|)-\frac{x\cdot y}{nk}\right)^2\right]_{\tilde{\Sigma}(\Phi),\tilde{\Sigma}(\Phi')}$$

$$\geq \frac{1}{4k^2}\int_{x,y\in\{0,1\}^n}\Big(\Delta_{+,+}(x,y)-\Delta_{+,-}(x,y)-\Delta_{-,+}(x,y)$$

$$+\Delta_{-,-}(x,y)\Big)^2\Phi(x)\Phi'(y)dxdy$$

$$\geq \frac{1}{4k^2}\Big(\Delta_{+,+}(\Phi,\Phi')-\Delta_{+,-}(\Phi,\Phi')-\Delta_{-,+}(\Phi,\Phi')+\Delta_{-,-}(\Phi,\Phi')\Big)^2 \quad (l1.1)$$

For any $I\in S_n$, we observe the 3 following stability properties:

(i) for any $(u,v)\in I^2, u\neq v$, $\{\sigma\in\mathfrak{S}_I|\sigma(u)=u'\ \&\ \sigma(v)=v'\}_{(u',v')\in I^2,u'\neq v'}$ is an equi-partition of $\mathfrak{S}_I$, which then implies:

$$\frac{1}{|\mathfrak{S}_I|}\sum_{\sigma\in\mathfrak{S}_I}M_\Phi(\sigma(u),\sigma(v)) = \frac{4}{n(n-2)}\sum_{\substack{r,s\in I\\r\neq s}}M_\Phi(r,s) = s_I(\Phi)$$

(ii) for any $(u,v)\in I\times\bar{I}$, $\{\sigma\in\mathfrak{S}_I|\sigma(u)=u'\ \&\ \sigma(v)=v'\}_{(u',v')\in I\times\bar{I}}$ is an equi-partition of $\mathfrak{S}_I$, which then implies :

$$\frac{1}{|\mathfrak{S}_I|} \sum_{\sigma \in \mathfrak{S}_I} M_\Phi(\sigma(u), \sigma(v)) = \frac{4}{n^2} \sum_{r,s \in I \times \bar{I}} M_\Phi(r,s) = c_I(\Phi)$$

(iii) for any $u \in I$, $\{\sigma \in \mathfrak{S}_I | \sigma(u) = u'\}_{u' \in I}$ is an equi-partition of $\mathfrak{S}_I$, which then implies:

$$\frac{1}{|\mathfrak{S}_I|} \sum_{\sigma \in \mathfrak{S}_I} M_\Phi(\sigma(u), \sigma(u)) = \frac{2}{n} \sum_{r \in I} M_\Phi(r,r) = d_I(\Phi)$$

Applying the 3 stability properties above to the calculation of $\Delta_{\xi,\xi'}(\Phi, \Phi')$, we obtain the following block matrix:

$$\Delta_{\xi,\xi'}(\Phi, \Phi') = \frac{1}{4n^2k^2} \left( \begin{pmatrix} S_{I_0}(\Phi) & c_\xi(\Phi)C_{n/2} \\ c_\xi(\Phi)C_{n/2} & S_{\bar{I}_0}(\Phi) \end{pmatrix} + \begin{pmatrix} S_{\bar{I}_0}(\Phi) & c_\xi(\Phi)C_{n/2} \\ c_\xi(\Phi)C_{n/2} & S_{I_0}(\Phi) \end{pmatrix} \right)$$

$$\cdot \left( \begin{pmatrix} S_{I_0}(\Phi') & c_{\xi'}(\Phi')C_{n/2} \\ c_{\xi'}(\Phi')C_{n/2} & S_{\bar{I}_0}(\Phi') \end{pmatrix} + \begin{pmatrix} S_{\bar{I}_0}(\Phi') & c_{\xi'}(\Phi')C_{n/2} \\ c_{\xi'}(\Phi')C_{n/2} & S_{I_0}(\Phi') \end{pmatrix} \right)$$

where $S_I(\Phi)$ and $C_{n/2}$ designate the $n/2$ block matrix:

$$S_I(\Phi) = \begin{pmatrix} d_I & s_I & \cdots & s_I \\ s_I & \ddots & \ddots & \vdots \\ \vdots & \ddots & & s_I \\ s_I & \cdots & s_I & d_I \end{pmatrix} \text{ and } C_{n/2} = \begin{pmatrix} 1 & \cdots & 1 \\ \vdots & & \vdots \\ 1 & \cdots & 1 \end{pmatrix}$$

The matrix scalar product above can also be written:

$$\Delta_{\xi,\xi'}(\Phi, \Phi') = \frac{1}{n^2} \left( \frac{n^2}{2} \left( \frac{s_{I_0}(\Phi) + s_{\bar{I}_0}(\Phi)}{2} \right) \left( \frac{s_{I_0}(\Phi') + s_{\bar{I}_0}(\Phi')}{2} \right) + \frac{n^2}{2} c_\xi(\Phi) c_{\xi'}(\Phi') + \triangleleft (n) \right)$$

then:

$$\left| \Delta_{+,+}(\Phi, \Phi') - \Delta_{+,-}(\Phi, \Phi') - \Delta_{-,+}(\Phi, \Phi') + \Delta_{-,-}(\Phi, \Phi') \right| =$$
$$\triangleleft \left( \frac{1}{n} \right) + \frac{1}{2} |c_+(\Phi) - c_-(\Phi)| |c_+(\Phi') - c_-(\Phi')|$$

which we can inject in $(l1.1)$ with definition of $\zeta(\alpha)$ to conclude the proof.

A similar result can be obtained with $E\left[ \left( g(|i|, |j|) - \frac{x \cdot j}{n} \right)^2 \right]_{\tilde{\Sigma}(\Phi), \tilde{\Sigma}(\Phi')}$ just using Proposition 7 and the fact that $E\left[ \left( \frac{x \cdot j}{n} - \frac{x \cdot y}{nk} \right)^2 \right]_{\forall \Phi, \Phi'} \leq \frac{2}{nk}$ as shown in Section II. $\blacksquare$

**Theorem 1.**

*Under the Deep Random assumption, $\mathcal{P}(\alpha, n, k, K)$ satisfies condition $(P)$.*

■ The proof is organized as follows: we will first determine a restricted form for the opponent's strategy by considering 3 symmetries applicable to the knowledge of $\xi$ and then applying accordingly the Deep Random Assumption with those 3 symmetries. Then, we will apply that restriction to determine the possible strategies of the opponent (phase 1). Then we will show technical inequalities applying to the error rates of respectively the legitimate receiver and the opponent (phase 2). At last, we will deduce Advantage Distillation from the said inequalities (phase 3).

The first symmetry $\Sigma_1$, and also the most important one, results from the fact that $\xi$ cannot distinguish $\sigma_\Phi$ within $(\mu_1, \mu_2)$, and neither $\sigma_{\Phi'}$ knowing the public information $\mathfrak{I} = \{i, j, (\mu_1, \mu_2), (\mu'_1, \mu'_2)\}$. But nevertheless, $\Phi_A$ designating the distribution potentially chosen by $A$ in the opponent's perspective, if $\xi$ consider $\mu_{t^b(1)}$ ($b \in \{0,1\}$) as the choice of $A$ for $\sigma_{\Phi_A}$ then obviously $\Phi_A$ should have $\mu_{t^b(1)}$ as its tidying permutation. The first symmetry thus results from the group $\{\mathrm{Id}, t\}$ applied to $(\mu_1, \mu_2)$ and $(\mu'_1, \mu'_2)$.

The second symmetry $\Sigma_2$ results from the fact that $\xi$ cannot distinguish the actual distribution among all distributions having $\mu_{t^b(1)}$ as synchronization permutation and composed with any permutation letting stable $i \cap \mu_{t^b(1)}(I_0)$, $i \cap \overline{\mu_{t^b(1)}(I_0)}$, $\bar{\imath} \cap \mu_{t^b(1)}(I_0)$ and $\bar{\imath} \cap \overline{\mu_{t^b(1)}(I_0)}$. The induced sub-group of $\mathfrak{S}_n$ is:

$$\mathfrak{S}_{\mu_{t^b(1)}(I_0), i} \triangleq \left\{ \sigma \in \mathfrak{S}_n \,\middle|\, \forall \{u, v\} \in \mu_{t^b(1)}(I_0) \times i, \{\sigma(u), \sigma(v)\} \in \mu_{t^b(1)}(I_0) \times i \right\}$$

We remind that the notation $\bar{I}$ designates the complement of $I$ in $\{0,1\}^n$, and the notation $|I|$ designates the cardinality of $I$. If $\Phi$ is a hidden distribution generated by $A$'s Deep Random Generator, and if the observer only knows $\mu_{t^b(1)}$ as the assumed synchronization permutation, and $i$ issued from a Bernoulli trial of parameter vector $x$ generated by $\Phi$, all happen for the observer as if $A$ would perform the sequence below:

$$\Phi \circ \sigma_\Phi : x'' \xrightarrow{\sigma \in \mathfrak{S}_{I_0, \mu_{t^b(1)}^{-1}(i)}} x' \xrightarrow{\mu_{t^b(1)}} x$$

At the first step $x''$ is generated by a certain synchronized distribution $\Phi \circ \sigma_\Phi$ (synchronized on $I_0$), at second step $x' = \sigma^{-1}(x'')$ corresponds to a mixing simultaneously within $I_0 \cap \mu_{t^b(1)}^{-1}(i)$, $I_0 \cap \overline{\mu_{t^b(1)}^{-1}(i)}$, $\bar{I}_0 \cap \mu_{t^b(1)}^{-1}(i)$, and $\bar{I}_0 \cap \overline{\mu_{t^b(1)}^{-1}(i)}$, and at third step, $\mu_{t^b(1)}$ produces the final $x$. The resulting distribution to be considered is denoted:

$$\Sigma_i\left(\Phi, \mu_{t^b(1)}\right) \triangleq \frac{1}{\left|\mathfrak{S}_{I_0, \mu_{t^b(1)}^{-1}(i)}\right|} \sum_{\sigma \in \mathfrak{S}_{I_0, \mu_{t^b(1)}^{-1}(i)}} \Phi \circ \sigma_\Phi \circ \sigma \circ \mu_{t^b(1)}^{-1}$$

$$= \frac{1}{\left|\mathfrak{S}_{\mu_{t^b(1)}(I_0), i}\right|} \sum_{\sigma \in \mathfrak{S}_{\mu_{t^b(1)}(I_0), i}} \Phi \circ \sigma_\Phi \circ \mu_{t^b(1)}^{-1} \circ \sigma$$

The second symmetry thus results from the group $\mathfrak{S}_{\mu_{t^b(1)}(I_0), i}$ applied to $(\Phi_A, \mu_1)$ or $(\Phi_A, \mu_2)$.

The third symmetry $\Sigma_3$ results from the fact that the distribution $\tilde{\Sigma}(\Phi)$ is entirely defined by a 2 variables distribution

$$p_{u,v} = P(|\sigma_\Phi(x) \cap I_0| = u, |\sigma_\Phi(x) \cap \overline{I_0}| = v)$$

Typically, prior the publication of $\{i, j, (\mu_1, \mu_2), (\mu'_1, \mu'_2)\}$, it is legitimate to say that $\xi$ has no ground to privilege $p_{u,v}$ rather than $p_{(u+t) \bmod (n/2+1),(v+t') \bmod (n/2+1)}$, $\forall (t, t') \in \mathbb{N}_{n/2}{}^2$ and then that group of translations drives to a uniform distribution of $p_{u,v}$. Posteriorly to the publication of $\{i, j, (\mu_1, \mu_2), (\mu'_1, \mu'_2)\}$, it remains legitimate, due to symmetry $\Sigma_1$, to say that $\xi$ should at least only consider distributions $p_{u,v}$ such that $p_{u_1,v_1} = p_{u_2,v_2}$ where:

$$\begin{pmatrix} u_1 = k|\mu_1(i) \cap I_0| \\ v_1 = k|\mu_1(i) \cap \overline{I_0}| \end{pmatrix}, \begin{pmatrix} u_2 = k|\mu_2(i) \cap I_0| \\ v_2 = k|\mu_2(i) \cap \overline{I_0}| \end{pmatrix}$$

Therefore, a couple of distributions $\Sigma_i(\Phi, \sigma_\Phi)$ and $\Sigma_i(\Psi \circ \sigma_\Psi \circ \sigma_d[i]^{-1}, \sigma_d[i])$ transformed by the symmetry $\Sigma_3$ to verify the above equality, are denoted $\hat{\Sigma}_i(\Phi, \sigma_\Phi)$ and $\hat{\Sigma}_i(\Psi \circ \sigma_\Psi \circ \sigma_d[i]^{-1}, \sigma_d[i])$. They verify the following equalities:

$$P\big(i, (\mu_1, \mu_2), (\mu'_1, \mu'_2)\big) = P(i, (\mu_1, \mu_2))$$
$$= \frac{1}{2}\left( \int_u \chi_i\left(\frac{u}{k}\right) \hat{\Sigma}_i(\Phi, \sigma_\Phi)(u) du + \int_u \chi_i\left(\frac{u}{k}\right) \hat{\Sigma}_i(\Psi \circ \sigma_\Psi \circ \sigma_d[i]^{-1}, \sigma_d[i])(u) du \right)$$
$$= \int_u \chi_i\left(\frac{u}{k}\right) \hat{\Sigma}_i(\Phi, \sigma_\Phi)(u) du = \int_u \chi_i\left(\frac{u}{k}\right) \hat{\Sigma}_i(\Psi \circ \sigma_\Psi \circ \sigma_d[i]^{-1}, \sigma_d[i])(u) du$$

and additionally, the transformations induced by symmetry $\Sigma_3$ maintains the symmetries between the components of the vector $x$ within the segments $I_0 \cap \mu_{t^b(1)}^{-1}(i)$, $I_0 \cap \overline{\mu_{t^b(1)}^{-1}(i)}$, $\overline{I_0} \cap \mu_{t^b(1)}^{-1}(i)$, and $\overline{I_0} \cap \overline{\mu_{t^b(1)}^{-1}(i)}$, and therefore $\hat{\Sigma}_i(\Phi, \sigma_\Phi)$ and $\hat{\Sigma}_i(\Psi \circ \sigma_\Psi \circ \sigma_d[i]^{-1}, \sigma_d[i])$ remains compatible with the calculations of Proposition 9. The third symmetry thus results from the group $\{\mathrm{Id}, t\}$ applied to $\left(\{p_{u,v}\}_{\{\Phi, \sigma_\Phi\}}, \{p_{u,v}\}_{\{\Psi \circ \sigma_\Psi \circ \sigma_d[i]^{-1}, \sigma_d[i]\}}\right)$.

Then, for $\xi$ willing to evaluate $V_A$ knowing the public information $\mathfrak{I}$, the knowledge of the probability distribution $P\big(x, (\sigma_{\Phi_A}, \sigma_A)|\mathfrak{I}\big)$ by $\xi$ is invariant by action of $\Sigma_1 \times \Sigma_2 \times \Sigma_3$ under Deep Random Assumption, and therefore we can write Bayes formula:

$$P\big(x, (\sigma_{\Phi_A}, \sigma_A)|\mathfrak{I}\big) = P\big(x, (\sigma_{\Phi_A}, \sigma_A)\big|i, (\mu_1, \mu_2), (\mu'_1, \mu'_2)\big)$$
$$= \frac{P\big(i, (\mu_1, \mu_2), (\mu'_1, \mu'_2)\big|x, (\sigma_{\Phi_A}, \sigma_A)\big) P\big(x, (\sigma_{\Phi_A}, \sigma_A)\big)}{P\big(i, (\mu_1, \mu_2), (\mu'_1, \mu'_2)\big)}$$

which results in the form $R_{\mathfrak{I}}(\Sigma_1 \times \Sigma_2 \times \Sigma_3)$ below (involving the above equalities for the $P\big(i, (\mu_1, \mu_2), (\mu'_1, \mu'_2)\big)$ denominator):

$$R_3(\Sigma_1 \times \Sigma_2 \times \Sigma_3)$$

$$= \left\{ \frac{1}{2} \left( \frac{\chi_i\left(\frac{x}{k}\right) \hat{\Sigma}_i(\Phi, \sigma_\Phi)(x) dx}{\int_u \chi_i\left(\frac{u}{k}\right) \hat{\Sigma}_i(\Phi, \sigma_\Phi)(u) du} \cdot \delta_{\sigma_\Phi}(\sigma_{\Phi_A}) + \frac{\chi_i\left(\frac{x}{k}\right) \hat{\Sigma}_i(\Psi \circ \sigma_\Psi \circ \sigma_d[i]^{-1}, \sigma_d[i])(x) dx}{\int_u \chi_i\left(\frac{u}{k}\right) \hat{\Sigma}_i(\Psi \circ \sigma_\Psi \circ \sigma_d[i]^{-1}, \sigma_d[i])(u) du} \right. \right.$$

$$\left. \left. \cdot \delta_{\sigma_d[i]}(\sigma_{\Phi_A}) \right) \cdot \left( \frac{1}{2} \left( \delta_{\sigma_{\Phi'}}(\sigma_A) + \delta_{\sigma'_d[j]}(\sigma_A) \right) \right) | \forall \Phi, \Psi \in \zeta(\alpha) \right\}$$

<u>Phase 1:</u>

Now that we have restricted the possible distributions $P\big(x, (\sigma_{\Phi_A}, \sigma_A)|\Im\big)$ for the opponent, let's examine the consequence of such restriction on the opponent's strategy. The best way for $\xi$ to estimate $\widetilde{e_A}|\Im$ is to compute $\tilde{e}_\xi(\Im) \in \{0,1\}$ verifying:

$$\inf_{\tilde{e}(\Im) \in \{0,1\}} E[\widetilde{e_A} \oplus \tilde{e}(\Im)] = \inf_{\tilde{e}(\Im) \in \{0,1\}} E[(\widetilde{e_A} - \tilde{e}(\Im))^2]$$

This is obtained as follows:

$$\int (\widetilde{e_A} - \tilde{e}(\Im))^2 P\big(x, (\sigma_{\Phi_A}, \sigma_A)|\Im\big) = \int \widetilde{e_A} P\big(x, (\sigma_{\Phi_A}, \sigma_A)|\Im\big) + \tilde{e}(\Im) \left( 1 - 2 \int \widetilde{e_A} P\big(x, (\sigma_{\Phi_A}, \sigma_A)|\Im\big) \right)$$

which gives, as expected:

$$(i) \int \widetilde{e_A} P\big(x, (\sigma_{\Phi_A}, \sigma_A)|\Im\big) < \frac{1}{2} \Longrightarrow \tilde{e}_\xi(\Im) = 0$$

$$(ii) \int \widetilde{e_A} P\big(x, (\sigma_{\Phi_A}, \sigma_A)|\Im\big) > \frac{1}{2} \Longrightarrow \tilde{e}_\xi(\Im) = 1$$

$$(iii) \int \widetilde{e_A} P\big(x, (\sigma_{\Phi_A}, \sigma_A)|\Im\big) = \frac{1}{2} \Longrightarrow \tilde{e}_\xi(\Im) = 0 \text{ or } 1 \text{ randomly}$$

We remind the notation

$$\Gamma_{\theta, \rho}(x) \triangleq \left\lceil \frac{x - \rho}{\theta} \right\rceil \bmod 2$$

In the following, we will denote for more simplicity:

$$\Gamma(x) \triangleq \Gamma_{K/\sqrt{nk}, \rho}(x)$$

$$\hat{\Sigma}_i(\Psi, \sigma_d[i]) \triangleq \hat{\Sigma}_i(\Psi \circ \sigma_\Psi \circ \sigma_d[i]^{-1}, \sigma_d[i])$$

The restriction of the possible distributions $P\big(x, (\sigma_{\Phi_A}, \sigma_A)|\Im\big)$ for the opponent enables us to compute $\int \widetilde{e_A} P\big(x, (\sigma_{\Phi_A}, \sigma_A)|\Im\big)$ as follows:

$$\int \widetilde{e_A} P\big(x, (\sigma_{\Phi_A}, \sigma_A)|\mathfrak{I}\big) = \Omega_{\mathfrak{I}}(\Sigma_1 \times \Sigma_2 \times \Sigma_3)$$

$$= \left\{ \frac{1}{4} \int_x \frac{\Big(\Gamma(\sigma_\Phi^{-1}(x) \cdot \sigma_{\Phi'}{}^{-1}(j)) + \Gamma(\sigma_\Phi^{-1}(x) \cdot \sigma'_d[j]^{-1}(j))\Big) \chi_i\left(\frac{x}{k}\right) \hat{\Sigma}_i(\Phi, \sigma_\Phi)(x) dx}{\int_u \chi_i\left(\frac{u}{k}\right) \hat{\Sigma}_i(\Phi, \sigma_\Phi)(u) du} \right.$$

$$\left. + \frac{1}{4} \int_x \frac{\Big(\Gamma(\sigma_d[i]^{-1}(x) \cdot \sigma_{\Phi'}{}^{-1}(j)) + \Gamma(\sigma_d[i]^{-1}(x) \cdot \sigma'_d[j]^{-1}(j))\Big) \chi_i\left(\frac{x}{k}\right) \hat{\Sigma}_i(\Psi, \sigma_d[i])(x) dx}{\int_u \chi_i\left(\frac{u}{k}\right) \hat{\Sigma}_i(\Psi, \sigma_d[i])(u) du} \, | \forall \Phi, \Psi \right.$$

$$\left. \in \zeta(\alpha) \right\}$$

<div align="right">(T1.0)</div>

We remind the notation corresponding to the canonical forms obtained in Proposition 9:

$$V_\xi(i, j, \sigma_{\xi,A}, \sigma_{\xi,B}) \triangleq \frac{2k\left(\Big(\sum_{r \in \sigma_{\xi,B}(I_0)} i_r\Big)\Big(\sum_{r \in \sigma_{\xi,A}(I_0)} j_r\Big) + \Big(\sum_{r \in \overline{\sigma_{\xi,B}(I_0)}} i_r\Big)\Big(\sum_{r \in \overline{\sigma_{\xi,A}(I_0)}} j_r\Big)\right)}{n^2}$$

$$\widetilde{e_\xi}(\sigma_{\xi,A}, \sigma_{\xi,B}) \triangleq \Gamma\big(V_\xi(i, j, \sigma_{\xi,A}, \sigma_{\xi,B})\big)$$

$$T_m = \left\{ \widetilde{e_\xi}(\mu'_1, \mu_1)_m, \widetilde{e_\xi}(\mu'_1, \mu_2)_m, \widetilde{e_\xi}(\mu'_2, \mu_1)_m, \widetilde{e_\xi}(\mu'_2, \mu_2)_m \right\}$$

where $\sigma_{\xi,B}$ is the permutation chosen by $\xi$ among $(\mu_1, \mu_2)$ as $\sigma_{\Phi_A}$ and $\sigma_{\xi,A}$ is the permutation chosen by $\xi$ among $(\mu'_1, \mu'_2)$ as $\sigma_A$.

We need now to justify that we can inverse the operator $\Gamma$ and the operator $\int$ in each of the terms of the equality (T1.0): for a given choice of $(\sigma_{\xi,A}, \sigma_{\xi,B})$, the respective estimations of $V_\xi$ and $V_A$ will be (i) or close compared to the filter width $K/\sqrt{nk}$ in which case they they will be close for most of the permutations found in the sum $\hat{\Sigma}_i$ and therefore it is justified to inverse the operators with approximation $o_n(1)$; or (ii) not close compared to the filter width $K/\sqrt{nk}$ in which case, by considering the independent random variable of the translation parameter $\rho$, we will obtain estimations close to $1/2$ for $\widetilde{e_\xi} \oplus \widetilde{e_A}$ in both cases of inversion or non-inversion of the operators (see Proposition 8 for an example of such reasoning), and thus again the inversion is justified with an approximation that can be made as close as desired from zero.

Then, due to the fact that, at each instantiation $m$ of the protocol, we have forced at step 5 the condition

$$|T_m| = 2, \ \forall m \qquad (CD2)$$

We are in the case $(iii)$ of the computation of $\int (\widetilde{e_A} - \tilde{e}(\mathfrak{I}))^2 P\big(x, (\sigma_{\Phi_A}, \sigma_A)|\mathfrak{I}\big)$, and the opponent cannot do better than to choose randomly its bit $\widetilde{e_\xi}(\sigma_{\xi,A}, \sigma_{\xi,B})$ within $T_m$. That means that the opponent cannot do better than picking its bit at 50% chance random, which means that its knowledge is null, at the approximation of Proposition 9.

The above reasoning is valid because the condition $(CD2)$ remains compatible with the symmetries $\Sigma_1, \Sigma_2, \Sigma_3$ and therefore, discarding the instances that do not verify $(CD2)$ does not affect the validity of the restriction $\Omega_{\mathfrak{I}}(\Sigma_1 \times \Sigma_2 \times \Sigma_3)$ allowed by the Deep Random Assumption.

<u>Phase 2:</u>

In this intermediate phase, we prove technical inequalities related to $P\big(\widetilde{e_A} \neq \widetilde{e_B}|S\big)$ and $P\big(\widetilde{e_A} \neq \widetilde{e_\xi}|S\big)$.

Let's consider the below situation that happens with a probability $\frac{1}{4}$ :

$$S \triangleq [(\sigma_A = \sigma_{\Phi'})\ \&\ (\sigma_B = \sigma_\Phi)]$$

Let's first consider the case $\Big[S\ \&\ \big((\sigma_{\xi,A}, \sigma_{\xi,B}) = (\sigma_{\Phi'}, \sigma_\Phi)\big)\Big]$. Thanks to Proposition 6 for the first inequality and Proposition 4 (ii) for the second one, we get:

$$E\Big[\big(V_\xi(i, j, \sigma_{\Phi'}, \sigma_\Phi) - V_A\big)^2 |S\Big] \leq E[(V_B - V_A)^2 |S] \leq \frac{2}{nk} \qquad (T1.1)$$

Let's secondly consider the case $\Big[S\ \&\ \big((\sigma_{\xi,A}, \sigma_{\xi,B}) = (\sigma'_d[j], \sigma_d[i])\big)\Big]$. We remark that $\forall \mu \in \mathfrak{S}_n$ $\sigma_d[\mu(i)]^{-1} \circ \mu = \sigma_d[i]^{-1}$, and therefore, $\sigma_d[i]^{-1}(i)$ is stable by action of $\mathfrak{S}_n$ on $i$. This also means that $\sigma_d[i]^{-1}(i)$ is of the form $\sigma_d[i]^{-1}(i) = f(|i|)$. Similarly, $\sigma'_d[j]^{-1}(j)$ is also stable by action of $\mathfrak{S}_n$ on $j$, and is therefore also of the same form. We also remark that $\forall \mu \in \mathfrak{S}_n$, $\mathfrak{S}_{\mu(I),\mu(i)} = \mu \circ \mathfrak{S}_{I,i} \circ \mu^{-1}$ and therefore, thanks to the above, we can deduce that the distribution $\Sigma_i(\Psi, \sigma_d[i]) \circ \sigma_d[i]$ is invariant by the transform $\forall \mu : i \to \mu(i)$.

The considerations above applied to $V_\xi(i, j, \sigma'_d[j], \sigma_d[i])$ imply that

$$\forall \sigma, \mu \in \mathfrak{S}_n, V_\xi(\sigma(i), \mu(j), \sigma_d[\sigma(i)], \sigma'_d[\mu(j)]) = V_\xi(i, j, \sigma_d[i], \sigma'_d[j])$$

Which also means that $V_\xi(i, j, \sigma_d[i], \sigma'_d[j])$ is of the form

$$V_\xi(i, j, \sigma_d[i], \sigma'_d[j]) = g(|i|, |j|)$$

In the situation $S$, the legitimate partners use both at step 4 the 2 tidying permutations $\sigma_\Phi$ and $\sigma_{\Phi'}$. We have seen in the definition of tidying permutations that picking a tidying permutation for $\Phi$ means choosing randomly and uniformly any possible tidying permutation for $\Phi$. Therefore, given the form $V_\xi(i, j, \sigma_d[i], \sigma'_d[j]) = g(|i|, |j|)$, the exact writing of $E\Big[\big(V_\xi(i, j, \sigma_d[i], \sigma'_d[j]) - V_A\big)^2|S\Big]$ is:

$$E\Big[\big(V_\xi(i, j, \sigma_d[i], \sigma'_d[j]) - V_A\big)^2|S\Big]$$

$$= \sum_{i,j} \int_{x,y} \Big(V_\xi(i, j, \sigma_d[i], \sigma'_d[j])$$

$$- \frac{\sigma_\Phi^{-1}(x) \cdot \sigma_{\Phi'}^{-1}(j)}{n}\Big)^2 \chi_i\Big(\frac{x}{k}\Big)\chi_j\Big(\frac{y}{k}\Big)\tilde{\Sigma}(\Phi)(\sigma_\Phi^{-1}(x))\tilde{\Sigma}(\Phi')(\sigma_{\Phi'}^{-1}(y))dxdy$$

$$= \sum_{i,j} \int_{x,y} \Big(V_\xi(i, j, \sigma_d[i], \sigma'_d[j]) - \frac{x \cdot j}{n}\Big)^2 \chi_i\Big(\frac{x}{k}\Big)\chi_j\Big(\frac{y}{k}\Big)\tilde{\Sigma}(\Phi)(x)\tilde{\Sigma}(\Phi')(y)dxdy$$

$$= E\Big[\Big(g(|i|, |j|) - \frac{x \cdot j}{nk}\Big)^2\Big]_{\tilde{\Sigma}(\Phi),\tilde{\Sigma}(\Phi')} \geq \frac{\alpha^2}{4k^2} \mathrel{-\!\triangleleft} \Big(\frac{2}{nk^2}\Big) = \frac{C(\alpha)}{k^2} \qquad (T1.2)$$

where the last inequality comes from Lemma 1.

Each experiment of the protocol $\mathcal{P}_{1\to5}$ leads to a binary encoding of $V_A$, $V_B$, and thus also $V_\xi$, to obtain

$$\widetilde{e_A} = \left\lfloor \frac{(V_A - \rho)\sqrt{nk}}{K} \right\rfloor \bmod 2, \widetilde{e_B} = \left\lfloor \frac{(V_B - \rho)\sqrt{nk}}{K} \right\rfloor \bmod 2, \widetilde{e_\xi} = \left\lfloor \frac{(V_\xi - \rho)\sqrt{nk}}{K} \right\rfloor \bmod 2$$

where $\rho$ is a translation parameter. If $\rho$ is chosen (publicly by $A$) such that

$$\inf_{q \in \mathbb{Z}} \left| V_A - \rho - \frac{qK}{\sqrt{nk}} \right| \geq \frac{\lambda K}{\sqrt{nk}} \qquad (CD1)$$

with $\frac{1}{K} \ll \lambda \leq \frac{1}{2}$ then the inequality $(T1.1)$ combined with Proposition 3, enables us to obtain:

$$P(\widetilde{e_A} \neq \widetilde{e_B}|S) \leq P\left((V_B - V_A)^2 \geq \frac{\lambda^2 K^2}{nk} \Big| S\right) \leq 2ne^{-\frac{\lambda^2 K^2}{2}} \qquad (T1.3)$$

The fact to choose $\rho$ publicly, does not give useful information to $\xi$ when $S$ is satisfied, because of $(T1.4)$ shown below.

Thanks to Proposition 8 (iv) combined with $(T1.2)$ we can determine a constant $C_2$ such that

$$P(\widetilde{e_A} \neq \widetilde{e_\xi}|S) \geq C_2 - O\left(K\sqrt{\frac{k}{n}}\right) \qquad (T1.4)$$

Due to the $(T1.4)$ bound compared to the $(T1.3)$ bound, there is a fair probability (controlled especially by the parameter $\lambda$) that the conditions $(CD1)$ and $(CD2)$ coexist. Discarding the instances that do not verify $(CD1)$ and $(CD2)$ is therefore reasonable.

<u>Phase 3:</u>

In this phase, we finally show that we create and Advantage for the partners compared to the opponent. Due to the exponentially small upper bound of $(T1.3)$ we can safely assume in the following that $\widetilde{e_A} = \widetilde{e_B}|S$, which we will also denote $\widetilde{e_A}(\sigma_{\Phi'}) = \widetilde{e_B}(\sigma_\Phi)$.

We obtained at phase 1 that the opponent cannot do better than to choose randomly its bit $\widetilde{e_\xi}(\sigma_{\xi,A}, \sigma_{\xi,B})$ within $\{0,1\}$ as soon as we impose $|T_m| = 2$. Therefore, it is sufficient to show that there exists a constant $C' > 0$ such that $P(\widetilde{e_A}(\sigma_{\Phi'}) = \widetilde{e_A}(\sigma'_d) = \widetilde{e_B}(\sigma_\Phi) = \widetilde{e_B}(\sigma_d) \big| |T_m| = 2) \geq C'$. Indeed, any instance where $\widetilde{e_A}(\sigma_{\Phi'}) \neq \widetilde{e_A}(\sigma'_d)$ or $\widetilde{e_B}(\sigma_\Phi) \neq \widetilde{e_B}(\sigma_d)$ will be non-contributive because in such case, we have exactly $P(\widetilde{e_A} = \widetilde{e_B}) = 1/2 = P(\widetilde{e_A} = \widetilde{e_\xi})$.

Let's thus show the above inequality:

From inequality $(T1.2)$, we can deduce that there exists a constant $C'_1 > 0$ such that

$$P\left(\left|V_\xi(i,j,\sigma_d[i],\sigma'_d[j]) - V_A(\sigma_{\Phi'})\right| \geq \frac{K}{\sqrt{nk}}\right) \geq C'_1$$

On the other hand, we know that $V_\xi(i,j,\sigma_{\Phi'},\sigma_\Phi) \approx V_A(\sigma_{\Phi'})$ compared to $\frac{K}{\sqrt{nk}}$ (this can be justified by using same reasoning than in Proposition 5). Therefore, due to the definition of $V_\xi(i,j,\mu,\mu')$, the two relations above imply necessarily that:

$$\sigma_d[i](i) \cdot I_0 \not\approx \sigma_\Phi(i) \cdot I_0$$
$$or$$
$$\sigma_d[i](i) \cdot \overline{I_0} \not\approx \sigma_\Phi(i) \cdot \overline{I_0}$$
$$or$$
$$\sigma'_d[i](i) \cdot I_0 \not\approx \sigma_{\Phi'}(i) \cdot I_0$$
$$or$$
$$\sigma'_d[i](i) \cdot \overline{I_0} \not\approx \sigma_{\Phi'}(i) \cdot \overline{I_0}$$

By observing that (i) the inversion of $I_0$ and $\overline{I_0}$ let the protocol unchanged, and (ii) that $(i, \sigma_\Phi, \sigma_d)$ and $(j, \sigma_{\Phi'}, \sigma'_d)$ are generated independently, we can conclude that there exists a probability bounded by a constant $> 0$ such that the elements in each pair within

$$\{V_A(\sigma_{\Phi'}), V_\xi(i, j, \sigma_\Phi, \sigma'_d[j]), V_\xi(i, j, \sigma_d[i], \sigma_{\Phi'}), V_\xi(i, j, \sigma_d[i], \sigma'_d[j])\}$$

are distant compared to $\frac{K}{\sqrt{nk}}$. Then, by considering the independent random variables of the translation parameter $\rho$ and the multiplicative factor $K$, we obtain as desired (similar reasoning than in Proposition 8) that there exists a constant $C'_2 > 0$ such that $P(\widetilde{e_A}(\sigma_{\Phi'}) = \widetilde{e_A}(\sigma'_d) \big| |T_m| = 2) \geq C'_2$. The same reasoning can stand by replacing $V_A(\sigma_{\Phi'})$ by $V_B(\sigma_\Phi)$, and thus finally, one can determine a constant $C' > 0$ such that $P(\widetilde{e_A}(\sigma_{\Phi'}) = \widetilde{e_A}(\sigma'_d) = \widetilde{e_B}(\sigma_\Phi) = \widetilde{e_B}(\sigma_d) \big| |T_m| = 2) \geq C'$.

We deduce from that inequality that

$$P(\widetilde{e_A} \neq \widetilde{e_B} \big| |T_m| = 2) \leq (1 - C') \frac{1}{2}$$

while

$$P(\widetilde{e_A} \neq \widetilde{e_\xi} \big| |T_m| = 2) = \frac{1}{2} + o(1)$$

We have therefore created an Advantage for the partners over the opponent. This is enough to conclude the proof of the Theorem. ∎

**Theorem 2.**

$\mathcal{P}(\alpha, n, k, K)$ *satisfies condition* $(P')$.

∎ If $\omega$ is the optimal strategy to estimate $V_A$ for a known distribution $\Phi_m$ in $D_{\mathfrak{I}_<}$, then the tidying permutation $\sigma_m$ of that distribution is known and the distribution can be assumed that $D_{\mathfrak{I}_<} \subset \mathfrak{S}_{\sigma_m(I_0)} = \mathfrak{S}_{\sigma_m(I_0), \{1, \ldots, 1\}}$ because the generator can choose either $\Phi_m$ or a permuted version $\Phi_m \circ \sigma$ provided that $\sigma$ keeps $\sigma_m$ as the tidying permutation. From Proposition 9, $\omega$ is under the form:

$$\omega(i, j) = \frac{2k}{n^2} \big( (\sigma_m^{-1}(i) \cdot I_0)(\sigma_m'^{-1}(j) \cdot I_0) + (\sigma_m^{-1}(i) \cdot \overline{I_0})(\sigma_m'^{-1}(j) \cdot \overline{I_0}) \big) + O\left(\frac{1}{k^2}\right)$$

(Typically, in a recursive DRG generation process, $\sigma_m$ represents a tidying permutation of the distribution generated at step $m$). We can justify the fact that $\omega$ depends only on $(i, j)$ and not on $(\mu_1, \mu_2, \mu'_1, \mu'_2)$ by remarking that when $\omega$ does know the tidying permutation $\sigma_m$ of the distribution, $(\mu_1, \mu_2, \mu'_1, \mu'_2)$ carry no additional useful information to elaborate $\omega$.

For more simplicity in writing, we will denote in the following:

$$\omega_{\sigma_m}^*(i,j) \triangleq \frac{2k}{n^2}\big((\sigma_m^{-1}(i) \cdot I_0)(\sigma_m'^{-1}(j) \cdot I_0) + (\sigma_m^{-1}(i) \cdot \overline{I_0})(\sigma_m'^{-1}(j) \cdot \overline{I_0})\big)$$

If the protocol now executes with a same given distribution $\Psi \in R_{\mathfrak{I}_<}(G)$ on both sides of $A$ and $B$, we can write:

$$\begin{aligned}
E[(\omega - V_A)^2|S] &= \sum_{i,j} \int_{x,y} \bigg(\omega(i,j) \\
&\quad - \frac{\sigma_{\Psi^{-1}}(x) \cdot \sigma_{\Psi^{-1}}(j)}{n}\bigg)^2 \chi_i\Big(\frac{x}{k}\Big)\chi_j\Big(\frac{y}{k}\Big)\tilde{\Sigma}(\Psi)(\sigma_{\Psi^{-1}}(x))\tilde{\Sigma}(\Psi)(\sigma_{\Psi^{-1}}(y))dxdy \\
&= \sum_{i,j}\int_{x,y}\bigg(\omega(\sigma_\Psi(i),\sigma_\Psi(j)) - \frac{x\cdot j}{n}\bigg)^2\chi_i\Big(\frac{x}{k}\Big)\chi_j\Big(\frac{y}{k}\Big)\tilde{\Sigma}(\Psi)(x)\tilde{\Sigma}(\Psi)(y)dxdy
\end{aligned}$$

It is easy to see that $\forall \mu \in \mathfrak{S}_n$, $\sigma_{\Psi\circ\mu} = \mu^{-1}\circ\sigma_\Psi$, and therefore, $\tilde{\Sigma}$ is invariant by the change $\Psi \to \Psi \circ \mu$, and thus $\forall \mu \in \mathfrak{S}_n$:

$$\begin{aligned}
E[(\omega - &V_A)^2|S]_{\Psi\circ\mu} \\
&= \sum_{i,j}\int_{x,y}\bigg(\omega(\mu^{-1}\circ\sigma_\Psi(i),\mu^{-1}\circ\sigma_\Psi(j)) - \frac{x\cdot j}{n}\bigg)^2\chi_i\Big(\frac{x}{k}\Big)\chi_j\Big(\frac{y}{k}\Big)\tilde{\Sigma}(\Psi)(x)\tilde{\Sigma}(\Psi)(y)dxdy
\end{aligned}$$

By summing in average over $\mu \in \mathfrak{S}_n$, and using Schwarz inequality, we get:

$$\begin{aligned}
\frac{1}{n!}\sum_{\mu\in\mathfrak{S}_n}&E[(\omega-V_A)^2|S]_{\Psi\circ\mu} \\
&\geq \sum_{i,j}\int_{x,y}\bigg(\frac{1}{n!}\sum_{\mu\in\mathfrak{S}_n}\omega_{\sigma_m}^*(\mu(i),\mu(j)) - \frac{x\cdot j}{n}\bigg)^2\chi_i\Big(\frac{x}{k}\Big)\chi_j\Big(\frac{y}{k}\Big)\tilde{\Sigma}(\Psi)(x)\tilde{\Sigma}(\Psi)(y)dxdy \\
&\quad - O\Big(\frac{1}{k^3}\Big)
\end{aligned}$$

By writing $\omega_{\sigma_m}^*(i,j) = \frac{2k}{n^2}\big(\sum_{r,s\in I_0} i_{\sigma_m^{-1}(r)}j_{\sigma_m^{-1}(s)} + \sum_{r,s\in \overline{I_0}} i_{\sigma_m^{-1}(r)}j_{\sigma_m^{-1}(s)}\big)$ and distinguishing the cases $r \neq s$ from $r = s$, we obtain with the stability properties introduced in proof of Lemma 1:

$$\frac{1}{n!}\sum_{\mu\in\mathfrak{S}_n}\omega_{\sigma_m}^*\big(\mu(i),\mu(j)\big) = \frac{k|i||j|}{n^2} + O\Big(\frac{k}{n}\Big)$$

And then eventually, via Lemma 1, we get:

$$\frac{1}{n!}\sum_{\mu\in\mathfrak{S}_n}E[(\omega-V_A)^2|S]_{\Psi\circ\mu} \geq \frac{C(\alpha)}{k^2} - O\Big(\frac{1}{k^3}\Big) - O\Big(\frac{1}{kn}\Big)$$

As the above inequality is true for the average summing over $\mu \in \mathfrak{S}_n$, it is in particular true for at least a particular $\Psi \circ \mu_M$. We can then finish the proof by using the same technique than for Theorem 1 from the step $(T1.2)$, with that particular distribution and the above inequality as a replacement of $(T1.2)$. ∎

## VII.    Annex 2: The Permutative Sleeking transform and its application to evaluate Deep Random generator maturity period

**The Permutative Sleeking transform**

The permutative sleeking transform is a tool that turn a given probability distribution into a « sufficiently regular » one, meaning avoiding brutal variation.

A permutative sleeking kernel $\gamma$ is a function $\mathbb{N}_n^* \to [0,1]$ (note that it is impossible that $|\sigma| = 1$ and thus the component for 1 can be ignored) that verifies:

$$\sum_{\sigma \in \mathfrak{S}_n} \gamma(|\sigma|) = 1$$

Then, for a given probability distribution $\Phi$, the permutative sleeking transformation of $\Phi$ is:

$$\Phi \mapsto T_\gamma[\Phi] \triangleq \sum_{\sigma \in \mathfrak{S}_n} \gamma(|\sigma|) \Phi \circ \sigma$$

A simple property of permutative sleeking transformations is the following:

**Proposition 11.**

*The space of permutative sleeking transformations, fitted with composition operation, is stable.*

■ Let's first remark that, for any $\sigma, \mu \in \mathfrak{S}_n$:

$$|\sigma| = |\mu \circ \sigma \circ \mu^{-1}|$$

Indeed, if $i \in \mu(supp(\sigma))$, then $i$ can uniquely be written $i = \mu(j)$ with $j \in supp(\sigma)$. Then $\mu \circ \sigma \circ \mu^{-1}(i) = \mu \circ \sigma(j) \neq i$ because $\sigma(j) \neq j$, which implies that $|\sigma| \leq |\mu \circ \sigma \circ \mu^{-1}|$. Reversely, if $i \notin \mu(supp(\sigma))$, then $i$ can uniquely be written $i = \mu(j')$ with $j' \notin supp(\sigma)$, which implies that $\sigma(j') = j'$ and following $\mu \circ \sigma \circ \mu^{-1}(i) = \mu \circ \sigma(j') = \mu(j') = i$ ; thus $n - |\sigma| \leq n - |\mu \circ \sigma \circ \mu^{-1}|$, and finally $|\sigma| = |\mu \circ \sigma \circ \mu^{-1}|$.

In addition, it is obvious that $\mu \circ \sigma \circ \mu^{-1} = \text{Id}_{\mathfrak{S}_n} \Leftrightarrow \sigma = \text{Id}_{\mathfrak{S}_n}$, and thus by cardinality $\{\sigma \, | \, |\sigma| = p\} = \{\mu \circ \sigma \circ \mu^{-1} \, | \, |\sigma| = p\} \; \forall \mu \in \mathfrak{S}_n$.

We can then write the composition of two permutative sleeking kernels :

$$T_\gamma \circ T_\delta[\Phi] = \sum_{\sigma \in \mathfrak{S}_n} \gamma(|\sigma|) \sum_{\mu \in \mathfrak{S}_n} \delta(|\mu|) \Phi \circ \sigma \circ \mu = \sum_{s \in \mathfrak{S}_n} \left( \sum_{\substack{\sigma, \mu \in \mathfrak{S}_n \\ \sigma \circ \mu = s}} \gamma(|\sigma|) \delta(|\mu|) \right) \Phi \circ s$$

Let's now show that the coefficient depends only on $|s|$ :

$$\sum_{\substack{\sigma, \mu \in \mathfrak{S}_n \\ \sigma \circ \mu = s}} \gamma(|\sigma|) \delta(|\mu|) = \sum_{\sigma \in \mathfrak{S}_n} \gamma(|\sigma|) \delta(|\sigma^{-1} \circ s|)$$

Then for any $\mu \in \mathfrak{S}_n$ :

$$\sum_{\sigma \in \mathfrak{S}_n} \gamma(|\sigma|)\delta(|\sigma^{-1} \circ \mu \circ s \circ \mu^{-1}|) = \sum_{\sigma \in \mathfrak{S}_n} \gamma(|\sigma|)\delta(|\mu^{-1} \circ \sigma^{-1} \circ \mu \circ s|)$$

$$= \sum_{\sigma \in \mathfrak{S}_n} \gamma(|\mu \circ \sigma \circ \mu^{-1}|)\delta(|\mu^{-1} \circ \sigma^{-1} \circ \mu \circ s|) = \sum_{\sigma \in \mathfrak{S}_n} \gamma(|\sigma'|)\delta(|\sigma'^{-1} \circ s|)$$

with the change of variable $\sigma' = \mu \circ \sigma \circ \mu^{-1}$, that courses the set $\{\sigma'||\sigma'| = |\sigma|\}$ as seen above. This shows that the coefficient only depends on $|s|$. In addition, obviously :

$$\sum_{s \in \mathfrak{S}_n} \sum_{\sigma \in \mathfrak{S}_n} \gamma(|\sigma|)\delta(|\sigma^{-1} \circ s|) = \left( \sum_{\sigma \in \mathfrak{S}_n} \gamma(|\sigma|) \right) \left( \sum_{s' \in \mathfrak{S}_n} \gamma(|s'|) \right) = 1$$

which means that $T_\gamma \circ T_\delta$ is still a permutative sleeking transformation.

$\blacksquare$

For $I$ a subset of $\mathbb{N}_n$ we define by $|I|$ the cardinality of $I$ and $\mathcal{S}_I$ the subgroup of $\mathfrak{S}_n$ including permutations for which $supp(\sigma) \subset I$.

We have the following inversion formula for the Permutative Sleeking transform:

**Proposition 12.**

*Let $f$ be a function $f : \mathfrak{S}_n \to \mathbb{R}$. Then, if we denote $F(s) \triangleq \sum_{\substack{\sigma \in \mathfrak{S}_n \\ |\sigma| = s}} f(\sigma)$ and $G(s) = \sum_{\substack{I \\ |I| = s}} \sum_{\sigma \in \mathcal{S}_I} f(\sigma)$, we have the following inversion formulas:*

$$G(s) = \sum_{u=0}^{s} \binom{n-u}{s-u} F(u)$$

$$F(s) = \sum_{u=0}^{s} (-1)^{u+s} \binom{n-u}{s-u} G(u)$$

■ We have the first relation by:

$$G(s) = \sum_{\substack{\sigma \in \mathfrak{S}_n \\ |\sigma| \leq s}} \left( \sum_{\substack{I \\ |I| = s \\ I \supset supp(\sigma)}} 1 \right) f(\sigma) = \sum_{\substack{\sigma \in \mathfrak{S}_n \\ |\sigma| \leq s}} \binom{n-|\sigma|}{s-|\sigma|} f(\sigma) = \sum_{u=0}^{s} \binom{n-u}{s-u} F(u)$$

This formula can be reversed by using the classical matrix inverse:

$$\begin{pmatrix} \vdots & \cdots & 0 \\ \binom{i}{j} & \cdots & \\ & & \end{pmatrix}_{i,j \in \mathbb{N}_n}^{-1} = \begin{pmatrix} \vdots & \cdots & 0 \\ (-1)^{i+j} \binom{i}{j} & \cdots & \\ & & \end{pmatrix}_{i,j \in \mathbb{N}_n}$$

(which can be easily seen by considering for instance the inverse of the $\mathbb{R}_n[X]$ endomorphism $P(X) \mapsto P(X+1)$).

By setting $j' = n - j$ and $i' = n - i$, we have $\binom{n-j}{i-j} = \binom{j'}{j'-i'} = \binom{j'}{i'}$, and thus :

$$\begin{pmatrix} & & \cdots & 0 \\ \vdots & \ddots & & \vdots \\ \binom{n-j}{i-j} & \cdots & & \end{pmatrix}_{i,j \in \mathbb{N}_n}^{-1} = \begin{pmatrix} & & \cdots & 0 \\ \vdots & \ddots & & \vdots \\ (-1)^{i+j}\binom{n-j}{i-j} & \cdots & & \end{pmatrix}_{i,j \in \mathbb{N}_n}$$

from which we take the reverse formula:

$$F(s) = \sum_{u=0}^{s} (-1)^{u+s} \binom{n-u}{s-u} G(u)$$

∎

### The Synchronized Strategy Lemma and its application to evaluate Deep Random generator maturity period

We introduce the operator $\Delta_0$ for 2 distributions $\Phi, \Phi'$:

$$\Delta_0(\Phi, \Phi') \triangleq \int_{x,y \in [0,1]^n} \left( \frac{|x||y|}{n^2} - \frac{x.y}{n} \right)^2 \Phi(x)\Phi'(y) dx dy$$

$\Delta_0$ gives a measure of synchronization between $\Phi$ and $\Phi'$.

**Proposition 13.**

*If $\Phi$ and $\Phi' \in \zeta(\alpha)$, then:*

$$\max_{\sigma \in \mathfrak{S}_n} \Delta_0(\Phi, \Phi' \circ \sigma) \geq \frac{\alpha^2}{4} - \frac{2}{n}$$

*$\Phi$ and $\Phi' \circ \sigma$ are then said 'synchronized'.*

∎ Let's simply write:

$$\max_{\sigma \in \mathfrak{S}_n} \Delta_0(\Phi, \Phi' \circ \sigma) \geq \Delta_0\big(\tilde{\Sigma}(\Phi), \tilde{\Sigma}(\Phi')\big)$$

And then, applying Lemma 1 (with $g(|x|, |y|) = \frac{|x||y|}{n^2}$ and $k = 1$), we get the result. ∎

The following lemma gives a useful calculation method when using distributions transformed by permutative sleeking.

**Lemma 2.**

*Let $\Phi$ and $\Phi'$ be two probability distributions, and $\gamma$ be a permutative sleeking kernel, then we have:*

$$\Delta_\gamma(\Phi, \Phi') \triangleq \sum_{\sigma \in \mathfrak{S}_n} \gamma(|\sigma|) \int_{x,y \in [0,1]^n} \left( x.y - x.\sigma(y) \right)^2 \Phi(x) \Phi'(y) dx dy$$

$$= \sum_{s=0}^{n} \gamma(s) A_n^s \left( \frac{s^2}{e} \Delta_0(\Phi, \Phi') + O(n) \right)$$

■ For $I$ a subset of $\mathbb{N}_n$ we define by $|I|$ the cardinality of $I$ and $\mathcal{S}_I$ the subgroup of $\mathfrak{S}_n$ including permutations for which $supp(\sigma) \subset I$. Let's set:

$$f(\sigma) = \int_{x,y \in [0,1]^n} \left( x.y - x.\sigma(y) \right)^2 \Phi(x) \Phi'(y) dx dy$$

$$F(s) = \sum_{\substack{\sigma \in \mathfrak{S}_n \\ |\sigma| = s}} f(\sigma)$$

$$G(s) = \sum_{\substack{I \\ |I| = s}} \sum_{\sigma \in \mathcal{S}_I} f(\sigma)$$

Thanks to Proposition 12, we can write $\Delta_\gamma$ depending on $G(u)$ as follows:

$$\Delta_\gamma = \sum_{\sigma \in \mathfrak{S}_n} \gamma(|\sigma|) f(\sigma) = \sum_{s=0}^{n} \gamma(s) \sum_{\substack{\sigma \in \mathfrak{S}_n \\ |\sigma| = s}} f(\sigma) = \sum_{s=0}^{n} \gamma(s) F(s)$$

$$= \sum_{s=0}^{n} \gamma(s) \sum_{u=0}^{s} (-1)^{u+s} \binom{n-u}{s-u} G(u) \qquad (l2.1)$$

Let's now calculate $G(u)$ by using the properties of $\mathcal{S}_I$ as a group (in the following, $C, C', C''$ are constants); we will use the notation $\Phi_{r,t} = M_\Phi(r, t)$ to simplify the writings:

$$G(u) = \sum_{|I|=s} \sum_{\sigma \in \delta_I} \left( \sum_{r,t \in \mathbb{N}_n^*} \left( \Phi_{r,t} \Phi'_{r,t} - 2\Phi_{r,t} \Phi'_{r,\sigma(t)} + \Phi_{r,t} \Phi'_{\sigma(r),\sigma(t)} \right) \right)$$

$$= \sum_{|I|=s} \left( u! \sum_{r,t \in \mathbb{N}_n^*} \Phi_{r,t} \Phi'_{r,t} - 2(u-1)! \sum_{r,t,t' \in I} \Phi_{r,t} \Phi'_{r,t'} - 2u! \sum_{r,t \in \mathbb{N}_n^* \times \bar{I}} \Phi_{r,t} \Phi'_{r,t} \right.$$

$$\left. + (u-2)! \sum_{r,r',t,t' \in I} \Phi_{r,t} \Phi'_{r',t'} + u! \sum_{r,t \in \bar{I}} \Phi_{r,t} \Phi'_{r,t} + \lhd (Cnu!) \right)$$

$$= u! \binom{n-2}{u-2} \sum_{r,t \in \mathbb{N}_n^*} \Phi_{r,t} \Phi'_{r,t} + (u-2)! \binom{n-4}{u-4} \sum_{r,r',t,t' \in \mathbb{N}_n^*} \Phi_{r,t} \Phi'_{r',t'}$$

$$- 2(u-1)! \binom{n-3}{u-3} \sum_{r,t,t' \in \mathbb{N}_n^*} \Phi_{r,t} \Phi'_{r,t'} + \lhd \left( C'nu! \binom{n}{u} \right)$$

$$= A_n^u (u^2 \Delta_0(\Phi, \Phi') + \lhd (C''n))$$

In addition, we have:

$$\sum_{u=0}^{s} (-1)^{u+s} u^2 A_n^u \binom{n-u}{n-s} = A_n^s \sum_{u=0}^{s} (-1)^{u+s} \frac{u^2}{(s-u)!} = A_n^s \left( \frac{s^2}{e} + \frac{2s}{e} + O(1) \right)$$

and:

$$\left| \sum_{u=0}^{s} (-1)^{u+s} A_n^u \binom{n-u}{n-s} \lhd (C''n) \right| \leq C''n \sum_{u=0}^{s} A_n^u \binom{n-u}{n-s} \leq eC''nA_n^s$$

By injecting the calculation of $G(u)$ in $(l2.1)$, we can conclude the proof with the two above relations.

∎

An immediate consequence of lemma 2 is that, if the probability distributions $\Phi$ and $\Phi' \in \zeta(\alpha)$ are synchronized, then stands the following inequality:

$$\Delta_\gamma(\Phi, \Phi') \geq \frac{\alpha^2}{4e} \sum_{s=0}^{n} \gamma(s) s^2 A_n^s + O(n)$$

We will also use the following corollary, that relies on the same calculation technique than presented in Lemma 2.

The « border polynomial » is defined by:

$$P_{u,v}(X) \triangleq X \left( \frac{u}{n} - X \right) \left( \frac{v}{n} - X \right) \left( 1 - \frac{u}{n} - \frac{v}{n} + X \right)$$

**Corollary 1 (treatment of border conditions).**

*For $\Phi$ and $\Phi'$ two synchronized probability distributions in $\zeta(\alpha)$, there exists a constant $K(\alpha) > 0$ (independent from $\Phi$ and $\Phi'$) and a permutative sleeking transformation $\delta$, depending on $\alpha$ but not on $\Phi$ and $\Phi'$, such that:*

$$\sum_{\sigma \in \mathfrak{S}_n} \int_{x,y \in [0,1]^n} P_{|x|,|y|}\left(\frac{x.y}{n}\right) P_{|x|,|y|}\left(\frac{x.\sigma(y)}{n}\right)\left(\frac{x.y}{n} - \frac{x.\sigma(y)}{n}\right)^2 \gamma(|\sigma|)\Phi(x)T_\delta[\Phi'](y)dxdy$$

$$\geq K(\alpha) \sum_{r=0}^{n} \frac{r^2}{n^2} A_n^r \gamma(r) - \frac{1}{\sqrt{e}}\sum_{r=0}^{n}\frac{r^3}{n^3}A_n^r\gamma(r) - O\left(\frac{1}{\sqrt{n}}\right)$$

∎

By using the same calculation technique than in Lemma 2 (we won't present the details here), we obtain more generally that for any polynomial $P$ with degree $p$, and $t \in [0,1]$, there exists a constant $K_p > 0$ such that :

$$\frac{1}{A_n^{nt}} \sum_{\substack{\sigma \in \mathfrak{S}_n \\ |\sigma| = nt}} P\left(\frac{x.\sigma(y)}{n}\right) = \frac{1}{e}P\left(t\frac{|x||y|}{n^2} + (1-t)\frac{x.y}{n}\right) + \triangleleft\left(\frac{K_p\|P\|_\infty}{n}\right) \quad (c1.1)$$

We have oviously (because $\left|P_{|x|,|y|}'(X)\right| \leq 4$ over $[0,1]$) :

$$P_{|x|,|y|}\left(\frac{x.\sigma(y)}{n}\right) = P_{|x|,|y|}\left(\frac{x.y}{n}\right) + \triangleleft\left(4\left|\frac{x.y}{n} - \frac{x.\sigma(y)}{n}\right|\right)$$

Let's apply a first time $(c1.1)$ to :

$$\frac{1}{A_n^{nt}} \sum_{\substack{\sigma \in \mathfrak{S}_n \\ |\sigma| = nt}} \int_{x,y \in [0,1]^n} P_{|x|,|y|}\left(\frac{x.y}{n}\right)P_{|x|,|y|}\left(\frac{x.\sigma(y)}{n}\right)\left(\frac{x.y}{n} - \frac{x.\sigma(y)}{n}\right)^2 \Phi(x)\Phi'(y)dxdy$$

$$\geq \frac{1}{A_n^{nt}} \sum_{\substack{\sigma \in \mathfrak{S}_n \\ |\sigma| = nt}} \int_{x,y \in [0,1]^n} P_{|x|,|y|}\left(\frac{x.y}{n}\right)^2\left(\frac{x.y}{n} - \frac{x.\sigma(y)}{n}\right)^2 \Phi(x)\Phi'(y)dxdy$$

$$- 4\left(\frac{1}{A_n^{nt}} \sum_{\substack{\sigma \in \mathfrak{S}_n \\ |\sigma| = nt}} \int_{x,y \in [0,1]^n}\left(\frac{x.y}{n} - \frac{x.\sigma(y)}{n}\right)^6 \Phi(x)\Phi'(y)dxdy\right)^{1/2}$$

$$\geq \frac{t^2}{e} \int_{x,y \in [0,1]^n} P_{|x|,|y|}\left(\frac{x.y}{n}\right)^2\left(\frac{|x||y|}{n^2} - \frac{x.y}{n}\right)^2 \Phi(x)\Phi'(y)dxdy +$$

$$\triangleleft\left(\frac{2K_2}{n}\right) - 4\left(\frac{t^6}{e} + \triangleleft\left(\frac{15K_6}{n}\right)\right)^{1/2} \qquad (c1.2)$$

Let's consider the probability distribution, obtained by the following « Dirac » permutative sleeking transformation:

$$\Phi'_l = \frac{1}{A_n^{nl}} \sum_{\substack{\mu \in \mathfrak{S}_n \\ |\mu| = nl}} \Phi' \circ \mu$$

It can be easily remarked that:

$$P_{|x|,|y|}\left(t\,\frac{|x||y|}{n^2}+(1-t)\,\frac{x.y}{n}\right)\geq t^4\left(\frac{|x|}{n}\,\frac{|y|}{n}\left(1-\frac{|x|}{n}\right)\left(1-\frac{|y|}{n}\right)\right)^2$$

By replacing $\Phi'$ by $\Phi'_l$ in $(c1.2)$ and applying a second time $(c1.1)$ to the first term of the right side of the inequality, with the above we obtain:

$$\int_{x,y\in[0,1]^n}P_{|x|,|y|}\left(\frac{x.y}{n}\right)^2\left(\frac{|x||y|}{n^2}-\frac{x.y}{n}\right)^2\Phi(x)\Phi'_l(y)dxdy$$

$$\geq\frac{1}{A_n^{nl}}\sum_{\substack{\mu\in\mathfrak{S}_n\\|\mu|=nl}}\int_{x,y\in[0,1]^n}P_{|x|,|y|}\left(\frac{x.\mu(y)}{n}\right)^2\left[\left(\frac{|x||y|}{n^2}-\frac{x.y}{n}\right)^2-2l\right]\Phi(x)\Phi'(y)dxdy$$

$$\geq\frac{l^8}{e}\int_{x,y\in[0,1]^n}\left(\frac{|x|}{n}\,\frac{|y|}{n}\left(1-\frac{|x|}{n}\right)\left(1-\frac{|y|}{n}\right)\right)^4\left[\left(\frac{|x||y|}{n^2}-\frac{x.y}{n}\right)^2\right.$$

$$\left.-2l\right]\Phi(x)\Phi'(y)dxdy+\lhd\left(\frac{28K_8}{n}\right)\qquad\qquad(c1.3)$$

In addition, as $\left(\frac{|x||y|}{n^2}-\frac{x.y}{n}\right)^2$ is unchanged by transformations $x\mapsto\bar{x}$ and $y\mapsto\bar{y}$ we can write :

$$\left(\frac{|x||y|}{n^2}-\frac{x.y}{n}\right)^2\leq\min\left(\left(\frac{|x||y|}{n^2}\right)^2,\left(\left(1-\frac{|x|}{n}\right)\left(1-\frac{|y|}{n}\right)\right)^2\right)\leq\frac{|x|}{n}\,\frac{|y|}{n}\left(1-\frac{|x|}{n}\right)\left(1-\frac{|y|}{n}\right)$$

And therefore we can extend $(c1.3)$ with Proposition 13 and the above:

$$\int_{x,y\in[0,1]^n}P_{|x|,|y|}\left(\frac{x.y}{n}\right)^2\left(\frac{|x||y|}{n^2}-\frac{x.y}{n}\right)^2\Phi(x)\Phi'_l(y)dxdy$$

$$\geq\frac{l^8}{e}\int_{x,y\in[0,1]^n}\left[\left(\frac{|x||y|}{n^2}-\frac{x.y}{n}\right)^{10}-2l\right]\Phi(x)\Phi'(y)dxdy+\lhd\left(\frac{28K_8}{n}\right)$$

$$\geq\frac{l^8}{e}[\Delta_0(\Phi,\Phi')^5-2l]+\lhd\left(\frac{28K_8}{n}\right)\geq\frac{l^8}{e}\left[\frac{\alpha^{10}}{2^{20}}-2l\right]+O\left(\frac{1}{n}\right)$$

We can take the maximum over $l\in[0,1]$ which is obtained for $l_0=\frac{\alpha^{10}}{9\cdot2^{18}}$ :

$$\int_{x,y\in[0,1]^n}P_{|x|,|y|}\left(\frac{x.y}{n}\right)^2\left(\frac{|x||y|}{n^2}-\frac{x.y}{n}\right)^2\Phi(x)\Phi'_{l_0}(y)dxdy\geq eK(\alpha)+O\left(\frac{1}{n}\right)$$

By injecting the above in $(c1.2)$ and remarking that $\sqrt{a+b}\leq\sqrt{a}+\sqrt{b},\forall a,b>0$, we obtain the result. The permutative sleeking transformation $T[\Phi']=\frac{1}{A_n^{nl_0}}\sum_{\substack{\mu\in\mathfrak{S}_n\\|\mu|=nl_0}}\Phi'\circ\mu$ is indeed only depending on $\alpha$.

∎

We introduce the set of strategies $\omega(i, j)$ that are invariant by application of a same permutation on $i$ and $j$:

$$\Omega_\# \triangleq \left\{ \omega \in [0,1]^{2^{2n}} \mid \omega(\sigma(i), \sigma(j)) = \omega(i, j) \ , \forall \sigma \in \mathfrak{S}_n \right\}$$

One can remark that:

$$\Omega_\# = \left\{ \omega \in [0,1]^{2^{2n}} \mid \omega_{i,j} = f(|i|, |j|, i.j), \forall f: \mathbb{N}_n^3 \mapsto [0,1] \right\}$$

Let's justify this equivalence:

■ it is clear that any strategy of the form $f(|i|, |j|, i.j)$ is in $\Omega_\#$. Let $\omega$ be in $\Omega_\#$, and let's consider $(i, j)$ and $(i', j')$ such that :

$$\begin{cases} |i| = |i'| \\ |j| = |j'| \\ i.j = i'.j' \end{cases}$$

Then, as $\{i \backslash j, i \cap j, j \backslash i, \overline{i \cup j}\}$ and $\{i' \backslash j', i' \cap j', j' \backslash i', \overline{i' \cup j'}\}$ are isomorphic partitions of $\mathbb{N}_n^*$, there exists $\sigma \in \mathfrak{S}_n$ such that :

$$\begin{cases} \sigma(i \backslash j) = \sigma(i' \backslash j') \\ \sigma(i \cap j) = \sigma(i' \cap j') \\ \sigma(j \backslash i) = \sigma(j' \backslash i') \\ \sigma(\overline{i \cup j}) = \sigma(\overline{i' \cup j'}) \end{cases}$$

and thus $\omega_{i',j'} = \omega_{\sigma(i),\sigma(j)} = \omega_{i,j}$, which means that $\omega_{i,j}$ only depends on $|i|, |j|, i.j$

■

We now demonstrate a technical Lemma:

Let's first establish the very simple Proposition:

**Proposition 14.**

With $a, b \in \mathbb{R}$ and $\delta > 0$, let $f : [a - \delta, b + \delta] \to \mathbb{R}^+$ having only one extremum on $[a - \delta, b + \delta]$ and such that this extremum is a maximum, reached for $c$. If :

$$|c - b| \geq D \ and \ |c - a| \geq D$$

then $S(\delta_1, \delta_2) \triangleq \sum_{n=a+\delta_1}^{b+\delta_2} f(n)$ verifies, for $|\delta_1|, |\delta_2| < \delta : \frac{S(\delta_1, \delta_2)}{S(0)} \leq 1 + \frac{\delta}{D}$.

■ We can suppose that $a, b, c$ are integers, which does not change the above argument. The worst case to upper bound $S(\delta_1, \delta_2)$ is $\delta_1 \leq 0, \delta_2 \geq 0$ :

$$S(\delta_1, \delta_2) \leq S(0) + \sum_{n=a-|\delta_1|}^{a} f(n) + \sum_{n=b}^{b+|\delta_2|} f(n) \leq S(0) + |\delta_1| f(a) + |\delta_2| f(b)$$

$$\leq S(0) + |\delta_1| \frac{\sum_{n=a}^{c} f(n)}{D} + |\delta_2| \frac{\sum_{n=c}^{b} f(n)}{D}$$

$$\leq \left(1 + \frac{|\delta_1|}{D}\right) \sum_{n=a}^{c} f(n) + \left(1 + \frac{|\delta_2|}{D}\right) \sum_{n=c}^{b} f(n) \leq S(0) \left(1 + \frac{\max(|\delta_1|, |\delta_2|)}{D}\right)$$

$$\leq S(0) \left(1 + \frac{\delta}{D}\right)$$

∎

**Lemma 3. (Synchronized Strategy lemma)**

*Let $\omega$ be a strategy in $\Omega_\#$, and $\Phi$ and $\Phi'$ be two probability distributions in $\zeta(\alpha)$ with at least one of them transformed by permutative sleeking (with a permutative sleeking transformation that will be explicited), say $\Phi' = T[\Phi''']$, then for $\sigma$ performing a synchronization of $\Phi$ and $\Phi'''$, there exists a constant $C(\alpha)$ such that :*

$$E\left[\left(\omega_{i,j} - \frac{x.y}{nk}\right)^2\right]_{\Phi, T[\Phi''' \circ \sigma]} \geq \frac{C_3(\alpha)}{n}$$

∎ The proof of lemma 3 is a bit long, we will sometimes give shortened explanations to avoid to present exhaustively some painful calculations. We divide the proof in 3 parts. In part 1, we will reduce the general Bernoulli distributions to simpler multi-Gaussian form. In part 2, we will discard the dependence in $\omega_{i,j}$. And in part 3, we will show that the resultant form from part 2 is not submitted to rapid variation in $x.y$ when $|x|$ and $|y|$ are fixed. In the conclusive part, we will deal with border conditions.

In the following, we will suppose $\Phi'''$ synchronized to $\Phi$ and thus we will denote the probability distribution directly $\Phi'''$ instead of $\Phi''' \circ \sigma$, and thus $\Phi' = T[\Phi''']$ instead of $\Phi' = T[\Phi''' \circ \sigma]$. $T$ will be explicited in the course of the proof. We will also restrict the probability distributions to the generation of experiment vectors in $\{0,1\}^n$ instead of $[0,1]^n$ because it eases the proof, and also because anyway, they are easier to manipulate with computers.

We will denote in the followings:

$$\langle \omega, \Phi, \Phi' \rangle \triangleq E\left[\left(\omega_{i,j} - \frac{x.y}{nk}\right)^2\right]_{\Phi, \Phi'}$$

**Part 1.**

Let's first set the following quantities, for $\{u, v, w\} \in \mathbb{N}_n^{\,3}$ :

$$\mathcal{J}(u) \triangleq \{x \in \mathbb{N}_n \,|\, |x| = u\}$$

$$\mathcal{K}(u,v,w) \triangleq \{x, y \in \mathbb{N}_n \,|\, |x| = u; |y| = v; x.y = w\}$$

$$\mathcal{\hat{K}}(u,v,w) \triangleq \{x, y \in \mathbb{N}_n \,|\, |x| \geq u; |y| \geq v; x.y \geq w\}$$

$$P(u,v,w) \triangleq \int_{\mathcal{K}(u,v,w)} \Phi(x)\Phi'(y)\,dx\,dy$$

$$\psi_{u,v,w}(x,y) \triangleq \sum_{i,j \in \mathcal{K}(u,v,w)} \chi_i(x)\chi_j(y)$$

$\{\psi_{r,s,t}\}$ is a linearly independant set that generates a vector subspace ; by considering dimension and because for any $\sigma \in \mathfrak{S}_n$,

$$\psi_{r,s,t}(\sigma(x),\sigma(y)) = \psi_{r,s,t}(x,y)$$

this subspace is thus isomorphic to the vector subspace of the multinomials :

$$\left\{ \sum_{i,j \in \{0,1\}^n} \omega_{i,j}\chi_i(x)\chi_j(y) \mid \omega \in \Omega_\# \right\}$$

This implies that the multinomial $\psi_{r,s,t}\left(\frac{x}{k},\frac{y}{k}\right)$ can be written as linear combination of $\{\psi_{u,v,w}(x,y)\}_{u,v,w}$. Let's then write this combination, with same method than seen in Proposition 1 :

$$\sum_{i,j \in \mathcal{K}(r,s,t)} \chi_i\left(\frac{x}{k}\right)\chi_j\left(\frac{x}{k}\right) = \sum_{i,j \in \mathcal{K}(r,s,t)} \frac{1}{k^{r+s}} \sum_{\substack{l \supset i \\ l' \supset j}} \left(1 - \frac{1}{k}\right)^{|l|+|l'|-r-s} \chi_l(x)\chi_{l'}(y)$$

$$= \sum_{l,l' \in \widehat{\mathcal{K}}(r,s,t)} \frac{1}{k^{r+s}} \left( \sum_{\substack{i,j \in \mathcal{K}(r,s,t) \\ i \subset l \\ j \subset l'}} 1 \right) \left(1 - \frac{1}{k}\right)^{|l|+|l'|-r-s} \chi_l(x)\chi_{l'}(y)$$

The drawing below:

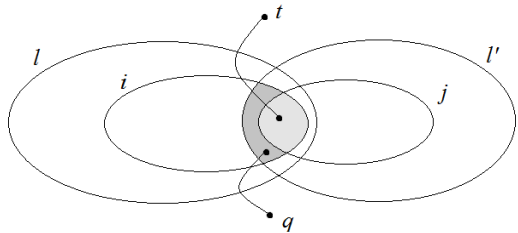

helps to understand the following expression of the above summing between brackets, where $u,v,w$ stand for $u = |l|, v = |l'|, w = l.l'$ :

$$\sum_{\substack{i,j \in \mathcal{K}(r,s,t) \\ i \subset l \\ j \subset l'}} 1 = \binom{w}{t} \sum_{q=0}^{\min(r-t,w-t,v-s)} \binom{w-t}{q}\binom{u-w}{r-t-q}\binom{v-t-q}{s-t}$$

Thanks to which we obtain:

$$\frac{1}{k^{r+s}}\left(\sum_{\substack{i,j\in\mathcal{K}(r,s,t)\\i\subset l\\j\subset l'}}1\right)\left(1-\frac{1}{k}\right)^{|l|+|l'|-r-s}$$

$$=\beta_{t,w}\left(\frac{1}{k^2}\right)\sum_{q=0}^{\min(r-t,w-t,v-s)}\beta_{q,w-t}\left(\frac{1}{k+1}\right)\beta_{s-t,v-t-q}\left(\frac{1}{k}\right)\beta_{r-t-q,u-w}\left(\frac{1}{k}\right)$$

$$=\beta_{t,w}\left(\frac{1}{k^2}\right)\sum_{q=t}^{\min(r,w,v-s+t)}\beta_{q-t,w-t}\left(\frac{1}{k+1}\right)\beta_{s-t,v-q}\left(\frac{1}{k}\right)\beta_{r-q,u-w}\left(\frac{1}{k}\right)\qquad(l3.1)$$

We remark that, at distance $O\left(\sqrt{\frac{n}{k}}\right)$ in the neighborhood of :

$$\begin{cases}r=u/k\\s=v/k\\t=w/k^2\end{cases}$$

the three maxima of the Gaussian filters under the sum in $(l3.1)$ are reached for an identical value of $q_m=\frac{w}{k}$ because

$$\frac{q_m-t}{w-t}=\frac{w/k-w/k^2}{w-w/k^2}=\frac{1}{k+1}$$

And by also remarking that :

$$\beta_{s-t,v-q_m}\left(\frac{1}{k}\right)=(1+o(1))\beta_{s-t+\frac{q_m}{k},v}\left(\frac{1}{k}\right)$$

we see that the width of the Gaussian filter $\beta_{q-t,w-t}\left(\frac{1}{k+1}\right)$ in $q$ is smaller than the width of the two other filters. Thus, by approximating the summing on the narrowest of the two other filters, we obtain that there exists a constant $C_1>0$ such that :

$$(l3.1)=\vartriangleright(C_1)\beta_{t,w}\left(\frac{1}{k^2}\right)\beta_{s-t,v-q_m}\left(\frac{1}{k}\right)\beta_{r-q_m,u-w}\left(\frac{1}{k}\right)\sum_{q=t}^{r}\beta_{q-t,w-t}\left(\frac{1}{k+1}\right)$$

And because $t\ll\frac{w}{k}\le\frac{u}{k}$ in the considered neighborhood, we have :

$$\sum_{q=t}^{r}\beta_{q-t,w-t}\left(\frac{1}{k+1}\right)\sim 1$$

Thus finally there exists a constant $C_2>0$ such that :

$$(l3.1)=\vartriangleright(C_2)\beta_{t,w}\left(\frac{1}{k^2}\right)\beta_{s,v}\left(\frac{1}{k}\right)\beta_{r,u}\left(\frac{1}{k}\right)$$

We saw that, as $\omega \in \Omega_{\#}$, it can be written as $\omega_{i,j} = f(|i|, |j|, i.j)$, thus, in the followings, we will write $\omega_{i,j} = \omega_{r,s}(t)$. It follows from the above that :

$$\langle \omega, \Phi, \Phi' \rangle = \triangleright (C_2) \sum_{u,v \in \mathbb{N}_n} \sum_{r,s \in \mathbb{N}_n} \left( \sum_{w \in \mathbb{N}_n} \left( \sum_{t=0}^{w} \left( \omega_{r,s}(t) \right. \right.\right.$$
$$\left.\left.\left. - \frac{w}{nk} \right)^2 \beta_{t,w}\left(\frac{1}{k^2}\right) \right) P(u,v,w) \right) \beta_{s,v}\left(\frac{1}{k}\right) \beta_{r,u}\left(\frac{1}{k}\right) \qquad (l3.2)$$

which ends Part 1.

Part 2.

From Proposition 2 (ii), $\beta_{\frac{w}{k^2}+u,w}\left(\frac{1}{k^2}\right)$ can be approximated (with $\sigma^2 = \frac{1}{k^2}\left(1 - \frac{1}{k^2}\right)$) by

$$\triangleright (C'') N(u, w\sigma^2)$$

under the condition that $|u| \leq \left(\frac{w}{k^2}\right)^{2/3}$. On the other hand, from Proposition 2 (i), if $|u| > \mathcal{K}(\alpha)\sqrt{\frac{w}{k^2}}$ for $\mathcal{K}(\alpha)$ sufficiently large, the remains of the binomial distribution can be kept bounded by $\frac{c(\alpha)}{2n}$. Thus, supposing that $n \gg k^2$, it is possible to ease calculations to replace the binomial distributions by gaussian distributions in ($l3.2$), with a multiplicative lower bound.

Let's now consider :

$$F_{r,s,u,v} \triangleq \int_{w \in \mathbb{R}} \left( \int_{t \in \mathbb{R}} \left( \omega_{r,s}(t) - \frac{w}{nk} \right)^2 N\left(t - \frac{w}{k^2}, w\sigma^2\right) dt \right) P(u,v,w) dw$$

where $\omega_{r,s}(t)$ is extended to $\mathbb{R}$ being 0 outside $[0,n]$ (note also that $P(u,v,w) = 0$ if $w \notin [0,n]$). The above argument means that there exists a constant $C_3 > 0$ such that :

$$\langle \omega, \Phi, \Phi' \rangle = \triangleright (C_3) \sum_{u,v \in \mathbb{N}_n} \sum_{r,s \in \mathbb{N}_n} F_{r,s,u,v} \beta_{s,v}\left(\frac{1}{k}\right) \beta_{r,u}\left(\frac{1}{k}\right) \qquad (l3.2')$$

We now study $F_{r,s,u,v}$ with objective to discard the dependence in $\omega_{r,s}(t)$.

On one hand, if $V$ and $V'$ are two random variables with identical probability distribution, and $K$ a constant, we have the obvious relations :

$$E[(V-K)^2] = E[(V - E[V])^2] + (E[V] - K)^2 = \frac{1}{2}E[(V-V')^2] + (E[V] - K)^2 \geq \frac{1}{2}E[(V-V')^2]$$

which, applied to $F_{r,s,u,v}$ gives :

$$F_{r,s,u,v} \geq \int_{w \in \mathbb{R}} \left( \int_{t,\delta \in \mathbb{R}} \left( \omega_{r,s}(t+\delta) - \omega_{r,s}(t) \right)^2 N\left(t + \delta - \frac{w}{k^2}, w\sigma^2\right) N\left(t\right.$$
$$\left. - \frac{w}{k^2}, w\sigma^2\right) dt \, d\delta \right) P(u,v,w) dw \qquad (l3.3)$$

We observe that :

$$N\left(t + \delta - \frac{w}{k^2}, w\sigma^2\right) N\left(t - \frac{w}{k^2}, w\sigma^2\right) = N(\delta, 2w\sigma^2) N\left(t + \frac{\delta}{2} - \frac{w}{k^2}, \frac{w\sigma^2}{2}\right)$$

We use the change of variable :

$$\begin{pmatrix} w \\ \delta \end{pmatrix} \mapsto \begin{pmatrix} 1 & -k^2/2 \\ 0 & 1 \end{pmatrix} \begin{pmatrix} w \\ \delta \end{pmatrix}$$

whose Jacobian is 1. Let's set, for $\in\ ]0,1[$ :

$$J(\delta, \lambda) = \{w | (w - k^2|\delta| \geq 2\lambda n) \ \& \ (n \geq w)\}$$

We have then, for $\in J(\delta, \lambda)$ :

$$N\left(\delta, 2\left(w + \frac{k^2\delta}{2}\right)\sigma^2\right) \geq \sqrt{\frac{2\lambda n}{w + \frac{k^2\delta}{2}}} N(\delta, 4\lambda n\sigma^2) \geq \sqrt{\lambda} N(\delta, 4\lambda n\sigma^2)$$

$$N\left(t - \frac{w}{k^2}, \frac{1}{2}\left(w + \frac{k^2\delta}{2}\right)\sigma^2\right) \geq \sqrt{\frac{2\lambda n}{w + \frac{k^2\delta}{2}}} N\left(t - \frac{w}{k^2}, \lambda n\sigma^2\right) \geq \sqrt{\lambda} N\left(t - \frac{w}{k^2}, \lambda n\sigma^2\right)$$

Applying the variable change to ($l3.3$) together with the above Gaussian relations, we obtain :

$$F_{r,s,u,v} \geq \lambda \int_{\delta \in \mathbb{R}} N(\delta, 4\lambda n\sigma^2) \int_{w \in J(\delta,\lambda)} \left( \int_{t \in \mathbb{R}} \left( \omega_{r,s}(t + \delta) - \omega_{r,s}(t) \right)^2 N\left(t - \frac{w}{k^2}, \lambda n\sigma^2\right) dt \right) P\left(u, v, w + \frac{k^2\delta}{2}\right) dw d\delta \qquad (l3.4)$$

On the other hand, by applying the change of variable, for any :

$$\begin{pmatrix} w \\ t \end{pmatrix} \mapsto \begin{pmatrix} w - k^2\delta \\ t + \delta \end{pmatrix}$$

we obtain the two equalities for $F_{r,s,u,v}$ :

$$F_{r,s,u,v} = \int_{w \in \mathbb{R}} \left( \int_{t \in \mathbb{R}} \left( \omega_{r,s}(t + \delta) - \frac{w}{nk} - \frac{k\delta}{n} \right)^2 N\left(t - \frac{w}{k^2}, (w + k^2\delta)\sigma^2\right) dt \right) P(u, v, w + k^2\delta) dw$$

$$= \int_{\delta \in \mathbb{R}} N(\delta, 4\lambda n\sigma^2) \int_{w \in \mathbb{R}} \left( \int_{t \in \mathbb{R}} \left( \omega_{r,s}(t + \delta) - \frac{w}{nk} - \frac{k\delta}{n} \right)^2 N\left(t - \frac{w}{k^2}, (w + k^2\delta)\sigma^2\right) dt \right) P(u, v, w + k^2\delta) dw \, d\delta \qquad (l3.5)$$

And trivially:

$$F_{r,s,u,v} = \int_{\delta \in \mathbb{R}} N(\delta, 4\lambda n\sigma^2) \int_{w \in \mathbb{R}} \left( \int_{t \in \mathbb{R}} \left( \omega_{r,s}(t) - \frac{w}{nk} \right)^2 N \left( t \right. \right.$$
$$\left. \left. - \frac{w}{k^2}, w\sigma^2 \right) dt \right) P(u,v,w) dw\, d\delta \qquad (l3.6)$$

As before we lower bound the involved Gaussian distributions, for $\in J(\delta, \lambda)$:

$$N \left( t - \frac{w}{k^2}, (w + k^2\delta)\sigma^2 \right) \geq \sqrt{\frac{\lambda n}{w + k^2\delta}} N \left( t - \frac{w}{k^2}, \lambda n\sigma^2 \right) \geq \sqrt{\frac{\lambda}{2}} N \left( t - \frac{w}{k^2}, \lambda n\sigma^2 \right)$$

$$N \left( t - \frac{w}{k^2}, w\sigma^2 \right) \geq \sqrt{\frac{\lambda n}{w}} N \left( t - \frac{w}{k^2}, \lambda n\sigma^2 \right) \geq \sqrt{\lambda} N \left( t - \frac{w}{k^2}, \lambda n\sigma^2 \right)$$

($l3.5$) and ($l3.6$) are thus lower-bounded in respectively:

$$F_{r,s,u,v} \geq \sqrt{\frac{\lambda}{2}} \int_{\delta \in \mathbb{R}} N(\delta, 4\lambda n\sigma^2) \int_{w \in J(\delta, \lambda)} \left( \int_{t \in \mathbb{R}} \left( \omega_{r,s}(t + \delta) - \frac{w}{nk} - \frac{k\delta}{n} \right)^2 N \left( t \right. \right.$$
$$\left. \left. - \frac{w}{k^2}, \lambda n\sigma^2 \right) dt \right) P(u,v,w + k^2\delta) dw\, d\delta \qquad (l3.7)$$

$$F_{r,s,u,v} \geq \sqrt{\lambda} \int_{\delta \in \mathbb{R}} N(\delta, 4\lambda n\sigma^2) \int_{w \in J(\delta, \lambda)} \left( \int_{t \in \mathbb{R}} \left( \omega_{r,s}(t) - \frac{w}{nk} \right)^2 N \left( t \right. \right.$$
$$\left. \left. - \frac{w}{k^2}, \lambda n\sigma^2 \right) dt \right) P(u,v,w) dw\, d\delta \qquad (l3.8)$$

For any $X, Y$ and $a, b > 0$, we have:

$$aX^2 + bY^2 \geq \frac{ab}{(a+b)} (X - Y)^2 \geq \min(a,b) (X - Y)^2$$

By applying this to ($l3.7$) and ($l3.8$) we obtain:

$$F_{r,s,u,v} \geq \sqrt{\frac{\lambda}{2}} \int_{\delta \in \mathbb{R}} N(\delta, 4\lambda n\sigma^2) \int_{w \in J(\delta, \lambda)} \left( \int_{t \in \mathbb{R}} \left( \omega_{r,s}(t + \delta) - \omega_{r,s}(t) - \frac{k\delta}{n} \right)^2 N \left( t \right. \right.$$
$$\left. \left. - \frac{w}{k^2}, \lambda n\sigma^2 \right) dt \right) \min(P(u,v,w), P(u,v,w + k^2\delta)) \, dw\, d\delta \qquad (l3.9)$$

Then we apply a second time the inequality, this time to ($l3.4$) and ($l3.9$) to finally discard $\omega$ in the lower bound of $F_{r,s,u,v}$ :

$$F_{r,s,u,v} \geq \min \left( \lambda, \sqrt{\frac{\lambda}{2}} \right) \int_{\delta \in \mathbb{R}} \frac{k^2\delta^2}{n^2} N(\delta, 4\lambda n\sigma^2) \int_{w \in J(\delta, \lambda)} \min \left( P(u,v,w), P \left( u,v,w \right. \right.$$
$$\left. \left. + \frac{k^2\delta}{2} \right), P(u,v,w + k^2\delta) \right) dw\, d\delta \qquad (l3.10)$$

which ends Part 2.

Part 3.

In this last but not least part, we will lower bound the minimum appearing in ($l3.10$), by showing that $P(u, v, w)$ is slowly varying in $w$, slowly compared to a variation of $k^2\delta$, when probability distributions $\Phi$ and $\Phi'$ verify the separable criteria (e.g. are in $\zeta(\alpha)$), with at least one of them, say $\Phi'$, being sufficiently sleek, as stated in the hypothesis of the lemma.

$\Phi'$ being result of a permutative sleeking transformation, there exists a probability distribution $\Phi'' \in \zeta(\alpha)$ and a permutative sleeking kernel $\gamma$ (that we will carefully choose) such that :

$$\Phi'(y) = \sum_{\sigma \in \mathfrak{S}_n} \gamma(|\sigma|)\Phi'' \circ \sigma(y)$$

And then (restricting $x, y$ to $\{0,1\}^n$ instead of $[0,1]^n$) :

$$P(u, v, w) = \sum_{x,y \in \mathcal{K}(u,v,w)} \Phi(x) \sum_{\sigma \in \mathfrak{S}_n} \gamma(|\sigma|)\Phi'' \circ \sigma(y) = \sum_{x,z \in \mathcal{J}(u)\times\mathcal{J}(v)} \left( \sum_{\substack{\sigma \in \mathfrak{S}_n \\ \sigma(x).z=w}} \gamma(|\sigma|) \right) \Phi(x)\Phi''(z)$$

$$= \sum_{x,z \in \mathcal{J}(u)\times\mathcal{J}(v)} \left( \sum_{l=0}^{n} \gamma(l) \sum_{\substack{\sigma \in \mathfrak{S}_n \\ \sigma(x).z=w \\ |\sigma|=l}} 1 \right) \Phi(x)\Phi''(z)$$

$$= \sum_{x,z \in \mathcal{J}(u)\times\mathcal{J}(v)} \left( \sum_{l=0}^{n} \gamma(l) \sum_{\substack{x' \in \mathcal{J}(u) \\ x'.z=w}} \left( \sum_{\substack{\sigma \in \mathfrak{S}_n \\ \sigma(x)=x' \\ |\sigma|=l}} 1 \right) \right) \Phi(x)\Phi''(z) \quad (l3.11)$$

It is easy to verify that, for any $\sigma, \mu \in \mathfrak{S}_n$ :

$$|\sigma| = |\mu \circ \sigma \circ \mu^{-1}|$$

as shown in the proof of Proposition 11. We set :

$$\theta_u(l, \|x - x'\|^2) = \sum_{\substack{\sigma \in \mathfrak{S}_n \\ \sigma(x)=x' \\ |\sigma|=l}} 1, \qquad \forall x, x' \in \mathcal{J}(u)$$

This writing can be justified thanks to the above by observing that, $\forall \mu \in \mathfrak{S}_n$ :

$$\sum_{\substack{\sigma \in \mathfrak{S}_n \\ \sigma \circ \mu(x)=\mu(x') \\ |\sigma|=l}} 1 = \sum_{\substack{\sigma \in \mathfrak{S}_n \\ \mu^{-1} \circ \sigma \circ \mu(x)=x' \\ |\mu^{-1}\circ\sigma\circ\mu|=l}} 1 = \sum_{\substack{\sigma' \in \mathfrak{S}_n \\ \sigma'(x)=x' \\ |\sigma'|=l}} 1$$

$\sigma \mapsto \mu^{-1} \circ \sigma \circ \mu$ being a bijective transformation in $\mathfrak{S}_n$. The explicit calculation of $\theta_u(l, s)$ can be obtained but is not necessary for the proof. We can then write ($l3.11$) as follows :

$$P(u,v,w) = \sum_{x,z \in \mathcal{J}(u) \times \mathcal{J}(v)} \left( \sum_{l=0}^{n} \gamma(l) \sum_{s=0}^{l} \theta_u(l,s) \left( \sum_{\substack{x' \in \mathcal{J}(u) \\ x'.z = w \\ \|x-x'\|^2 = s}} 1 \right) \right) \Phi(x) \Phi''(z) \qquad (l3.12)$$

The quantity in the internal brackets of the above can be calculated by considering the drawing below :

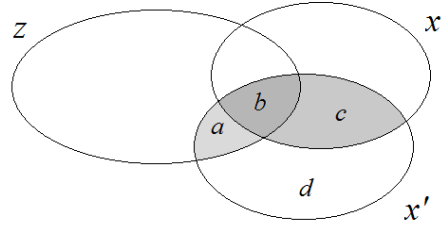

$x'$ can be considered as partitionned in zones $a, b, c, d$, with the following constraints :

$$\begin{cases} 0 \le a = w - b \le v - x.z \\ 0 \le c = u - \dfrac{s}{2} - b \le u - x.z \\ 0 \le d = b + \dfrac{s}{2} - w \le n - u - v + x.z \\ 0 \le b \le x.z \end{cases}$$

Let's denote :

$$b^-(v,w,x.z,s) \triangleq \max\left(0, w - v + x.z, x.z - \frac{s}{2}, w - \frac{s}{2}\right)$$

$$b^+(u,v,w,x.z,s) \triangleq \min\left(x.z, w, u - \frac{s}{2}, n - u - v + x.z + w - \frac{s}{2}\right)$$

$$F_{u,v,w,x.z,s}(b) \triangleq \binom{v - x.z}{w - b} \binom{x.z}{b} \binom{u - x.z}{u - s/2 - b} \binom{n - u - v + x.z}{b + s/2 - w}$$

Then, the counting of the quantity in the internal brackets of $(l3.12)$ gives, with the help of the above drawing :

$$M(u,v,w,x.z,s) \triangleq \sum_{\substack{x' \in \mathcal{J}(u) \\ x'.z = w \\ \|x-x'\|^2 = s}} 1 = \sum_{b = b^-(v,w,x.z,s)}^{b^+(u,v,w,x.z,s)} F_{u,v,w,x.z,s}(b)$$

It can be shown with no particular difficulty that $b^-(v,w,x.z,s) \le b^+(u,v,w,x.z,s)$. The boundaries being non empty imply that :

$$\max(x.z - s/2, w - s/2) \le \min(x.z, w) \Rightarrow s/2 \ge |w - x.z| \qquad (l3.13)$$

Let's set also :

$$K_\gamma(u,s) \triangleq \sum_{l=s}^{n} \gamma(l) \theta_u(l,s)$$

$$\Delta_\gamma(\Phi, \Phi'') = E[(w - x.z)^2] = \sum_{\sigma \in \mathfrak{S}_n} \gamma(|\sigma|) \int_{x,y \in [0,1]^n} (x.y - x.\sigma(y))^2 \Phi(x)\Phi''(y)dxdy$$

$$= \sum_{s=0}^{n} \gamma(s) A_n^s \left( \frac{s^2}{e} \Delta_0(\Phi, \Phi'') + O(n) \right)$$

The last relation coming from Lemma 2.

$P(u, v, w)$ can be expressed with those notations :

$$P(u, v, w) = \sum_{x,z \in \mathcal{J}(u) \times \mathcal{J}(v)} \sum_{s=0}^{n} K_\gamma(u, s) M(u, v, w, x.z, s) \Phi(x) \Phi''(z)$$

Injecting this relation into $(l3.10)$, and denoting :

$$M_p = M_p(u, v, w, x.z, s) \triangleq M\left( u, v, w + p\frac{k^2\delta}{2}, x.z, s \right), \ p \in \mathbb{N}$$

we obtain the intermediary lower bound (from $(l3.2')$ and $(l3.10)$) :

$$\langle \omega, \Phi, \Phi' \rangle \geq \ C_3 \min\left( \lambda, \sqrt{\frac{\lambda}{2}} \right) \sum_{\substack{u,v,w \in \mathbb{N}_n \\ w \in J(\lambda)}} \sum_{x,z \in \mathcal{J}(u) \times \mathcal{J}(v)} \sum_{s=0}^{n} K_\gamma(u, s) \left( \int_{|\delta| \leq \lambda n/k^2} \frac{k^2\delta^2}{n^2} N(\delta, \right.$$

$$4\lambda n\sigma^2) \min(M_0, M_1, M_2) \, d\delta \right) \Phi(x)\Phi''(z) \tag{$l3.14$}$$

where $\hat{J}(\lambda) = \{w | w \geq 3\lambda n\}$ ; indeed, if $|\delta| \leq \lambda n/k^2$, then $J(\delta, \lambda) \supset \hat{J}(\lambda)$.

We will now concentrate on min $(M_0, M_1, M_2)$, to discard the minimum with lower bound by a constant under the hypothesis of the Lemma.

We set the following conditions under simplex form (their role will appear later) :

$$(\Pi) \begin{cases} P_0(\lambda) = \{w | w \geq 3\lambda n\} & (1) \\[4pt] P_1(\theta) = \left\{ w, x, z \, | (w - x.z)^2 \geq \theta^2 \Delta_\gamma \right\} & (2) \\[4pt] P_2(\varepsilon) = \left\{ u, v, w, x, z, s \, \left| \left( x.z - w - \frac{s}{2} \left( \frac{x.z}{u} - \frac{v - x.z}{n - u} \right) \right)^2 \leq \varepsilon \Delta_\gamma \right. \right\} & (3) \\[4pt] P_3(\theta') = \left\{ u, v, w, x, z \, \left| \min\left( P_{u,v}\left( \frac{w}{n} \right), P_{u,v}\left( \frac{x.z}{n} \right) \right) \geq \theta' \right. \right\} & (4) \\[4pt] P_4(n\varepsilon', \theta'') = \left\{ u, v, w, x, z \, \left| \eta\left( u, v, x.z, \frac{n\varepsilon'}{2} \right) \geq \theta'' \right. \right\} & (5) \end{cases}$$

where :

$$\eta(u, v, x.z, s/2) \triangleq \min\left( \frac{(u - s/2)(u - x.z)x.z}{u^3}, \frac{(n - u - s/2)(v - x.z)(n - u - v + x.z)}{(n - u)^3} \right)$$

From $(l3.14)$, we can write the weaker lower bound :

$$\langle \omega, \Phi, \Phi' \rangle \geq C_3 \min\left(\lambda, \sqrt{\frac{\lambda}{2}}\right) \sum_{\substack{u,v,w \in \mathbb{N}_n \\ x,z \in \mathcal{J}(u) \times \mathcal{J}(v) \\ P_0, P_1, P_2, P_3, P_4}} \sum_{s=0}^{n} K_\gamma(u,s) \left( \int_{|\delta| \leq \lambda n/k^2} \frac{k^2 \delta^2}{n^2} N(\delta, \right.$$

$$\left. 4\lambda n \sigma^2 \right) \min(M_0, M_1, M_2) \, d\delta \right) \Phi(x) \Phi''(z) \qquad (l3.15)$$

The condition $P_3$ comes from the examination of the distance between the unique pick of $F_{u,v,w,x.z,s}(b)$ and the boundaries of the summing $b^-(v, w, x.z, s)$ and $b^+(u, v, w, x.z, s)$. The extrema of $F$ verify (with error lesser than a unit) :

$$\frac{(w-b)(x.z-b)(u-b-\frac{s}{2})(n-u-v+x.z+w-b-\frac{s}{2})}{b(v-x.z-w+b)(b+\frac{s}{2}-x.z)(b+\frac{s}{2}-w)} = 1 \qquad (l3.16)$$

obtained by writing $\frac{F_{u,v,w,x.z,s}(b+1)}{F_{u,v,w,x.z,s}(b)} = 1$.

The boundaries of the summing $M$ imply that :

$$b \in \left[ w - \frac{s}{2}, w \right] \cap \left[ x.z - \frac{s}{2}, x.z \right]$$

One observes that there exists only one unique zero of ($l3.16$) within the boundaries. Indeed, the numerator is a decreasing function within the boundaries, because its four monomials are positive and decreasing within. On the other hand the numerator is increasing because its four monomials are positive and increasing within. Thus there exists a unique zero. In addition, that zero is a maximum of $F$ because $F$ is increasing at righthand neighborhood of the lower boundary $b^-$ (the fraction is there $+\infty > 1$).

Let $b_0$ be the maximum. It can be shown without any particular difficulty (we won't present the useless details here) that :

$$\min_{\epsilon \in \{+,-\}} d(b_0, b^\epsilon) \geq \frac{s}{16} \min\left( R_{u,v}(w, x.z), R_{u,v}(x.z, w), R_{u,v}^{-1}(w, x.z), R_{u,v}^{-1}(x.z, w) \right)$$

$$(l3.16')$$

where :

$$R_{u,v}(X, Y) \triangleq \frac{(u-Y)(n-u-v+X)}{Y(v-X)}$$

It is easy to remark that :

$$P_3(\theta') \Rightarrow \min\left( R_{u,v}(w, x.z), R_{u,v}(x.z, w), R_{u,v}^{-1}(w, x.z), R_{u,v}^{-1}(x.z, w) \right) \geq \theta'$$

The strategy for examining $\min(M_0, M_1, M_2)$ is to put $F$ under the shape :

$$F(b) = G(b - \rho w) H(w)(1 + o(1))$$

where $G$ is a gaussian distribution with $\rho$ independant from $b$ and $w$, and $H$ is a gaussian distribution whose filter's width is large compared to $k^2\delta \sim k\sqrt{\lambda n}$. Indeed then $H$ varies slowly by translation of lengh $k^2\delta$ and a variation of $w$ in $G$ causes an indicial translation of length $\sim \rho k^2 \delta$ in the summing in $M$, and this length can be shown as remaining small compared to $\min_{\epsilon \in \{+,-\}} d(b_0, b^\epsilon)$.

To that extent, we consider the following grouping in :

$$F_{u,v,w,x,z,s}(b) = \left( \binom{x.z}{b}\binom{u-x.z}{u-s/2-b} \right)\left( \binom{v-x.z}{w-b}\binom{n-u-v+x.z}{b+s/2-w} \right)$$

Let's first remark that, for any integers $m, n, p, q, t$, such that :

$$\max(-m+p, -q) \leq t \leq \min(p, n)$$

we have, $\forall \zeta \in \,]0,1[\,$ :

$$\binom{m}{p-t}\binom{n}{q+t} = \frac{\zeta^{p-t}(1-\zeta)^{m-p+t}\zeta^{q+t}(1-\zeta)^{m-q-t}}{\zeta^{p+q}(1-\zeta)^{m+n-(p+q)}}\binom{m}{p-t}\binom{n}{q+t}$$

$$= \frac{1}{\zeta^{p+q}(1-\zeta)^{m+n-(p+q)}}\beta_{p-t,m}(\zeta)\beta_{q+t,n}(\zeta)$$

We choose $\zeta$, in such a manner that the picks in $t$ of the above binomial distributions coincide in a common pick in $t_0$. This leads to :

$$\begin{cases} p - t_0 = \alpha m \\ q + t_0 = \alpha n \end{cases}$$

which results in $\zeta = \frac{p+q}{m+n}$ and $t_0 = \frac{np-mq}{m+n}$.

Then, from Proposition 2 (ii) at the neighborhood of $t_0$ we get :

$$\binom{m}{p-t}\binom{n}{q+t} \sim \frac{1}{2\pi\zeta^{p+q+1}(1-\zeta)^{m+n-(p+q)-1}\sqrt{mn}}e^{-\frac{m+n}{2\zeta(1-\zeta)mn}\left(t-\frac{np-mq}{m+n}\right)^2}$$

We apply this approximation to both grouping in $F$, and we denote $b_1$ and $b_2$ the respective picks of both grouping. We have :

$$b_1 - b_2 = x.z - w - \frac{s}{2}\left(\frac{x.z}{u} - \frac{v-x.z}{n-u}\right)$$

We determine now the permutative sleeking kernel as follows :

$$\gamma(l) \triangleq \frac{1}{A_n^l}\Gamma(\varepsilon'n, l)$$

where $\Gamma(a,b) = \begin{cases} 1 \text{ if } a = b \\ 0 \text{ if } a \neq b \end{cases}$. By using the same technique than in Lemma 2 to perform the calculations, we obtain (calculations are a bit painful but present no higher difficulty than in Lemma 2, it is required that $x, z \in \{0,1\}^n$ to get the result) that there exists a constant $C_4 > 0$ such that :

$$E\left[\left(x.z - w - \frac{s}{2}\left(\frac{x.z}{u} - \frac{v-x.z}{n-u}\right)\right)^2\right] = \lhd \left(C_4 \varepsilon'^2 n\right) \qquad (l3.17)$$

This important relation justifies the choice of condition $P_2$, that thus implies :

$$(b_1 - b_2)^2 \leq \varepsilon \Delta_\gamma$$

$\varepsilon$ is chosen such that :

$$C_4 \varepsilon'^2 n \ll \varepsilon \Delta_\gamma \ll \theta \theta'' \sqrt{\Delta_\gamma} \qquad (l3.17')$$

which can be made possible (see discussion on parameters at the end of the proof), and enables us to obtain, from Proposition 2, the simultaneous validity of the Gaussian approximations for the two grouping within $F$. Indeed, the filter's width for them are respectively :

$$\begin{cases} \dfrac{s(u-s/2)(u-x.z)x.z}{u^3} & (\triangleq a^{-1}) \\ \dfrac{s(n-u-s/2)(v-x.z)(n-u-v+x.z)}{(n-u)^3} & (\triangleq a'^{-1}) \end{cases}$$

that are both of superior to $\theta \theta'' \sqrt{\Delta_\gamma}$ under condition $P_1(\theta)$ and $P_4(\theta'')$.

We then use the identity :

$$a(t-c)^2 + a'(t-c')^2 = (a+a')\left(t - \frac{ac+a'c'}{a+a'}\right)^2 + \frac{aa'(c-c')^2}{a+a'}$$

to reconstruct the two Gaussian under the shape :

$$G(b - \rho w)H(w)$$

We find :

$$\rho = \left(1 + \frac{u^3(n-u-s/2)(v-x.z)(n-u-v+x.z)}{(n-u)^3(u-s/2)(u-x.z)x.z}\right)^{-1}$$

In addition, the filter's width of $H$ equals to :

$$\left(a^{-1} + a'^{-1}\right) = s\left(\frac{(u-s/2)(u-x.z)x.z}{u^3} + \frac{(n-u-s/2)(v-x.z)(n-u-v+x.z)}{(n-u)^3}\right)$$

$$\geq 2\eta(u,v,x.z,s/2)\theta\sqrt{\Delta_\gamma}$$

due to $(l3.13)$ and (for last inequality above) under condition $P_1$.

As $(c-c')^2 = (b_1 - b_2)^2 \leq \varepsilon \Delta_\gamma \ll \theta \theta'' \sqrt{\Delta_\gamma}$, we have :

$$H(w) = H(w,u,v,x.z,s) = H_0(u,v,x.z,s)e^{-\frac{aa'(c-c')^2}{a+a'}} \in \left[H_0 e^{-\frac{1}{2}}, H_0\right]$$

When we apply the indicial translation of length $\sim \rho k^2\delta \sim \rho k\sqrt{\lambda n}$ in the summing in $M$ (corresponding to the variation of $w$ in $M_0, M_1, M_2$), it remains very small compared the distance from the pick to the boundaries, indeed, under $P_1$ and $P_3$ (and considering $(l3.13)$ and $(l3.16')$) :

$$\min_{\epsilon \in \{+,-\}} d(b_0, b^\epsilon) \geq \frac{1}{8}\theta\theta'\sqrt{\Delta_\gamma} \triangleq D$$

We must choose $\varepsilon'$ so that :

$$k\sqrt{\lambda n} \ll \theta\theta'\sqrt{\Delta_\gamma} \qquad (l3.17'')$$

and remaining compatible with $(l3.17')$.

From Proposition 14 applied to the sub-part $G$ of $F_{u,v,w,x.z,s}(b)$ within the boundaries of the summing in $M$ we obtain, for $\in \{0,1,2\}$ :

$$\sum_{b=b^-(w+p\frac{k^2\delta}{2})}^{b^+(w+p\frac{k^2\delta}{2})} G\left(b - \rho\left(w + p\frac{k^2\delta}{2}\right)\right) \geq \left(1 - \frac{(1+\rho)k^2\delta}{D}\right)\sum_{b=b^-(w)}^{b^+(w)} G(b - \rho w)$$

And finally :

$$M_p \geq e^{-\frac{1}{2}}\left(1 - \frac{(1+\rho)k^2\delta}{D}\right)M_0 \geq e^{-\frac{1}{2}}\left(1 - \frac{2k^2\delta}{D}\right)M_0$$

which also means that :

$$\min(M_0, M_1, M_2) \geq e^{-\frac{1}{2}}\left(1 - \frac{2k^2\delta}{D}\right)M_0 \qquad (l3.18)$$

We note that, by choice of kernel $\gamma$, we have necessarily $s \leq n\varepsilon'$ (because otherwise $K_\gamma(u,s) = 0$), and thus $P_4(n\varepsilon', \theta'') \Rightarrow P_4(s, \theta'')$.

This ends Part 3.

In order to finish the proof, we will now deal with the conditions $P_0, \dots, P_4$ and discuss the values of the parameters $\varepsilon, \varepsilon', \theta, \theta', \theta'', \lambda$.

First, we can inject $(l3.18)$ in $(l3.15)$ to obtain :

$$\langle \omega, \Phi, \Phi'\rangle \geq \ C_3 \min\left(\lambda, \sqrt{\frac{\lambda}{2}}\right)e^{-\frac{1}{2}}\left(\int_{|\delta|\leq\lambda n/k^2}\left(1 - \frac{2k^2\delta}{D}\right)\frac{k^2\delta^2}{n^2}N(\delta,\right.$$

$$\left. 4\lambda n\sigma^2)d\delta\right)\sum_{\substack{u,v,w\in\mathbb{N}_n \\ x,z\in\mathcal{J}(u)\times\mathcal{J}(v) \\ P_0,P_1,P_2,P_3,P_4}}\sum_{s=0}^{n}K_\gamma(u,s)M(u,v,w,x.z,s)\,\Phi(x)\Phi''(z)$$

$$(l3.19)$$

We have :

$$\sum_{\substack{u,v,w\in\mathbb{N}_n \\ x,z\in\mathcal{J}(u)\times\mathcal{J}(v) \\ P_0,P_1,P_2,P_3,P_4}} \sum_{s=0}^{n} K_\gamma(u,s)M(u,v,w,x.z,s)\,\Phi(x)\Phi''(z)$$

$$\geq \sum_{\substack{u,v,w\in\mathbb{N}_n \\ x,z\in\mathcal{J}(u)\times\mathcal{J}(v) \\ P_0,P_1,P_3,P_4}} \sum_{s=0}^{n} K_\gamma(u,s)M(u,v,w,x.z,s)\,\Phi(x)\Phi''(z)$$

$$-\sum_{\substack{u,v,w\in\mathbb{N}_n \\ x,z\in\mathcal{J}(u)\times\mathcal{J}(v) \\ \overline{P_2}}} \sum_{s=0}^{n} K_\gamma(u,s)M(u,v,w,x.z,s)\,\Phi(x)\Phi''(z) \qquad (l3.19')$$

We can thus deal with $P_0, P_1, P_3, P_4$ all together on one hand, and $P_2$ (that is the only one of the 5 conditions depending on $s$) on the other hand. As already explained above, the choice of kernel $\gamma$ gives :

$$K_\gamma(u,s) = \begin{cases} \theta_u(n\varepsilon',s) & \text{if } s \leq n\varepsilon' \\ 0 & \text{otherwise} \end{cases}$$

We can first remark the following dependence between the conditions $P_0, P_1, P_3, P_4$. As already noticed :

$$P_3(\theta') \Rightarrow \min\left(R_{u,v}(w,x.z), R_{u,v}(x.z,w), R_{u,v}{}^{-1}(w,x.z), R_{u,v}{}^{-1}(x.z,w)\right) \geq \theta'$$

In addition, still under $P_3(\theta')$, it can easily be seen that :

$$\eta(u,v,x.z,n\varepsilon') \geq \theta'^2 - \varepsilon'$$

which implies that,

$$\text{if } \theta'^2 - \varepsilon' \geq \theta'', \text{ then } P_3(\theta') \Rightarrow P_4(n\varepsilon',\theta''). \qquad (l3.20)$$

At last, under $P_1(\theta)$ and $P_3(\theta')$, $\frac{x.z}{n} \geq \theta' \Rightarrow \frac{w}{n} \geq \theta' - \left|\frac{w}{n} - \frac{x.z}{n}\right| \geq \theta' - \theta\frac{\sqrt{\Delta_\gamma}}{n}$

which implies that,

$$\text{if } \theta' - \theta\frac{\sqrt{\Delta_\gamma}}{n} \geq \lambda, \text{ then } P_1(\theta) \,\&\, P_3(\theta') \Rightarrow P_0(\lambda). \qquad (l3.20')$$

This means that under conditions $(l3.20)$ and $(l3.20')$ on $\varepsilon', \theta, \theta', \theta'', \lambda$, we have :

$$\sum_{\substack{u,v,w \in \mathbb{N}_n \\ x,z \in \mathcal{J}(u) \times \mathcal{J}(v) \\ P_0, P_1, P_3, P_4}} \sum_{s=0}^{n} K_\gamma(u,s) M(u,v,w,x.z,s) \, \Phi(x) \Phi''(z)$$

$$= \sum_{\substack{u,v,w \in \mathbb{N}_n \\ x,z \in \mathcal{J}(u) \times \mathcal{J}(v) \\ P_1, P_3}} \sum_{s=0}^{n} K_\gamma(u,s) M(u,v,w,x.z,s) \, \Phi(x) \Phi''(z)$$

$$= \sum_{\substack{u,v \in \mathbb{N}_n \\ x,z \in \mathcal{J}(u) \times \mathcal{J}(v) \\ P_3}} \sum_{\substack{w \in \mathbb{N}_n \\ (w-x.z)^2 \geq \theta^2 \Delta_\gamma}} \sum_{\substack{\sigma \in \mathfrak{S}_n \\ \sigma(x).z = w}} \gamma(|\sigma|) \, \Phi(x) \Phi''(z)$$

$$= \sum_{\substack{u,v \in \mathbb{N}_n \\ x,z \in \mathcal{J}(u) \times \mathcal{J}(v) \\ P_3}} \sum_{\substack{\sigma \in \mathfrak{S}_n \\ (\sigma(x).z - x.z)^2 \geq \theta^2 \Delta_\gamma}} \gamma(|\sigma|) \, \Phi(x) \Phi''(z)$$

$$\geq \sum_{\substack{u,v \in \mathbb{N}_n \\ x,z \in \mathcal{J}(u) \times \mathcal{J}(v) \\ P_3}} \sum_{\substack{\sigma \in \mathfrak{S}_n \\ (\sigma(x).z - x.z)^2 \geq \theta^2 \Delta_\gamma}} \frac{4(\sigma(x).z - x.z)^2}{n^2 \varepsilon'^2} \gamma(|\sigma|) \, \Phi(x) \Phi''(z)$$

$$\geq \sum_{\substack{u,v \in \mathbb{N}_n \\ x,z \in \mathcal{J}(u) \times \mathcal{J}(v) \\ P_3}} \sum_{\substack{\sigma \in \mathfrak{S}_n \\ (\sigma(x).z - x.z)^2 \geq \theta^2 \Delta_\gamma}} \min\left( P_{u,v}\left( \frac{\sigma(x).z}{n} \right), P_{u,v}\left( \frac{x.z}{n} \right) \right) \frac{4(\sigma(x).z - x.z)^2}{n^2 \varepsilon'^2} \gamma(|\sigma|) \, \Phi(x) \Phi''(z)$$

$$- \theta' \sum_{\substack{u,v \in \mathbb{N}_n \\ x,z \in \mathcal{J}(u) \times \mathcal{J}(v) \\ (\sigma(x).z - x.z)^2 \geq \theta^2 \Delta_\gamma}} \sum_{\sigma \in \mathfrak{S}_n} \frac{4(\sigma(x).z - x.z)^2}{n^2 \varepsilon'^2} \gamma(|\sigma|) \, \Phi(x) \Phi''(z)$$

$$\geq \sum_{\substack{u,v \in \mathbb{N}_n \\ x,z \in \mathcal{J}(u) \times \mathcal{J}(v)}} \sum_{\sigma \in \mathfrak{S}_n} P_{u,v}\left( \frac{\sigma(x).z}{n} \right) P_{u,v}\left( \frac{x.z}{n} \right) \frac{4(\sigma(x).z - x.z)^2}{n^2 \varepsilon'^2} \gamma(|\sigma|) \, \Phi(x) \Phi''(z) - \frac{4\theta^2 \Delta_\gamma}{n^2 \varepsilon'^2}$$

$$- \theta' \sum_{\substack{u,v \in \mathbb{N}_n \\ x,z \in \mathcal{J}(u) \times \mathcal{J}(v)}} \sum_{\substack{\sigma \in \mathfrak{S}_n \\ (\sigma(x).z - x.z)^2 \geq \theta^2 \Delta_\gamma}} \frac{4(\sigma(x).z - x.z)^2}{n^2 \varepsilon'^2} \gamma(|\sigma|) \, \Phi(x) \Phi''(z)$$

where the first inequality comes from the fact that $2|w - x.z| \leq s \leq n\varepsilon'$ as expressed in ($l3.13$). In the second inequality, the first term can be lower-bounded thanks to Corollary 1, and the third term can be upper bounded by $4\theta'$.

To deal with the first term, we replace $\Phi''$ by $\Phi'''$ as follows :

$$\Phi''(y) = \sum_{\sigma \in \mathfrak{S}_n} \delta(|\sigma|) \Phi''' \circ \sigma(y)$$

where $\delta$ is the permutative sleeking transformation introduced in Corollary 1.

We thus get then from Corollary 1:

$$\sum_{\substack{u,v,w \in \mathbb{N}_n \\ x,z \in \mathcal{J}(u) \times \mathcal{J}(v) \\ P_0, P_1, P_3, P_4}} \sum_{s=0}^{n} K_\gamma(u,s) M(u,v,w,x.z,s)\, \Phi(x)\Phi''(z)$$

$$\geq K(\alpha) - \frac{\varepsilon'}{\sqrt{e}} - O\left(\frac{1}{\sqrt{n}}\right) - \frac{4\theta^2 \Delta_\gamma}{n^2 \varepsilon'^2} - 4\theta' \qquad (l3.21)$$

It remains now to upper bound the term conditioned by $P_2$ as appearing in $(l3.19')$, this can be easily done thanks to Markov inequality :

$$\sum_{\substack{u,v,w \in \mathbb{N}_n \\ x,z \in \mathcal{J}(u) \times \mathcal{J}(v) \\ \overline{P_2}}} \sum_{s=0}^{n} K_\gamma(u,s) M(u,v,w,x.z,s)\, \Phi(x)\Phi''(z)$$

$$= \sum_{\substack{u,v,w \in \mathbb{N}_n \\ x,z \in \mathcal{J}(u) \times \mathcal{J}(v) \\ \overline{P_2}}} \sum_{s=0}^{n\varepsilon'} \theta_u(n\varepsilon',s) M(u,v,w,x.z,s)\, \Phi(x)\Phi''(z)$$

$$\leq \sum_{\substack{u,v,w \in \mathbb{N}_n \\ x,z \in \mathcal{J}(u) \times \mathcal{J}(v) \\ \overline{P_2}}} \sum_{s=0}^{n\varepsilon'} \frac{\left(x.z - w - \frac{s}{2}\left(\frac{x.z}{u} - \frac{v - x.z}{n-u}\right)\right)^2}{\varepsilon \Delta_\gamma} \theta_u(n\varepsilon',s) M(u,v,w,x.z,s)\, \Phi(x)\Phi''(z)$$

$$\leq \frac{C_4 \varepsilon'^2 n}{\varepsilon \Delta_\gamma}$$

where last inequality comes from $(l3.17)$ ; we know that $\frac{C_4 \varepsilon'^2 n}{\varepsilon \Delta_\gamma} \ll 1$ from $(l3.17')$.

As long as $\lambda \gg \frac{k^2}{n}$, we can write :

$$\int_{|\delta| \leq \lambda n/k^2} \frac{k^2 \delta^2}{n^2} N(\delta, 4\lambda n\sigma^2) d\delta = \int_{\mathbb{R}} \frac{k^2 \delta^2}{n^2} N(\delta, 4\lambda n\sigma^2) d\delta - \triangleleft \left(e^{-\frac{\lambda n}{8k^2}}\right) = \frac{4\lambda}{n}\left(1 - \frac{1}{k^2}\right) -$$

$$\triangleleft \left(e^{-\frac{\lambda n}{8k^2}}\right)$$

Finally, we get from $(l3.19)$, $(l3.19')$, $(l3.21)$ and the above :

$$\langle \omega, \Phi, \Phi' \rangle \geq C_3 \min\left(\lambda, \sqrt{\frac{\lambda}{2}}\right) e^{-\frac{1}{2}}\left(1 - \frac{2k\sqrt{\lambda n}}{D}\right)\left(\frac{4\lambda}{n}\left(1 - \frac{1}{k^2}\right) - e^{-\frac{\lambda n}{8k^2}}\right)\left(K(\alpha) - \frac{\varepsilon'}{\sqrt{e}} - O\left(\frac{1}{\sqrt{n}}\right)\right.$$

$$\left. - \frac{4\theta^2 \Delta_\gamma}{n^2 \varepsilon'^2} - 4\theta' - \frac{C_4 \varepsilon'^2 n}{\varepsilon \Delta_\gamma}\right) \qquad (l3.22)$$

And we obtained from Lemma 2 and Corollary 1 the lower/upper bound for $\Delta_\gamma$ :

$$\frac{n^2\varepsilon'^2}{e} + O(n) \geq \Delta_\gamma = \sum_{s=0}^{n} \gamma(s) A_n^s \left( \frac{s^2}{e} \Delta_0(\Phi, \Phi'') + O(n) \right) \geq \frac{n^2\varepsilon'^2}{e} \left( \Delta_0(\Phi, \Phi''') - 2l_0(\alpha) \right) + O(n)$$

$$\geq \frac{n^2\varepsilon'^2}{e} \left( \frac{\alpha^2}{16} - 2l_0(\alpha) \right) + O(n)$$

This inequality $(l3.22)$, that gives the result of Lemma 3, can only stand if the following simplex conditions can be satisfied $((l3.17'), ((l3.17''), (l3.20), (l3.20'))$ :

$$(S): \begin{cases} \varepsilon' \ll K(\alpha) & (1) \\[2mm] \dfrac{1}{\sqrt{n}} \ll K(\alpha) & (2) \\[2mm] \theta^2 \ll K(\alpha) & (3) \\[2mm] \theta' \ll K(\alpha) & (4) \\[2mm] n^2\varepsilon'^2 \gg n & (5) \\[2mm] \dfrac{n\varepsilon'^2}{\varepsilon\Delta_\gamma} \ll K(\alpha) & (6) \\[2mm] \theta'^2 - \varepsilon' \geq \theta'' & (7) \\[2mm] \theta' - \theta\dfrac{\sqrt{\Delta_\gamma}}{n} \geq \lambda & (8) \\[2mm] k\sqrt{\lambda n} \ll \theta\theta'\sqrt{\Delta_\gamma} & (9) \\[2mm] \varepsilon\sqrt{\Delta_\gamma} \ll \theta\theta''\sqrt{\Delta_\gamma} & (10) \end{cases}$$

Replacing $\Delta_\gamma$ by its lower/upper bound above, it can be easily seen that those conditions can be satisfied. $\theta, \theta', \theta''$ can be chosen to satisfy (3), (4) and (7) (with $\varepsilon'$ small compared to them), then $\lambda$ is chosen to satisfy (8), then $\varepsilon'$ is chosen sufficiently small so that $\varepsilon$ can exist (it is sufficient that $\varepsilon' \ll \theta\theta'' \left( \frac{\alpha^2}{16} - 2l_0(\alpha) \right) K(\alpha)$ by gathering (6) and (10), thanks to which $\varepsilon$ can be found in $\frac{1}{n\left(\frac{\alpha^2}{16} - 2l_0(\alpha)\right)K(\alpha)} \ll \varepsilon \ll \frac{\theta\theta''}{n\varepsilon'}$), and finally $n$ is chosen sufficiently large to match (2), (5), (8) and (9).

The permutative sleeking transformation to adopt is (with notation of Corollary 1) :

$$\Phi' = T_\gamma \circ T_\delta[\Phi''']$$

and Proposition 11 confirms that such composition is still a permutative sleeking transformation.

<div align="right">■</div>

The Synchronized Strategy Lemma enables us to get the following general result:

**Theorem 3.**

*Let $\Phi$ be a sleeked distribution (under the meaning of Proposition 12 in Annex 1) in $\zeta(\alpha)$, then:*

*(i)     For any strategy $\omega$, there exist $\sigma_\omega \in \mathfrak{S}_n$ such that $\langle \omega, \Phi \circ \sigma_\omega \rangle \geq \frac{c_3}{n}$*

*(ii)    For any strategy $\omega$, $\frac{1}{n!}\sum_{\sigma \in \mathfrak{S}_n} \langle \omega, \Phi \circ \sigma \rangle \geq \frac{c_3}{n}$*

∎

(iii) is an immediate consequence of the Synchronized Strategy Lemma:

$$\frac{1}{n!}\sum_{\sigma \in \mathfrak{S}_n} \langle \{\omega_{i,j}\}, \Phi \circ \sigma \rangle = \frac{1}{n!}\sum_{\sigma \in \mathfrak{S}_n} \langle \{\omega_{\sigma(i),\sigma(j)}\}, \Phi \rangle \geq \langle \left\{ \frac{1}{n!}\sum_{\sigma \in \mathfrak{S}_n} \omega_{\sigma(i),\sigma(j)} \right\}, \Phi \rangle$$

And because $\frac{1}{n!}\sum_{\sigma \in \mathfrak{S}_n} \omega_{\sigma(i),\sigma(j)} \in \Omega_\#$, the right hand term of the above inequality can be lower bounded by $\frac{c_3}{n}$ as stated by Lemma 3.

And (ii) is an immediate consequence of (iii) as the average summing is lesser than the maximum element.

∎